\documentclass{unswthesis}
\usepackage{mythesis}
\begin{document}
\frontmatter
\maketitle
\newpage

\begin{abstract}
Physical exercise has significant benefits for humans in improving the health and quality of their lives, by improving the functional performance of their cardiovascular and respiratory systems. However, it is very important to control the workload, e.g. the frequency of body movements, within the capability of the individual to maximise the efficiency of the exercise. The workload is generally represented in terms of heart rate (HR) and oxygen consumption (VO$_2$). We focus particularly on the control of HR and VO$_2$ using the workload of an individual body movement, also known as the exercise rate (ER), in this research.

The first part of this report deals with the modelling and control of HR during an unknown type of rhythmic exercise. A novel feature of the developed system is to control HR via manipulating ER as a control input. The relation between ER and HR is modelled using a simple autoregressive model with unknown parameters. The parameters of the model are estimated using a Kalman filter and an indirect adaptive H$_\infty$ controller is designed. The performance of the system is tested and validated on six subjects during rowing and cycling exercise. The results demonstrate that the designed control system can regulate HR to a predefined profile.

The second part of this report deals with the problem of estimating VO$_2$ during rhythmic exercise, as the direct measurement of VO$_2$ is not realisable in these environments. Therefore, easy-to-use sensors are used to non-invasively measure HR, RespR, and ER to estimate VO$_2$. The developed approach for cycling and rowing exercise predicts the percentage change in maximum VO$_2$ from the resting to the exercising phases, using a Hammerstein model. These VO$_2$ estimators were validated on six subjects by comparing the measured and estimated values of VO$_2$ for quality of fit. Results show that the average quality of fit in both exercises is improved as the intensity of exercise is increased. This shows that the relation between ER, RespR, HR and VO$_2$ is highly correlated during high-intensity exercise. Consequently, a self-biofeedback (SBF) control of VO$_2$ is implemented in real time and the experimental results show the effectiveness of the proposed approach. Moreover, the efficiency of the SBF system is also analysed to show that it is more efficient in controlling VO$_2$ during low-intensity exercise.

This research can facilitate the efficacy of rhythmic exercise for gaited and cardiovascular patients.
\end{abstract}

\newpage

\tableofcontents
\mainmatter

\chapter{Materials and Methods}\label{ch:mm}
\section{Overview}
This Chapter presents the materials and methods used throughout this research work, in its Section \ref{equip}. Section \ref{method} introduces the methodologies that are used to carry out this study.
\section{Equipment}\label{equip}
A brief description of the hardware and software involved in the proposed project is given in the following subsections.
The following equipments were used in the experimental study.
\begin{enumerate}
\item{Treadmill machine: T-Track Tunturi Model Gamma 300}
\item{Cycling Ergo-meter:Tunturi Stationary Recumbent 6}
\item{Rowing machine: Concept PM3}
\item{Computer and Software}
\item{Heart Polar Belt: TM 310 and its receiver}
\item{Wireless Triaxial Accelerometer}
\item{Respiratory Belt: MLT1132}
\item{Gas Analyser: COSMED Kb4}
\item{National Instrument data acquisition system: DAQ62E}
\end{enumerate}
\subsection{Treadmill machine: T-Track Tunturi Model Gamma 300}
A commercial motor driven treadmill (T-Track Tunturi Model Gamma 300) was utilised in this research.
 The treadmill running surface is 150 cm long by 50 cm. The treadmill features speeds of 0.5 - 16km/hr with 0.1 km/hr increments and gradients (or slope) of 0 - 10\% with 1\% increments. Parallel handrails are attached to the treadmill's front vertical beam, on which a digital control panel is located.

The treadmill is equipped with an RS-232C serial port interface through which connection to the personal computer (PC) is integrated for the computer controlled system.
\subsection{Cycling Ergo-meter:Tunturi Stationary Recumbent 6}
A commercial cycling ergo-meter (Tunturi Stationary Recumbent 6) was utilised in this research. Perfectly suitable for recuperation and gives a comfortable workout by giving good support in the back. The cycling ergo-meter $length \times width \times height$:$1872 \times 640 \times 727$mm ($73.7 \times 25.2 \times 28.6$"), this allows maximum exerciser weight approx. 180kg (400lbs). Parallel handstands are attached to the cycling ergo-meter, and a digital control panel is located at the front of the cycling ergo-meter seat.

The cycling ergo-meter is equipped with an RS-232C serial port interface through which connection to the computer station is integrated for the computer controlled system.
\subsection{Rowing machine: Concept2 PM3}
A commercial indoor rower (Concept2 PM3) was utilised in this research. The ergonomic equips with digital monitor.
This digital display monitor provides the information about the exercising rate in terms of stroke/min, exercise duration. Its handle allows for a natural arm and hand position while rowing. Its spiral damper controls airflow to the flywheel allowing exercisers to choose the feel of a sleek, fast boat or a slow, heavy boat.

The treadmill is equipped with an RS-232C serial port interface through which connection to the personal computer (PC) is integrated for the computer controlled system.
\subsection{Computer and Software}\label{Csoftware}
A PC was the main interface with all the experimental apparatus for a computer controlled system. Matlab and Simulink were standard software tools that were used for analysis, design and simulations. The LabVIEW software from well-known National Instrument was used for data acquisition and implementation in conjunction with Matlab/Simulink to acquire real-time data.
\subsection{Heart Polar Belt: TM-31 and R-to-R receiver: PN-1185}
The R-to-R pulses are detected and transmitted using TM-310 Polar belt transmitter.  Subject requires wearing the Polar belt while he/she is engaging in exercising and resting phase, in order to measure HR at the receiver end.

PN-1185 is used as a receiver for TM-310 Polar belt, it receives the R-to-R pulses in terms of T-31 coded HR data in order to achieve the HR measurement.
\subsection{Triaxial Accelerometer}
TA is used to detect the accelerations along x-, y- and z- axes. The TA converts the continuous time data into discrete time data, this discrete data thereafter encoded into binary and transform into the data packet. These binary coded data packets are transmitted with the sampling frequency of 50-Hz and receive via blue-tooth receiver at the computer station.
\subsection{Respiratory Belt MLT1130}
The MLT1132 Respiratory Belt Transducer contains a piezo-electric device that responds linearly to changes in length. It measures changes in thoracic or abdominal circumference during respiration. These measurements can indicate inhalation, expiration and breathing strength and can be used to derive breathing rate. The transducer may be used to characterize breathing patterns. The transducer is a solid-state device that requires no excitation. The MLT1132 connects directly to a BNC Input on any PowerLab or MatLab , and NI data acquisition system. It is comfortable to wear, rugged, reliable and washable. It is suitable for use on animals of various sizes as well as humans.
\subsection{COSMED KB4 Analyzer (S.R.I, Italy)}
The breath by breath measurement of VO$_{2}$ was recorded through a portable Cosmed KB4 analyser (S.R.I, Italy). This measurement was used for estimator validation. The KB4 analyser was calibrated before and after each session of an exercise.

\section{Heart Rate Measurement System}\label{HRmeasured}
The HR of the subject is continuously monitored by the remote computer station in real-time using a wireless HR monitoring system (POLAR, T-31) at a sampling frequency of 100-Hz. The pulses generated by the polar system are received by remote computer station and are used further to extract R-R intervals and in turn the HR. The recorded HR data is down sampled to 1 sample/sec and filtered using a standard moving average filter with a 2 sec window.

\section{Exercise Rate Estimation Using the Measurement of TA}\label{TAmeasured}

To measure the current ER of the subject, a wireless TA is deployed for the detection of the body accelerations, i.e., a$_{x}$, a$_{y}$ and a$_{z}$ along x, y and z axes, respectively. These accelerations reflect the periodicity of the human body movements during rhythmic exercises. The TA is always mounted on the right thigh of the subject for the lower extremity exercises. The time period of these accelerations are used to predict the exercise period (EP) along each axis in real-time. The EP is used to compute the frequency of rhythmic movements, i.e., ER along each axis. However, the accurate ER can be reflected on the TA signal where rhythmic body movements influence the TA measured on one axis and other two axes may be influenced by noise. During waking exercise, the acceleration along x  axes, represents good signal strength as compared to y and z axes. Similarly, during rowing and cycling, the body rhythmic movements represent good quality measurements at y and z axes, respectively. For system identification, EP is calculated along each axis and the minimum time period is used to estimate the fundamental frequency of the rhythmic movement and considered as ER. However, in order to estimate the ER in real-time, the universal algorithm for the estimation of ER is described by \cite{Cheng2009}. \\
The following subsection describes the recursive algorithm for the estimation of ER.
\subsection{Universal Algorithm for Exercise Rate  Estimation}\label{unialgo}
\textbf{AVERAGE MAGNITUDE DIFFERENCE FUNCTION}

A universal algorithm based on the Average Magnitude Difference Function (AMDF) is used to detect the fundamental frequency \cite{Cheveigne2002} and its selection is mainly due to ease in hardware implementation. This AMDF formulation is considered as simple and efficient for estimating the fundamental frequency and defined as a discrete signal $x_K$  in \ref{AMDF1}.
\begin{equation}\label{AMDF1}
e_k(d)= \frac{1}{N}\sum^{N}_{i=1}|x_k(i)-x_k(i-d)|
\end{equation}
where, $e_k(d)$ is the AMDF of lag $d$ calculated at time index $k$ and $N$ is the summation window size in terms of samples.\\
\textbf{A CAUSAL MEDIAN FILTER}

To improve the reliability of the estimate further, a causal median filter with window length $(L=2)$ is employed for removing spikes in the time period estimates since the above search may not remove the sub harmonic errors completely \cite{Cheveigne2002}. The procedure detects the fundamental period for each of the three acceleration measurements at time instant $k$. The TA acceleration measurements are in fact the projections of the total body acceleration vector $a_k$ on x-, y- and z-axes. Hence, we propose to obtain an extra period estimate by considering the acceleration vector of x-, y- and z-axes, and the AMDF defined as follows:
\begin{equation}\label{AMDF2}
e_k(d)= \frac{1}{N=10}\sum^{N=10}_{i=1}|a_k(i)-a_k(i-d)|_1
\end{equation}
where, $|.|$  is the 1-norm of a vector.
The above procedure gives $y(k)=~1/4 [t_x(k)~ t_k(k) ~t_z(k)]^T$  as the four fundamental period estimates at time instant $k$.\\
\textbf{KALMAN DATA FUSSION}

Now the kalman data fusion technique gives the estimated exercise rate (ER$_{est}$); see \cite{Cheng2009}.

\subsection{Real-time Estimation of Exercise Rate}
The remote computer station (CS) receives the TA data in real-time, where the received data packet from TA is decoded in LabVIEW. The data packet consists of a synchronous acknowledgement of the received packet, time stamp and three acceleration channels along x-, y- and z-axis, with a sampling frequency of 50Hz. These three acceleration signals are filtered using Butter-worth filter of order 8 and a Median filter of order 5 to completely eliminate the noise. The filtered data of these accelerations are used for the estimation of ER by using a universal algorithm as discussed in \ref{unialgo}. This algorithm uses the initial 5 minutes recording of TA to initialise the window for the AMDF formula and another 1 minute recording is used by shifting the window for the purpose of using AMDF formula in real-time. After the estimation of the EPs along each axis, the KF data fusion technique is employed for the estimation of the fundamental frequency, i.e., ER.
\section{Respiratory Rate Measurement}\label{RespRmeasured}
The respiratory belt (MLT1132) was fixed firmly around the lower chest to measure RespR and it was connected with computer station (CS) using BNC connector via data acquisition system (DAQ62E). The frequency of inhale and exhale data was measured through National Instrument LabVIEW software and was recorded every 10 seconds and was used to calculate the RespR in (Breath/min).
\section{Methods}\label{method}
The main approaches of this research are: the method of least squares (LS) and recursive LS, a Hammerstein model and an indirect adaptive H$_\infty$ control for modelling, estimation and control respectively.  The focus of our discussion here is on these methods involving identification of the HR model in open loop and the design of a controller. Subsequent chapters will apply these general methods to each group of experiments with detailed description of the protocols, subjects and experiments.
\subsection{Modelling}\label{arxmodel} %
Model is essential for a controller design and it gives the relationship between the  plant input and output. The model can be obtained mathematically using physical laws and principles. Similarly, a model can also be obtained experimentally. Furthermore, model can be obtained by the combination of physical laws/principles and experiments. In this section and throughout this report, models are derived from the experiments. In addition, the purpose of a model will affect the choice and selection of that model. In this research, the model is intended to be used for designing a controller to achieve a certain closed-loop performance using an indirect adaptive H$_\infty$ controller, and the model is also intended for the purpose of VO$_2$ estimation using the Hammerstein system. \\
Our implemented method is based on a well-known parametric ARX (Auto Regressive with exogenous input) model for HR control and non-parametric Hammerstein model for VO$_2$ estimation. The procedure involves the identification of the transfer function by computing the parameters via least square algorithm \cite{Ljung1999}.\\
\textbf{Model Structure}\\
A model structure for ARX and ARMAX  has been selected from \cite{Ljung1999}. of the choice of this research ARX model structure is based on its simplicity and least square structure  \cite{Ljung1999}. Nevertheless, other suitable model structures  were also searched for by testing a number of different structures and then comparing/evaluating the resulting models.  Equation \ref{Eq11a} describes the dynamic properties and also represents the standard linear model representation with single input variable $u$, and measured output $y$ and measurement noise $e$.

\begin{eqnarray}\label{Eq11a}
y(t)+a_1 y(t-1)+a_2 y(t-2)+ ... + a_{n_a} y(t-n_a)\nonumber \\=~ b_1 u(t-nk)+b_2u(t-nk-1)...+b_{t-n_b-1} u(t-nk-n_b)+e(t)\nonumber
\end{eqnarray}
this equation can be written as:
\begin{equation}\label{Eq11b}
A(q^{-1})y(t) = B(q^{-1})u(t - nk) + e(t)
\end{equation}
where $q^{-1}y(t) = y(t - 1)$
Equation \ref{Eq11b} can be expressed in this generic form:
\begin{equation}\label{Eq11c}
y(t) = q^{-nk}\frac{B(q^{-1})}{ A(q^{-1})}+ \frac{1}{ A(q^{-1})}e(t)
\end{equation}

$A(q_1) = 1 + a_1q_1 + ... + a_{n_a}q_{n_a}$\\
\\
$B(q_1) = b_1 + b_2q_1 + … + b_{n_b}q_{n_b}+1$\\
and generally written in this generic form: ARX(n$_a$,n$_b$,nk).\\
Here,\\
A(q$_1$) = Auto-Regressive (AR) polynomial in the backward shift operator, q$^{-1}$\\
B(q$_1$) = polynomial for external input, u(t) in the backward shift operator, , q$^{-1}$\\
n$_a$ = order of A(q$_1$)\\
n$_a$ = order of B(q$_1$)\\
n$_k$ = number of delays from input to output\\
e(t) = white noise\\
q$_1$ = unit delay operator\\
\textbf{Parametric Estimation}\\
The collected data was divided into two datasets, one dataset is used for estimation of the models parameters and the other dataset is used for validating the estimated models. The estimated model was validated with the validation dataset. The various ARX models structures were configured for the purposes of the model estimation and then compared with the validation dataset in terms of the loss function. These model structures were selected via varying the degree of the polynomial, and delay time. The best fit model structure which gave less loss function than other model was used as an initial model. This initial model is now again estimated with entire dataset to obtain the estimated model. Since the noise disturbance $e$ is assumed to be uncorrelated zero-mean stochastic signal, the best 1-step-ahead moment prediction is given by: \\
$\hat{y}(t) =\theta_T\phi (t-1)$\\
where,\\
the parameter vector $\theta_T$ and regression vector $phi (t - 1)$ are defined as\\
$\theta = (a_1…..a_{n_a};  b{_1}……b_{n_b})^T$\\
$\phi(t - 1) = (-y(t - 1).... -y(t - n_a); u(t - nk).....u(t - nk - n_b + 1))^T$\\
A measure of model accuracy is provided by the least-square criterion expressed as:\\
\begin{equation}\label{MSEeq}
J_N(\theta)=\frac{1}{N-1}\sum_{t=1}^{N}(y(t)-\hat{y}(t-1))^{2})
\end{equation}
where, $N$ is the number of data points. The analytical solution for the optimal parameter estimate is derived by [70] as:\\
\begin{equation}\label{thetapre1}
\hat{\theta}_T =(\sum_{t=1}^{N}(\phi(t-1) \phi^T(t-1))^{-1} (\phi(t-1) y(t))
\end{equation}
\textbf{Recursive Least Square: Kalman Filter Approach}
In order to track the time variations of the parameters, the model vector $\hat{\theta}$(t) is described by a random walk:
\begin{equation}\label{Eq14a}
\hat{\theta}(t)= \hat{\theta}_{t-1}+ w_t
\end{equation}

where, $w_t$ is Gaussian white noise with covariance matrix $E[w(t)~w(t)^T]=R_1$.  KF estimates $\hat{\theta}_t$ for parameter vector $\theta_t$ and is given in   \ref{Eq14a}
\begin{eqnarray}\label{Eq15A}
e_t = y_t - \hat{y}_t,~~~\hat{y}_t = \phi_t\hat{\theta}_{t-1},~~~\hat{\theta}_t =\hat{\theta}_{t-1} + K_te_t\nonumber\\
\hat{y}_t = \phi^T_t\hat{\theta}_{t-1}\nonumber,~~~\phi_t = [y_{t-1}~y_{t-2}~u_{t-1}~u_{t-2}]^T\nonumber\\
K_t = Q_t\phi_t \nonumber,~~~Q_t =\frac{P_{t-1}}{R_2+\phi^T_tP_{t-1}\phi_t}\nonumber\\
P_t = P_{t-1}+ R_1 -\frac{P_{t-1}\phi^T_t\phi_tP_{t-1}}{R_2+\phi^T_tP_{t-1}\phi_t}\nonumber\\
\end{eqnarray}

where $y_t$ is the measured value,\\
 $\hat{y}_t$  is the estimated value ,\\
 $R_2$ is the covariance of the innovations of $y_t$,
and \\
$P_t$ is the covariance matrix that indicates parameter estimation errors. \\
These parameters are selected as a guess at the covariance of the initial state error, at the measurement noise intensity and a rough value for the added process noise intensity. Such a guess or rough value approach is standard in applied control theory; see, e.g., \cite{Good2001}.
Equations \ref{thetapre1} and \ref{Eq15A} are used for parameter estimation which are incorporated in the Matlab System Identification Toolbox using the  commands arx and rarx. \\
The fitness (expressed in \%) is given by:\\
\begin{equation}\label{Req}
R^2=1-\frac{J_N(\theta)}{ \frac{1}{N}\sum_{t=1}^{N}|y(t)|^2}
\end{equation}
\textbf{Model Validation}

Finally, the various ARX models  are compared for quality of the fit, and the model with the best fitness is chosen using:
\begin{enumerate}
 \item{ The best fit models are Simulated and used for the purpose of the cross validation}
\item{ Cross-validation (The simulated response is compared with the measurements in terms of quality of fit).}
\end{enumerate}
\subsection{Hammerstein System}\label{hammodel}
\begin{figure}
\begin{centering}
\includegraphics[scale=1]{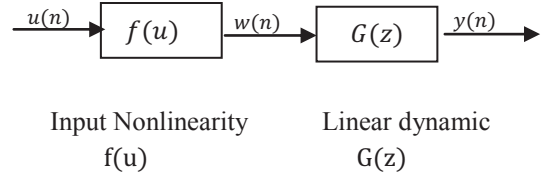}
\caption{The Hammerstein model.}\label{Fig2}
\end{centering}
\end{figure}

The non-parametric System Identification Tool as a Hammerstein model is used for developing a VO$_2$ estimator.  A brief description of a Hammerstein system is described as follows:

The Hammerstein system consists of two blocks, a static input nonlinearity function $(f(u))$  and  a linear dynamic system as shown in Fig.\ref{Fig2}. These blocks were identified for both types of exercises.
\begin{equation}\label{Eq4}
w(n)= f(u(n)) 	
\end{equation}
The input static nonlinearity of the Hammerstein system is defined by the function $f(u)$ as  in  \ref{Eq4} transforming input $u(n)$ into nonlinear space. The dimensions of the $f(u)$ are $(nu\times1)$ ($nu$ is number of inputs) which is the transpose of input vector $u(n)$. However, the dimensions of $w(n)$ are equivalent to $u(n)$. The function $f(u)$ can be represented as sigmoid-net, piecewise linear, tree, and wave-net.
\begin{equation}\label{Eq5}
y(n)=G(z)w(n)  	
\end{equation}
$G(z)$ is the linear transfer function in \ref{Eq5}. It consists of two polynomials, $G(z)=Bii/Fjj$, where $Bii$ is the polynomial that is dependent on $nu$, $Fjj$ is dependent on number of outputs $(ny)$. $y(n)$ is the transformation of non-linear space $f(u(n))$ into linear space.
\begin{equation}\label{Eq6}
V_N (G,f)=\sum^{N}_{(t=1)}(e(n))^2 \\
=\sum^{N}_{(t=1)}((y_{meas}-y_{est}))^2\\
\end{equation}	
The numerical optimisation algorithm \cite{Nocedal1999} was used to minimise the cost function that was given in \ref{Eq6}.
The goal of the optimisation is to find the optimal parameters of the function $f(u)$ in  \ref {Eq4} and the optimal parameters of the linear system $G(z)$ by minimising the error signal $(e(n))$ between measured output $(y_{meas})$ and estimated output $(y_{est})$.  The estimated model was validated as described in the subsection \ref{arxmodel}.
\subsection{Indirect Adaptive H$_\infty$ Controller}\label{controller}
This section presents a controller design method that was used for HR regulation system for an unknown type of rhythmic activity. The control strategy is based on an indirect adaptive control technique \cite{Gene1997}. 
An indirect adaptive control technique uses on-line system identification method to estimate the parameters of the plant and subsequently a "controller designer module" to specify the parameters of the controller. If the plant parameters vary, the identifier will provide estimates of these variations and the controller designer will subsequently re-tune the controller. It is inherently assumed that the estimated plant parameters are equivalent to the actual ones at all times (this is called the "certainty equivalence principle"). The plant will be controlled successfully if the controller designer can specify a controller for each set parameter estimates of the plant.  The overall approach is called "indirect adaptive control" since it tunes the controller indirectly by first estimating the plant parameters.

The function of the control designer is to tune the controller automatically based on the estimates of the plant parameters $\theta$(t).  $H_\infty$ control is one of the most effective modern control techniques. It provides effective control for linear systems ensuring high robustness and stability even in adverse operating conditions such as parameter variations, high noise disturbance environment, actuator saturations and the model uncertainty; see e.g., \cite{PUS00,VaUgr2000,Savkin2002}. Therefore, H$_\infty$ control approach is integrated with indirect adaptive control scheme and described as follows:

\textbf{H$_\infty$ Controller}
A mixed sensitivity H$_\infty$ controller is usually designed for the identified model to achieve robust tracking. The plant model $G$ is augmented with certain weight functions such as the sensitivity weight function $W_1$, the control sensitivity weight function $W_2$ and the complementary weight function $W_3$, so that the desired performance can be achieved from the closed loop transfer function of the plant. The transfer function of the augmented plant model $F$ is given by in \ref{Eq16}.
\begin{equation}\label{Eq16}
F= \left [
\begin{array}{cc}
W_1 &  -GW_1\\
0 & W_2\\
0 & W_3G\\
I & -G
\end{array} \right].
\end{equation}
\begin{figure}
\centering
\includegraphics[scale=0.5]{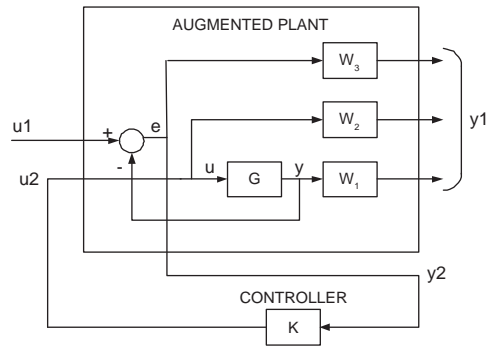}
\caption{The mixed sensitivity $H_\infty$ controller design.} \label{Figcloop}
\end{figure}
The augmented plant model includes three additional blocks representing three complementary weighting functions. The order of this augmented plant is higher than the order of the original plant. A controller designed for the augmented plant delivers good performance for original plant. This is a well-known trick in H$_\infty$ control, see, e.g., \cite{SP96}. The closed loop structure of mixed sensitivity H$_\infty$ design is shown in Fig.  \ref{Figcloop}.%
The controller transfer function $K$ is designed based on the following three criteria:
\begin{enumerate}
\item{\textbf{Stability Criteria}: states that the roots of the characteristic equation $1+G(z)K(z)$ should lie inside the unit circle.}
\item{\textbf{Performance Criteria}: states that the sensitivity $S(z)=1/(1+G(z)K(z))$ must be small where disturbance and set point changes are large.}
\item{\textbf{Robustness Criteria}: demands that stability and performance must be maintained not only for the nominal model but also for the set of neighbouring plant models resulting from the unavoidable presence of modelling errors.}
\end{enumerate}
In general, the H$_\infty$ norm of transfer function $F$ of the closed loop system is defined as:
$\|F\|=\textrm{sup}\sigma(F)\leq\gamma$
where, $\sigma$ is the maximum singular value of $F$. According to the small gain theorem $\gamma$ should be less than 1 to achieve the desired closed loop performance \cite{PUS00,VaUgr2000}.

\chapter{Modelling and Control of Heart Rate During Rhythmic Exercises}\label{ch:hrc}

\section{Introduction}\label{section4.2}

A rhythmic exercise is a kind of exercise in which large muscles group are engaged in a physical exertion, and the individual maintains the HR within 35-85\% from its maximum limits known as aerobic zone. It is important to control HR within the aerobic zone to ensure the sustainability of rhythmic exercise for a longer duration. The control of HR during any type of rhythmic activity is achieved via manipulating ER as control input because of versatility of its estimation using TA ; see \cite{Cheng2009}. However, limitation of ER as a universal measure of exercise intensity is its own bandwidth during rhythmic activities \cite{Cheng2009}.  This exercise bandwidth depends on the individual body's biomechanics and body's posture that is normally engaged during the rhythmic activities. For example during stationary bicycling, frequency of leg movements has its own bandwidth in a stationary sitting position and hence HR response also varies with the type of rhythmic activity. This Chapter presents the modelling and control of HR during an unknown type of rhythmic exercises taking into the account the ER bandwidth.
In general, real-time control of HR during exercises is useful in developing training and rehabilitation exercising protocols. These protocols are useful for athletes, healthy individuals and cardiac patients for a safer exercise \cite{Su2007a, Cheng2008, Su2010a}.  To date, control of HR has been achieved via manipulating the machine workload on the exercising individual \cite{Su2007a,Cheng2008,Hajek1980,Cooper1998,Kawada1999a,Coyle2001,Mazenca2011}; e.g., treadmill speed. In the case of a rhythmic exercise, ER is used as a control input to control HR and the control of HR is possible by giving the measurement of HR (HR$_{meas}$) as a self biofeedback. In this way human engages his/her brain to achieve a targeted ER (ER$_{T}$) which is then used to achieve the targeted HR (HR$_{T}$).  This biofeedback technique for controlling the HR depends on the individual's perception and motivation to achieve HR$_{T}$. As the brain knows the measurements of HR, it is in fact allows the heart to increase or decrease the HR due to some psychological factors \cite{Dardik1991a}.  Therefore, HR may include non-metabolic factors in its response and leads to inaccuracy in predicting the actual demand of the workload. Literature reveals that the efficiency of biofeedback control is limited only for low intensity exercises \cite{Goldstein1977}.  The control of HR during unknown type of rhythmic activity is also possible by generating the ER$_{T}$ from an automatic controller. 
This approach requires the analyses of the HR response in relation to the ER during various types of rhythmic activities.  In this research, the HR response has been keenly analysed and a control methodology has been developed and simulated and then tested in a real-time platform.
\section{Modelling of Heart Rate During Various Types of Rhythmic Exercises}\label{section4.3}
Modelling of the HR response in relation to workload of an exercise machine has been carried out by various researchers \cite{Brodan1971, Hajek1980, Hughson2003, Su2007a,Cheng2008}.  These modelling results reveal that the relation between workload and steady state HR is nonlinear and the behaviour of HR varies differently during onset and offset activity of the exercise.  Also, the heavy workout (high intensity) alters the transient response due to cardiovascular drift \cite{Rowell1993,Johnson1975,Coyle2001}.  It has been shown that the variations in HR response are dependent on the individual physical fitness, and hence the response of HR varies from subject to subject.  Therefore, the behaviour of HR response is more complicated during multiple rhythmic exercises due to the bandwidth of ER in a single rhythmic activity. This ER bandwidth is directly related with the body posture, and the active muscle mass, i.e., individual body biomechanics that involves in a particular rhythmic activity. For example, movements of legs limit the body workout in a stationary bicycling exercise. The focus of this study is to develop a model that can interpret the variations of HR response due to interpersonal variations and capture the exercise variations. The experimental studies were carried out for three types of exercises, walking, cycling and rowing, to develop the ER and HR relation for each type of exercise. These rhythmic workouts for walking, cycling and rowing exercises were achieved through treadmill machine, cycling ergo-meter and rowing machine by keeping the machine resistance constant against the body weight. In this scenario, the HR is exclusively dependent on the value of the ER.
\subsection{Experimental Procedure}
In this study, experimental studies were carried out for walking, cycling and rowing on a treadmill, a cycle ergo-meter and a rowing machine, respectively.
Before staring the experiments, the minimum and maximum achievable limits of the ER are measured for walking, cycling and rowing exercises of all participating subjects. Practically the average bandwidth for treadmill walking exercise is  $0.36~\leq~ER(t)~\leq~1.4$, for stationary bike cycle is $0.36~\leq~ER(t)~\leq~1.4$ and for rowing is $0.27~\leq~ER(t)~\leq~0.666$ and these experimentally determined bandwidth are considered as the desired bandwidth of the exercise. The controller requires to manipulate its output i.e., ER$_T$ within these specified range of the ER.\\

\textbf{Subject}

Two healthy males participated in this experimental study and both were free of cardiac and pulmonary diseases. The subjects were clearly familiarised with the experimental procedure and possible risks involved before their participation in the activity and the consent form was duly signed by each of them.  Physical characteristics of the subjects are tabulated in Table \ref{phychar1}.\\
\begin{ourtable}
\centering
\caption{Physical characteristics of the subjects.}\label{phychar1}
\begin{center}
\tabcolsep 2.0pt
\begin{tabular}{c c c c c}
\hline
\hline
Subject& Age (years) & Weight (Kg) & Height(cm) & BMI\\
\hline
\hline
S1&25&75&175&22.9\\
S2&38&60&168&21.0\\
\hline
\hline
\end{tabular}
\end{center}
\end{ourtable}
\textbf{Experimental Equipment}

The experimental setup consists of the following exercise machines:
\begin{enumerate}
\item{A treadmill machine}
\item{A cycling ergo-meter}
\item{A rowing machine}
\end{enumerate}

The measurements of ER and HR were obtained using the TA and Polar belt (TM 310). For further details for the measurement of ER and HR; see section \ref{equip}.\\
\textbf{Experimental Protocols}
\begin{ourtable}
\centering
\caption{Exercise intensities.}\label{intlevel}
\begin{center}
\tabcolsep 2.0pt
\begin{tabular}{c c c c}
\hline
\hline
~~~~~~ & Intensity~1 & Intensity~2 & Intensity~3\\
\hline
\hline
Walking (km/hr)   	& 4.5&  5.5 & 6.5\\
Cycling	(pedals/min)& 48 &	60  & 72\\
Rowing (strokes/min	& 20 &	25  & 30 \\
\hline
\hline
\end{tabular}
\end{center}
\end{ourtable}

The experimental exercises like walking, cycling and rowing were performed in three different distinct sessions based on the exercise intensity levels as given in Table \ref{intlevel}.  Each exercising session was further classified into three different phases of an experiment, i.e., resting, exercising and recovery.\\
\textbf{Resting Phase}

During the resting phase, the recordings of the HR were obtained while the subject was resting for 5 minutes. Subsequently, these measurements were used to calculate a mean value of resting HR (HR$_{rest}$).\\
\textbf{Exercising Phase}

During this phase, the subject was exercising for 15 minutes at the desired intensity level and the HR and ER were recorded for the purpose of modelling the relation between ER and HR.\\
\textbf{Recovery Phase}

Recovery phase starts immediately after the exercising phase where the subject has to be at rest for at least 10 minutes in order to recover his/her HR back to its HR$_{rest}$  value.\\

\subsection{Data Processing}
As mentioned earlier, the experimental data was processed to model the relation between ER and change in HR from resting to the exercising phase ($\Delta$HR). The $\Delta$HR was obtained by subtracting HR$_{rest}$ from the measured of value of HR (HR$_{meas}$) obtained during the exercising phase. As each exercise was performed at 3 intensity levels, therefore 18 input/output (I/O) datasets for 2 subjects were the outcomes of the experimental study.\\
\subsection {Linear Time Invariant Model}\label{lti}
\begin{figure}
\begin{centering}
\includegraphics[scale=1.25]{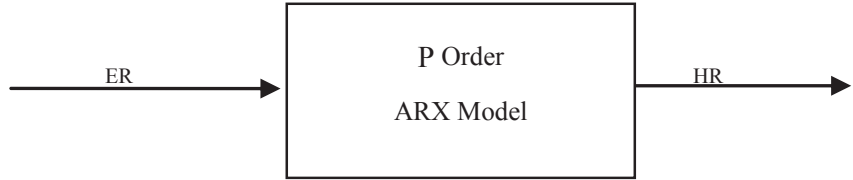}
\caption{A block diagram of ER and HR  model.}\label{modelblk}
\end{centering}
\end{figure}
The relation between ER and $\Delta$HR was modelled using a linear time invariant (LTI) p-order model. The block diagram of the LTI p-order model is shown in Fig. \ref{modelblk} and given in \ref{Eq11} where y(t)  represents $\Delta$HR in (bpm), and u(t) represents ER (Hz), $a_i$ and $b_i$ represent the constant model parameters that fits the I/O data into a model structure.
\begin{equation}\label{Eq11}.
y(t)=a_1 y(t-1)+a_2 y(t-2)+ ... + a_p y(t-p)+b_1 u(t-1)+b_2 u(t-2)...+b_p u(t-p)
\end{equation}

\begin{figure}
\centering
\includegraphics[scale=0.55]{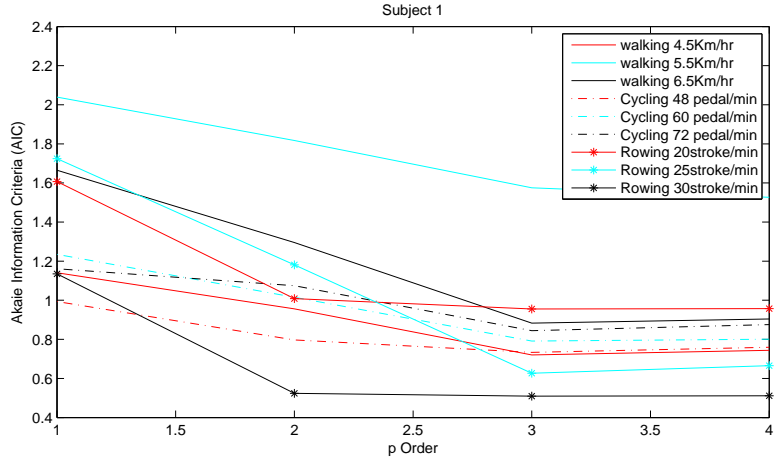}
\caption{AIC values for p-order LTI model for walking, cycling and rowing exercises.} \label{AICS1}
\end{figure}

\begin{figure}
\centering
\includegraphics[scale=0.55]{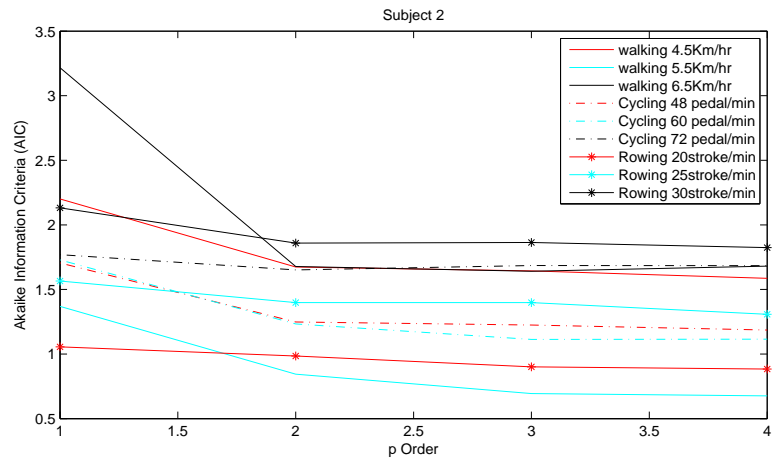}
\caption{AIC value for p-order LTI model for walking, cycling and rowing exercises.} \label{AICS2}
\end{figure}

It is important to identify the proper order of the LTI model in order to get a maximum quality of fit between measured and estimated values of $\Delta$HR. Akaike's information criterion (AIC) was used to define the model order.  Various $p$ order models were estimated for all experimental data sets, and were assigned into the AIC criteria using the Matlab command $aic$.  According to Akaike's theory, the most accurate model has the smallest AIC value. The computed AIC values of the LTI models obtained during walking, cycling and rowing exercises are presented in Fig. \ref{AICS1} for Subject S1 and Fig. \ref{AICS2} for Subject S2.  In these figures, the LTI model orders were varied from 1 to 5.
It has been observed from the Figures \ref{AICS1} and \ref{AICS2} that the second order LTI model (i.e., p=2) is suitable for modelling the experimental data, whereas four cases suggested p=3 order model.  These four cases were mainly related to the high intensity exercises. Since this study required a universal model structure that can model the HR response during various rhythmic activities at various intensity levels. Therefore, the model order $p=2$ was finally selected for all I/O datasets of walking, cycling and rowing exercises. This 2$^{nd}$-order LTI model structure has four parameters, i.e., $a_1$, $a_2$, $b_1$ and $b_2$. The LTI parameters a$_1$ and a$_2$ indicate that the dynamic response of the HR during the onset activity, which contributes to a rapid increase in HR, and this initial HR response during onset activity is represented by the parameters $a_{1}$ and $a_{2}$. For all constant intensity levels, the HR achieves the steady state which represents that the blood circulation meets the metabolic demands at particular ER. This steady state value in $\Delta$HR is represented by 2$^{nd}$- order LTI model by the interactions of the parameters $a_{1},~a_{2},~b_{1}$~and~$b_{2}$.

\begin{figure}
\centering
\includegraphics[scale=0.6]{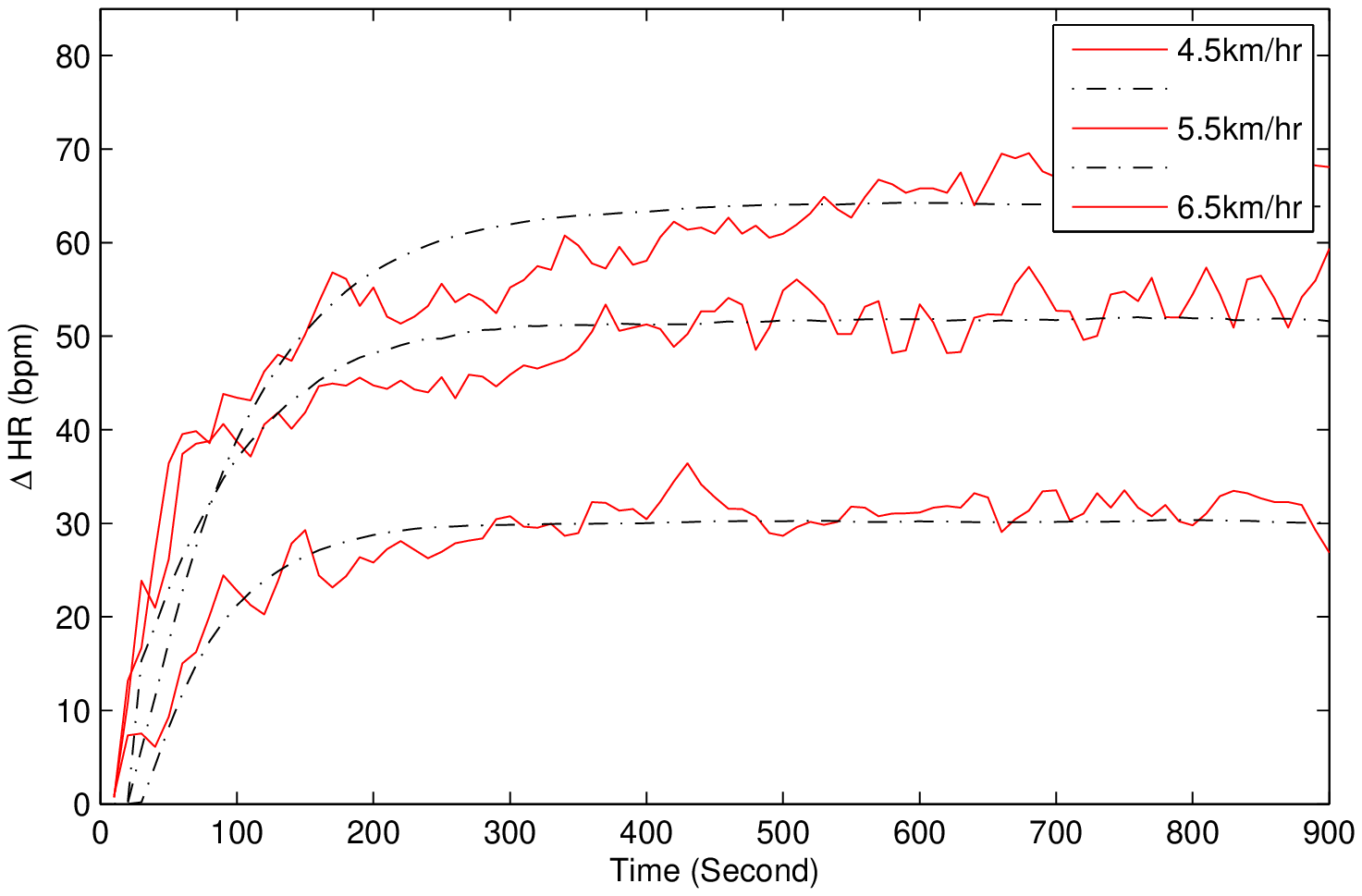}
\caption{Subject S1: Measured and estimated $\Delta$HR response using LTI model of a walking exercise.}\label{hoawalkmod}
\end{figure}

\begin{figure}
\centering
\includegraphics[scale=0.6]{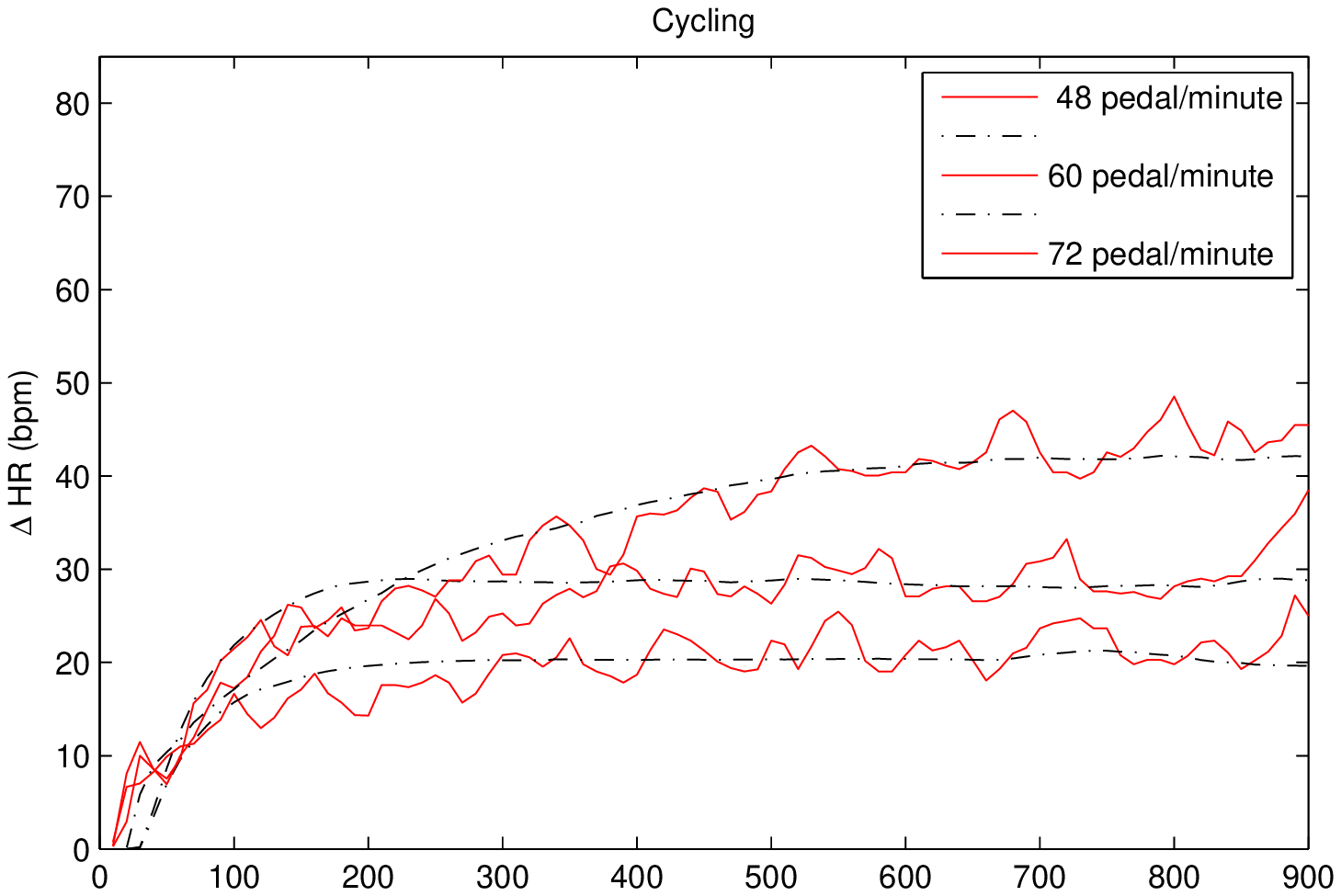}
\caption{Subject S1: Measured and estimated $\Delta$HR response using LTI model of a cycling exercise.}\label{hoacycmod}
\end{figure}

These parameters of LTI models were identified using the method of  least squares (LS) for each dataset of a particular exercise. The identified parameters for each subject for each type of exercise at three different intensity levels are given Table \ref{arxparameter}. The performance of 2$^{nd}$-order LTI model is evaluated based on mean square error (MSE) and correlation coefficient (R) and these performance measures are also given in Table \ref{statmodel}. The measured and estimated $\Delta$HR  are shown in Figs.\ref{hoawalkmod}, \ref{hoacycmod} and \ref{hoacycmod} for Subject S1 . Similarly, Figures \ref{hoawalkmod}, \ref{hoacycmod} and \ref{hoacycmod} present the same results for Subject S2.

The analysis of generalised model structure, which was based on AIC criteria and was suitable to model the HR response during various rhythmic activities at various intensity levels, it has been observed that The estimated parameters of this generalise LTI structure during walking, cycling and rowing exercises at various intensity levels show significant parameter variations amongst subjects, the type of exercises and intensity level.  These parameter variations are interpreted as a standard deviation (SD)  as given in Table \ref{statpara}. The SD for parameters $a_1$ (SD=0.2311) and $a_2$  (SD=0.1727) do not change significantly among subjects and types of exercise as compared to parameters $b_1$ and $b_2$ (SD=9.2870, SD=10.0628, respectively). This suggests that the parameters $b_1$ and $b_2$ are more sensitive to the change in subject and the type of exercise.  The performance of the estimated model was computed in terms of mean square error (MSE) as in \ref{MSEeq} and correlation coefficient $R$ as in \ref{Req} and these statistical properties of the linear 2$^{nd}$-order model  for each subject are given in Table \ref{statpara}.

\begin{figure}
\centering
\includegraphics[scale=0.6]{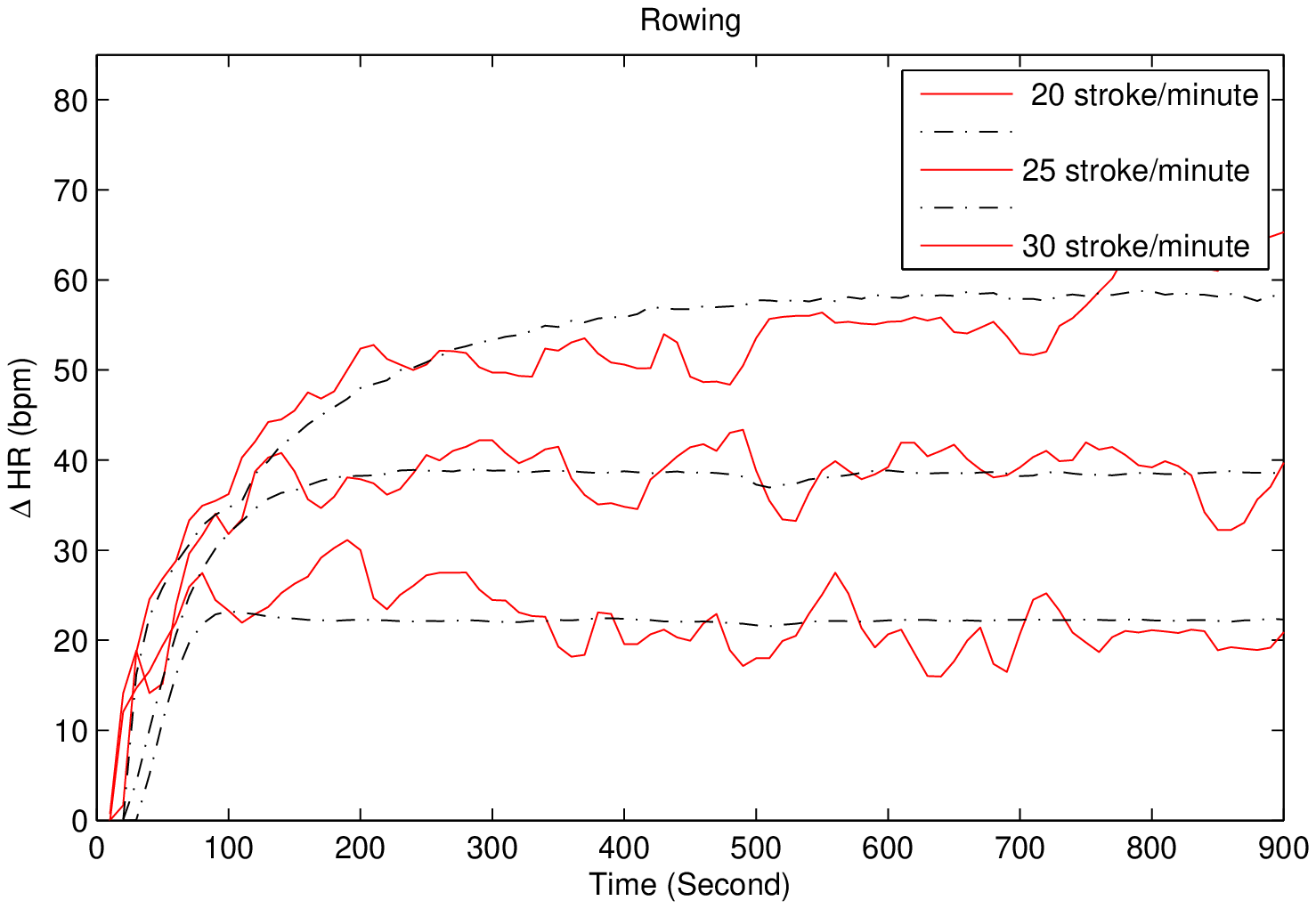}
\caption{Subject S1: Measured and estimated $\Delta$HR response using LTI model of a rowing exercise.}\label{hoarowmod}
\end{figure}
\newpage

\begin{ourtable}
\centering
\caption{2$^{nd}$-order LTI model.}\label{arxparameter}
\begin{center}
\tabcolsep 2.0pt
\begin{tabular}{c c c c c}
\hline
\hline	
		
Subject S1\\	
	&$a_1$&	$a_2$&$b_1$& $b_2$\\			
Walking\\
(km/hr)\\	
4.5& 	-1.0323&    0.1640&    0.1172&    4.1769\\
5.5& 	-0.8034 &   -0.0700&   15.1587&   -8.7281\\
6.5& 	   -0.9046&    0.0142&    6.0322&    0.4490\\
Cycling	 (Pedals/min)\\   -		
48 &      -1.1005&    0.2361&    0.1023&    3.3537\\
60 &     -1.1143&    0.1795&    2.0004&    0.0938\\
72 & -1.0917&    0.1300&    6.0985&   -4.6899\\
Rowing\\
(strokes/min)	\\		
20& -1.2193&    0.4385&   -0.4637&   14.7858\\
25&    -1.1367&    0.2768&    9.7000&    3.2166\\
30& -1.1703 &   0.2243&   33.7415& -27.3794\\
Subject S2\\			
Walking\\
(km/hr)\\	
4.5 &-0.4696&   -0.0737&    8.1702&    6.5464\\
5.5 &   -0.7084  &  0.0100&   13.3070 &  -1.3237\\
6.5 &   -0.4836 & -0.3032 &  2.9962&    7.3352\\
Cycling\\
(Pedals/min)\\
48     &-1.1084&    0.1821&    0.7262&    2.1152\\
60&	   -0.9604 &   0.0327&   13.6206&  -10.8870\\
72&	   -0.8887&   -0.0285&    7.8862 & -2.8568\\
Rowing\\
	(strokes/min)\\
20 &	   -0.9212&    0.1268 &  22.9027&  -11.2361\\
25 &	   -1.2824 &   0.3249&   19.6925&  -13.1839\\
30 &	-1.0602 &   0.1497 &    2.0681 &   11.2466\\
\hline
\hline
\end{tabular}
\end{center}
\end{ourtable}
\begin{figure}
\centering
\includegraphics[scale=0.61]{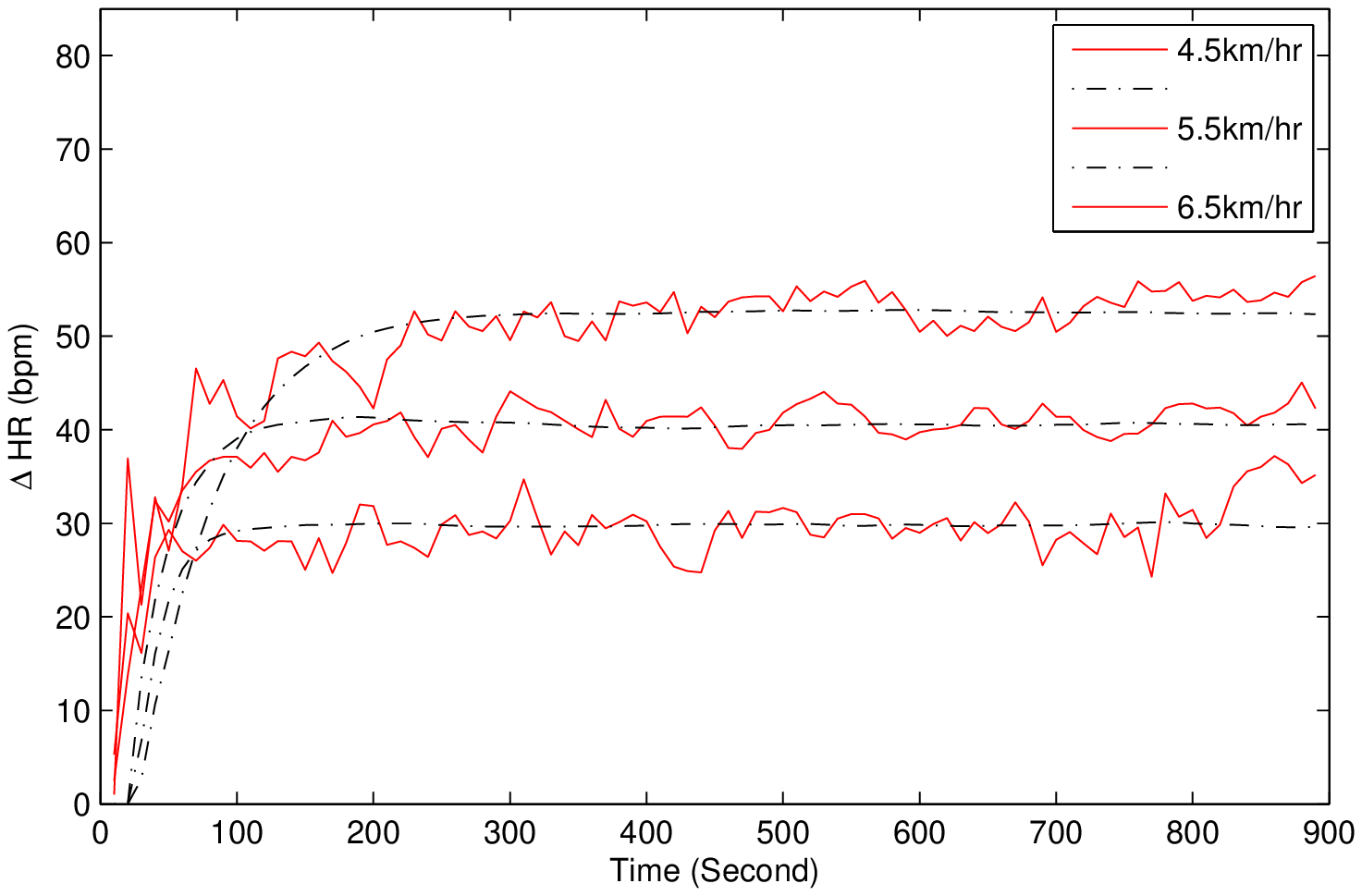}
\caption{Subject S2: Measured and estimated $\Delta$HR response using LTI model of a walking exercise.} \label{tedwalkmod}
\end{figure}
\begin{figure}
\centering
\includegraphics[scale=0.61]{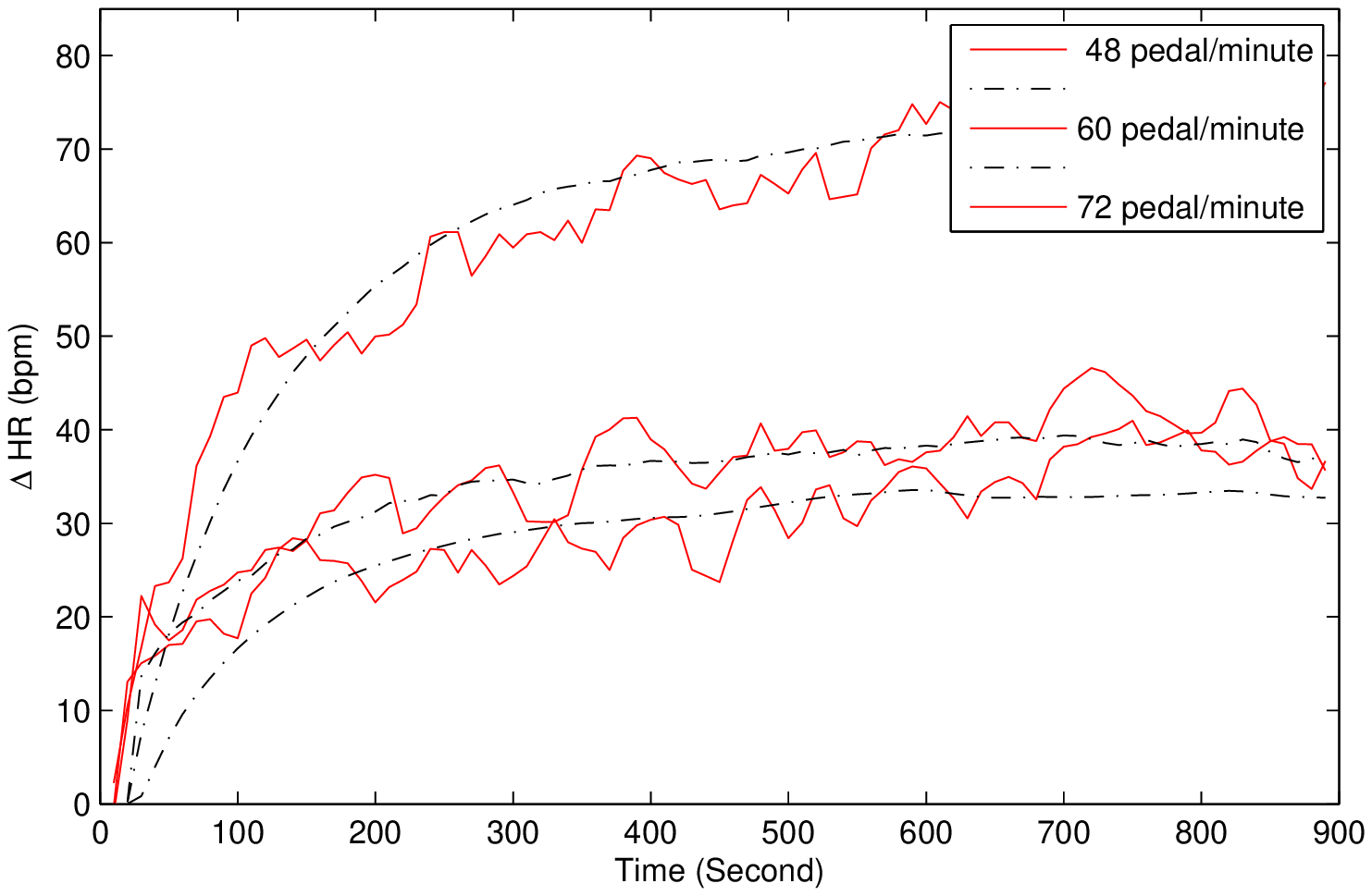}
\caption{Subject S2: Measured and estimated $\Delta$HR response using LTI of a cycling exercise.} \label{tedcycmod}
\end{figure}
\begin{figure}
\centering
\includegraphics[scale=0.61]{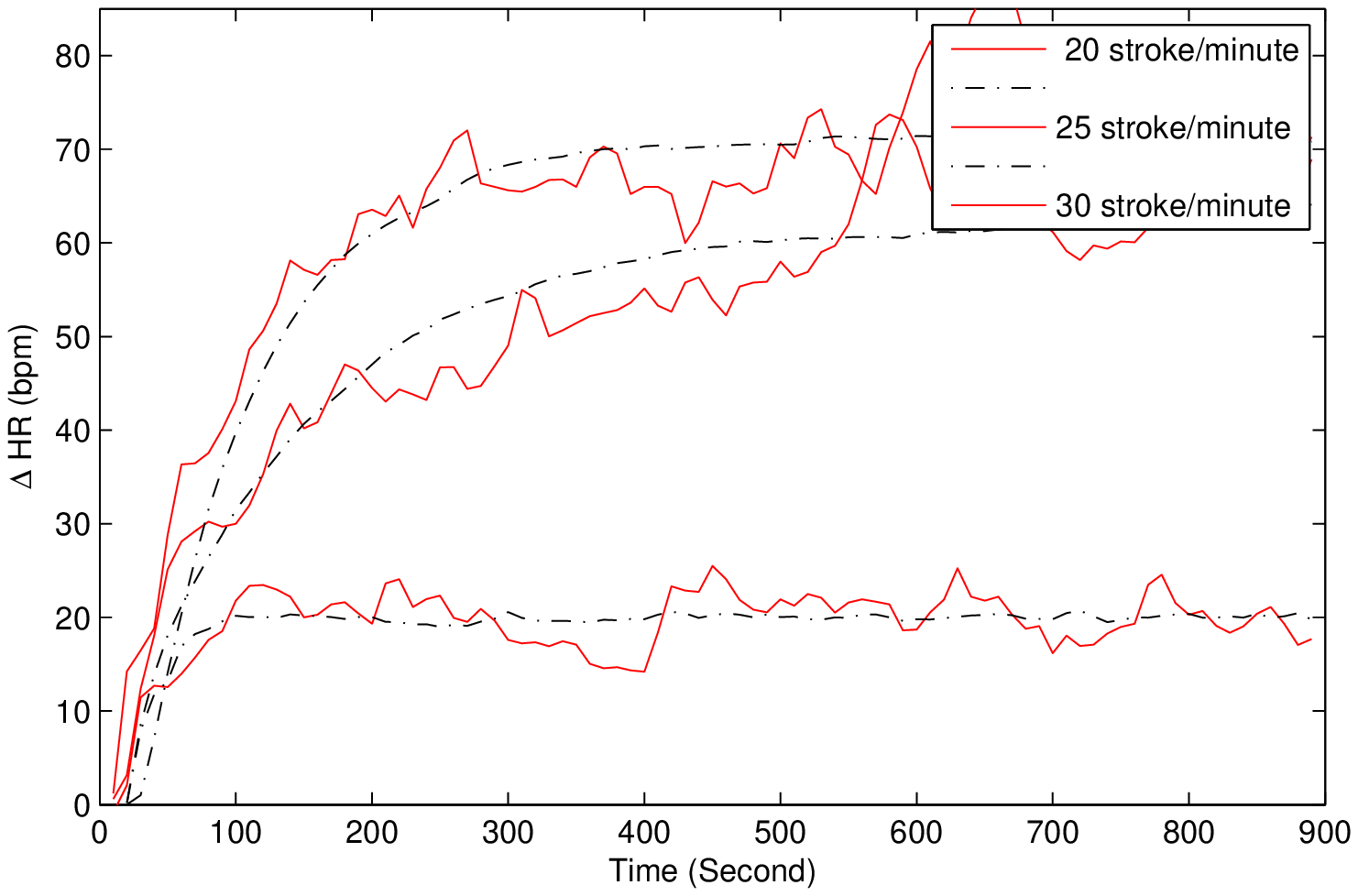}
\caption{Subject S2: Measured and estimated $\Delta$HR response using LTI model of a rowing exercise.}\label{tedrowmod}
\end{figure}
These results clearly indicate that the 2$^{nd}$-order system gives the adequate performance  of the HR response in terms of the transient and steady state characteristics. However, in all cases of high intensity exercises, i.e., 6.5 km/hr, 72 pedal/min and 32 stroke/min for walking, cycling and rowing exercises, the HR response perturbed from its previous steady state values just after 10 minutes of exercise and it adapts the higher steady state value for both subjects. This change in HR response represents the accumulation of non-metabolic factors such as the body temperature and dehydration and this phenomena represents the cardiovascular drift \cite{Rowell1968}.  Based on these facts, we can conclude that the HR response using 2$^{nd}$-order LTI model gives good performance for low and moderate intensity exercises. However, during the prolong exercises, the 2$^{nd}$-order LTI model may not be adequate to interpret the HR response for the entire duration of the exercise. Therefore, the model parameters vary along the period of time as HR changes its behaviour.
\begin{ourtable}
\centering
\caption{Statistical properties of the parameters.}\label{statpara}
\begin{center}
\tabcolsep 2.0pt
\begin{tabular}{c c c c c}
\hline
\hline
&$a_1$&	$a_2$&$b_1$& $b_2$\\
\hline
\hline
Mean&   -0.9698&    0.1119 &   9.1032&   -1.4981\\
Standard deviation (SD)&	  0.2311&    0.1727  &  9.2870&   10.0628\\
\hline
\hline
\end{tabular}
\end{center}
\end{ourtable}
\subsection{Linear Time Variant Model}\label{ltv}
The experimental data analyses show that the HR response does not depend on exercise intensity  or ER only,  but is also affected by other non-metabolic factors such as temperature, heat, age, over training and health conditions. These non-metabolic factors cause variations in HR response along the period of time and can be modelled using linear time variant models. Therefore the parameters $a_{1,t}, ~a_{2,t}, ~b_{1,t}, ~b_{2,t}$  are considered as time-varying. Thus, the estimation of these parameters ($a_{1,t}, ~a_{2,t}, ~b_{1,t}, ~b_{2,t}$) is performed recursively, and the corresponding HR model is given in \ref{Eq12}.

\begin{equation}\label{Eq12}
\Delta HR_t= a_{1,t}\Delta HR_{t-1}+a_{2,t}\Delta HR_{t-2}+b_{1,t}ER_{t-1}+ER_{2,t}u_{t-2}
\end{equation}
In \ref{Eq12},  $\Delta$HR$_t$ is the change in HR from its resting value at time instants $t$, $t-1$ and $t-2$,  respectively.  Similarly, $ER_{t-1}$ and $ER_{t-2}$ represent the ER at time instants $t-1$ and $t-2$ respectively, and $a_{1,t},~a_{2,t},~b_{1,t},$ and $b_{2,t}$ are the time varying parameters.\\
\textbf{Parameter Identification}

The unknown parameters $a_{1,t},~a_{2,t},~b_{1,t},~b_{2,t}$ of the recursive model presented in \ref{Eq12} were estimated recursively using a recursive least square algorithm with well-known Kalman Filter as an adaptation gain. We assumed that the $\theta_t$ be the vector of time varying parameters and is given as follow:
\begin{equation}\label{Eq13}
\theta_t=[a_{1,t}~a_{2,t}~b_{1,t}~b_{2,t}]^T
\end{equation}
In order to track the time-varying behaviour of the parameters, the model parameter vector is described by a random walk as follows:
\begin{equation}\label{Eq14}
\hat{\theta}_t= \hat{\theta}_{t-1}+ w_t
\end{equation}
where the estimate of $\theta_t$ is represented by $\hat{\theta}_t$, $w_t$ is a Gaussian white noise with covariance matrix $E[w(t)~w(t)^T]=R_1$, Kalman Filter estimates the $\hat{\theta}_t$ for parameter vector $\theta_t$.
\begin{eqnarray}\label{Eq15}
e_t = HR(t) -\hat{HR}(t),~~~\hat{HR}(t) = \phi_t\hat{\theta}_{t-1},~~~\hat{\theta}_t =\hat{\theta}_{t-1} + K_te_t\nonumber\\
\hat{HR}(t) = \phi^T_t\hat{\theta}_{t-1}\nonumber,~~~\phi_t = [\Delta HR(t-1)~\Delta HR(t-2)~ER(t-1)~ER(t-2)]^T\nonumber\\
K_t = Q_t\phi_t \nonumber,~~~Q_t =\frac{P_{t-1}}{R_2+\phi^T_tP_{t-1}\phi_t}\nonumber\\
P_t = P_{t-1}+ R_1 -\frac{P_{t-1}\phi^T_t\phi_tP_{t-1}}{R_2+\phi^T_tP_{t-1}\phi_t}\nonumber\\
\end{eqnarray}
 $\hat{\Delta HR}(t)$ is the estimate of $\Delta HR$, R$_2$ is the covariance of the innovations $\Delta HR_{meas}$, and $P_t$ is the covariance matrix that indicates parameter estimation errors. The KF was initialised with the following variables, i.e., $\hat{\theta}(0)$, R$_1$, R$_2$: see Table \ref{Kfpara}.\\
 \begin{ourtable}
\centering
\caption{KF Initialisation}\label{Kfpara}
\begin{center}
\tabcolsep 2.0pt
\begin{tabular}{c c}
\hline
\hline
  $\hat{\theta}(0)$&   $[-0.9698~0.1119~ 9.1032~  -1.4981]$\\
  $R_1$~~&$      diag(2\times10^{-6}, 2\times 10^{-6}, 2\times10^{-6}, 2×10^{-6})$\\
 $R_2$~~~& 1 \\
 P(0)~~~&diag(0.2311,   0.1727,  9.2870,   10.0628)\\
\hline
\hline
\end{tabular}
\end{center}
\end{ourtable}
These parameters were selected as an initial guess at the covariance of the initial state error, a guess at the measurement noise intensity and a rough value for the added process noise intensity. Since a guess/rough value approach is standard in applied control; see, e.g., \cite{Good2001}.  For the purpose of of LTV model estimation, $\hat{\theta}(0)$  and P(0) were selected as mean value and SD of LTI models which are given in Table \ref{statpara}.
The simulation performance of the 2$^{nd}$-order LTI models and LTV models was compared for  walking, cycling and rowing exercises at different level of intensities.  Performance comparison in terms of MSE and $R^2$ is given in Table \ref{statmodel}. The comparative analysis of the results suggests that the LTV model performs better than the LTI model and gives the best fit. \\
\begin{ourtable}
\centering
\caption{Statistical analysis of LTI and LTV models.}\label{statmodel}
\begin{center}
\tabcolsep 2.0pt
\begin{tabular}{c c c c c}
\hline
\hline	
&MSE LTI&MSE LTV& R$^{2}$LTI&R$^{2}$ LTV\\
Subject S1\\				
Walking	\\
(km/hr)\\			
4.5&	0.0133&	0.0072&	0.7527&	0.8550\\
5.5 &	0.4890&	0.1556&	0.7759&	0.9321\\
6.5 &	0.1155&	0.0024&	0.8533&	0.9723\\
Cycling	\\
(Pedals/min)\\			
48 &	0.0586&	0.0115&	0.6400&	0.8199\\
60 &	0.4519&	0.3754&	0.8307&	0.8544\\
72 &	0.1733&	0.0266&	0.8903&	0.9628\\
Rowing\\
 (strokes/min)\\					
20 &	0.0072&	0.0012&	0.3232&	0.5842\\
25 &	0.0125&	0.0465&	0.8187&	0.8032\\
30 &	0.5632&	0.0115&	0.8010&	0.9058\\
Subject S2\\				
Walking\\
 (km/hr)\\				
4.5  &0.2130&	0.0648&	0.4927&	0.6208\\
5.5 &0.1155&	0.0613&	0.8289&	0.9046\\
6.5 &0.1141&	0.0138&	0.8436&	0.8930\\
Cycling \\
(Pedals/min)\\				
48&	0.0143&	0.1663&	0.7113&	0.8796\\
60&	0.0055&	0.0680&	0.7320&	0.8832\\
72&	0.0894&	0.0001&	0.9080&	0.9762\\
Rowing\\
(strokes/min)\\
20& 	0.0011&	0.0030&	0.6101&	0.6106\\
25&     0.4079&	0.0993&	0.8413&	0.9507\\
30& 	0.0062&	0.0016&	0.8829&	0.9589\\
\hline
\hline
\end{tabular}
\end{center}
\end{ourtable}
\newpage
\textbf{Walking}

\begin{figure}
\centering
\includegraphics[scale=0.8]{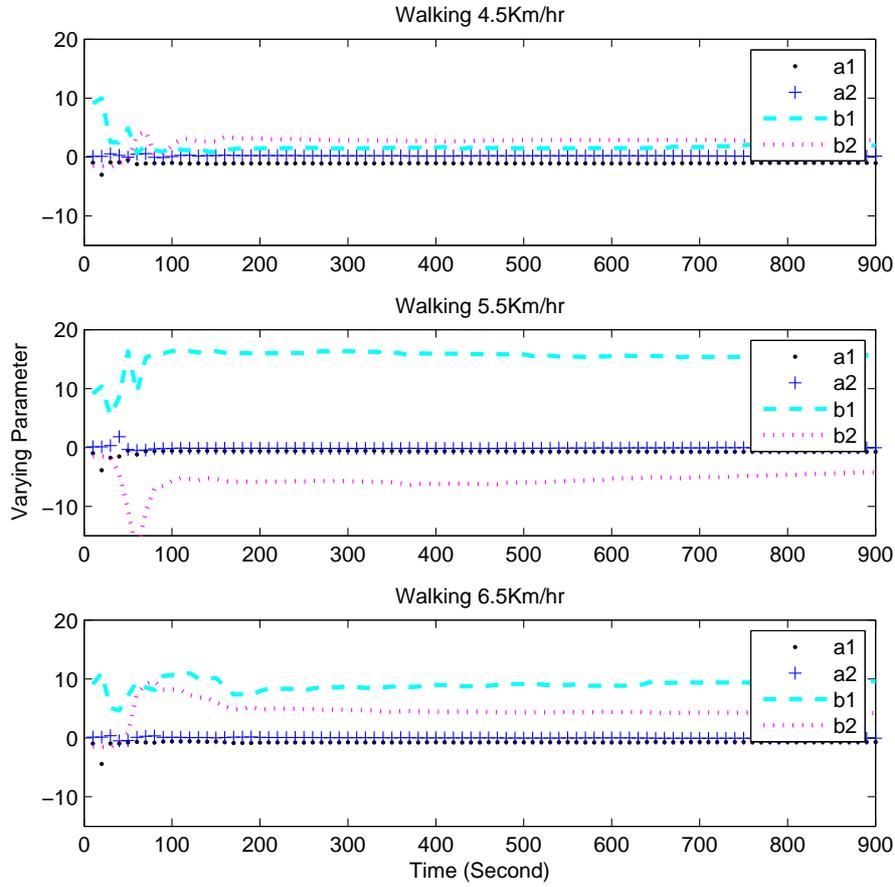}:
\caption{Subject S1: Estimated LTV parameters for a walking exercise.} \label{hoawalkLTVpa}
\end{figure}

\begin{figure}
\centering
\includegraphics[scale=0.7]{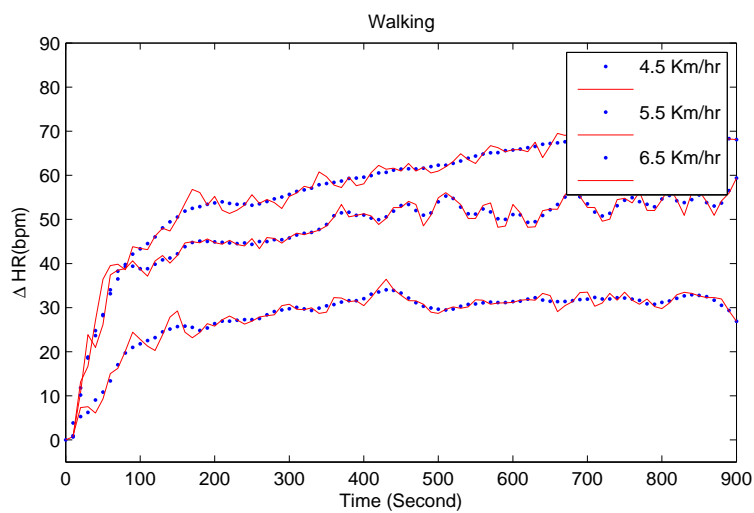}
\caption{Subject S1: Measured and estimated $\Delta$HR response for a walking exercise.} \label{hoawalkLTV}
\end{figure}

Fig. \ref{hoawalkLTVpa} shows the time evolutions of the LTV model parameters at ER of 4.5, 5.5 and 6.5 km/hr for Subject S1. The simulated HR response using LTV models of Subject S1 is shown in Fig. \ref{hoawalkLTV}. The simulated $\Delta$HR response improves using the LTV model as compared LTI model for all three intensity levels of walking exercise for Subject S1, as shown in Fig.  \ref{hoawalkLTV}. The MSE between the measured and estimated values of $\Delta$HR using LTI model is 0.1155 bpm where  Subject S1 is engaged in walking exercise at the speed of 6.5 km/hr with R =~0.8533, while LTV model reduces the MSE to 0.0024 bpm with $R =0. 9723$. Similar findings were also observed at the ER of 48 and 60 pedal/min for  Subject S1 as shown in Fig. \ref{hoawalkLTV} and Table \ref{statmodel}.

\newpage

\begin{figure}
\centering
\includegraphics[scale=0.8]{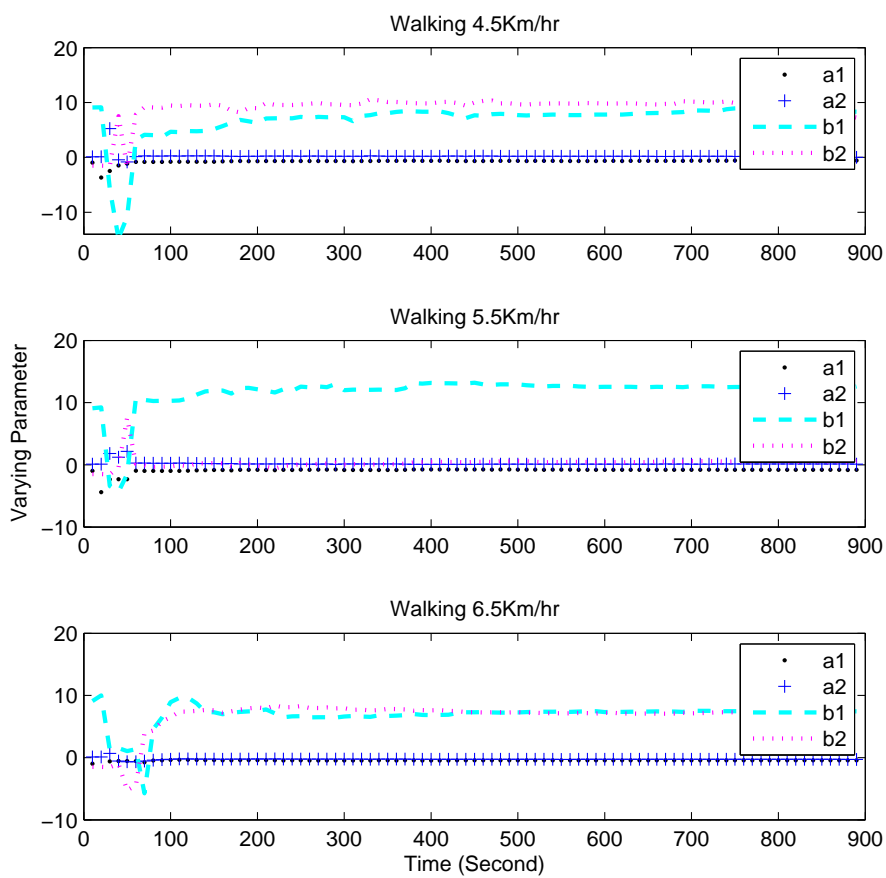}
\caption{Subject S2: Estimated LTV parameters for a walking exercise.} \label{tedwalkLTVpa}
\end{figure}

\begin{figure}
\centering
\includegraphics[scale=0.7]{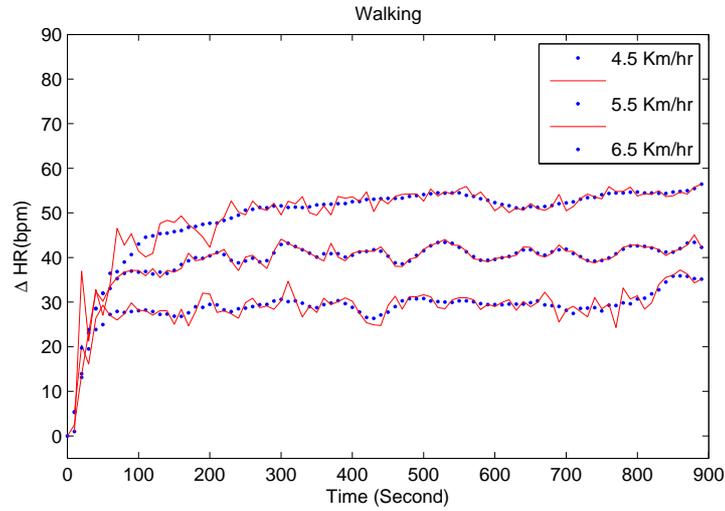}
\caption{Subject S2: Measured and estimated $\Delta$HR response for a walking exercise.} \label{tedwalkLTV}
\end{figure}

Fig. \ref{tedwalkLTVpa} shows the time evolutions of the LTV model parameters at ER of 4.5, 5.5 and 6.5 km/hr for Subject S2. The simulated HR response using LTV models of Subject S2 is shown in Fig.  \ref{tedwalkLTV}.  Again, LTV model enhances the performance of the simulated output of the $\Delta$HR response as compared with LTI models for all three intensity levels of walking exercise for Subject S2, as shown in Fig. \ref{tedwalkLTV}. The MSE between the measured and estimated values of $\Delta$HR using LTI model is 0.1141 bpm where  Subject S2 is engaged in walking exercise at the speed of 6.5 km/hr with R =~0.843, while MSE gets reduced to 0.0138 bpm with $R =0.8930$ in case of LTV model. Similar findings were also observed at the treadmill speeds of 4.5 and 5.5 km/hr for Subject S2 as presented in Fig. \ref{hoawalkLTV} and Table \ref{statmodel}.\\

\newpage
\textbf{Cycling}

\begin{figure}
\centering
\includegraphics[scale=0.8]{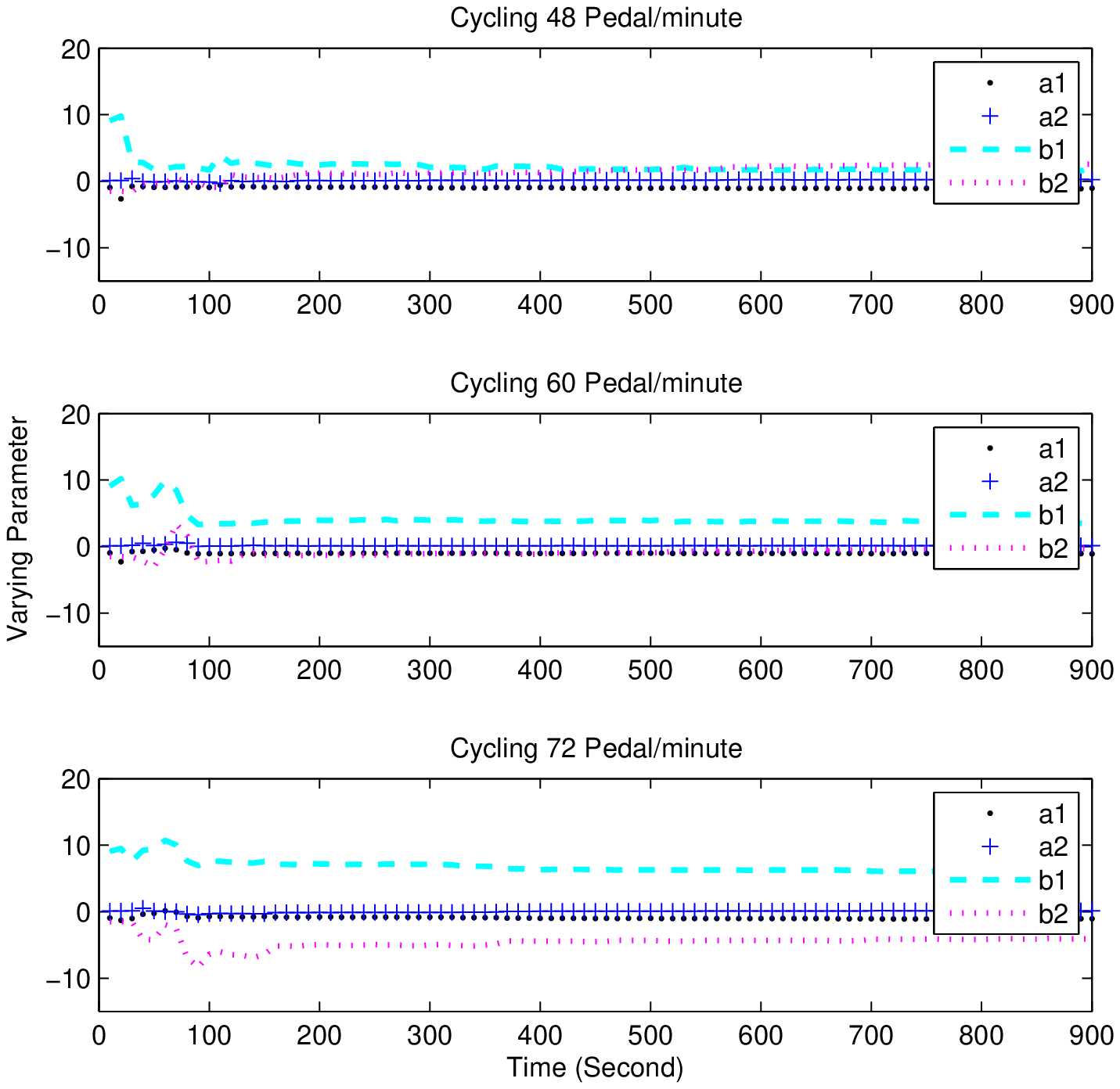}
\caption{Subject S1: Estimated LTV parameters for a cycling exercise.} \label{hoacycLTVpa}
\end{figure}

\begin{figure}
\centering
\includegraphics[scale=0.7]{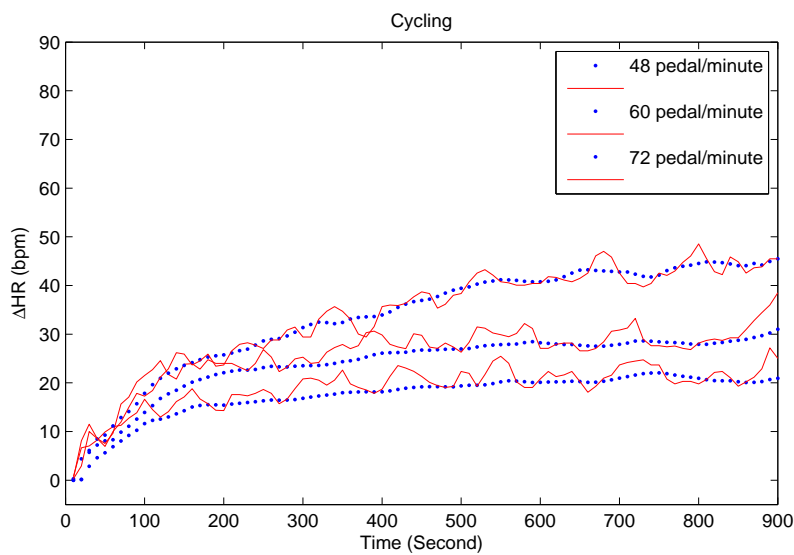}
\caption{Subject S1: Measured and estimated $\Delta$HR response for a cycling exercise.} \label{hoacycLTV}
\end{figure}

Fig. \ref{hoacycLTVpa} shows the time evolutions of the LTV model parameters at ER of 48, 60 and 72 pedal/min for Subject S1. The simulated HR response using LTV models of Subject S1 is shown in Fig.  \ref{hoacycLTV}.  In all three cases of cycling exercise, the performance of the LTV model is better than LTI model as shown in Fig. \ref{hoacycmod}. The MSE between the measured and estimated values of $\Delta$HR using LTI model is 0.1733 bpm when the  Subject S1 is kept engaged in cycling exercise at the ER of 72 pedal/min with $R =~0.8903$. The use of LTV model improved the performance and MSE gets reduced to 0.00266 with $R =0.9723$. Similar findings were also observed at the ER of 48 and 60 pedal/min for  Subject S1 as shown in Fig. \ref{hoacycLTV} and Table \ref{statmodel}.

\begin{figure}
\centering
\includegraphics[scale=0.8]{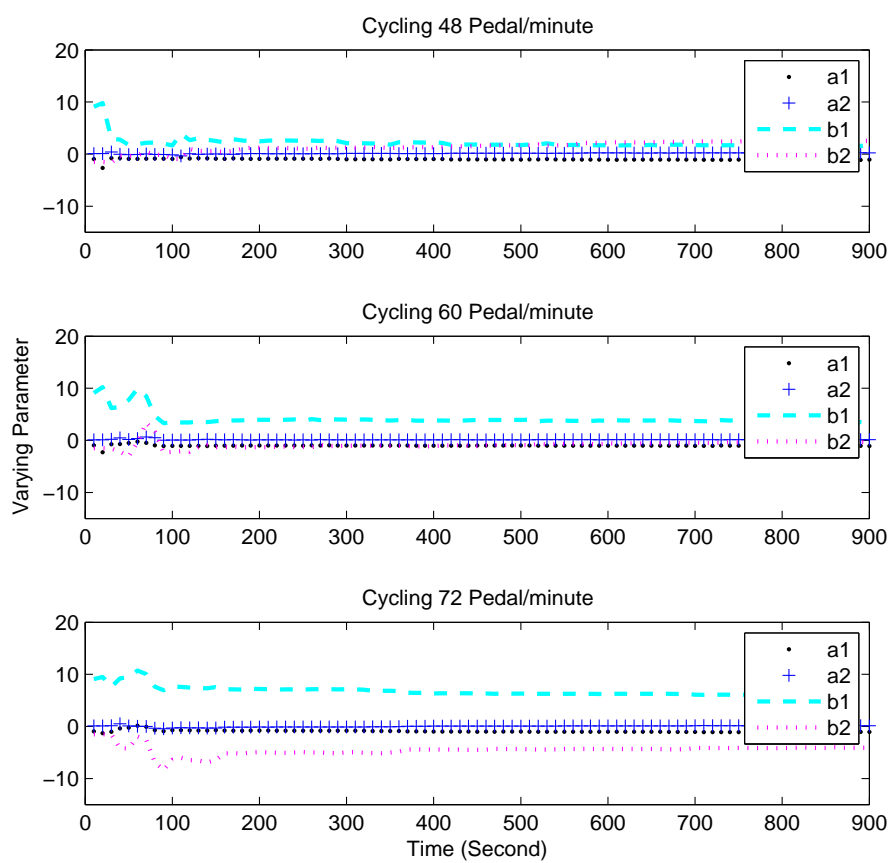}
\caption{Subject S1: Estimated LTV parameters for a cycling exercise.} \label{tedcycLTVpa}
\end{figure}

\begin{figure}
\centering
\includegraphics[scale=0.7]{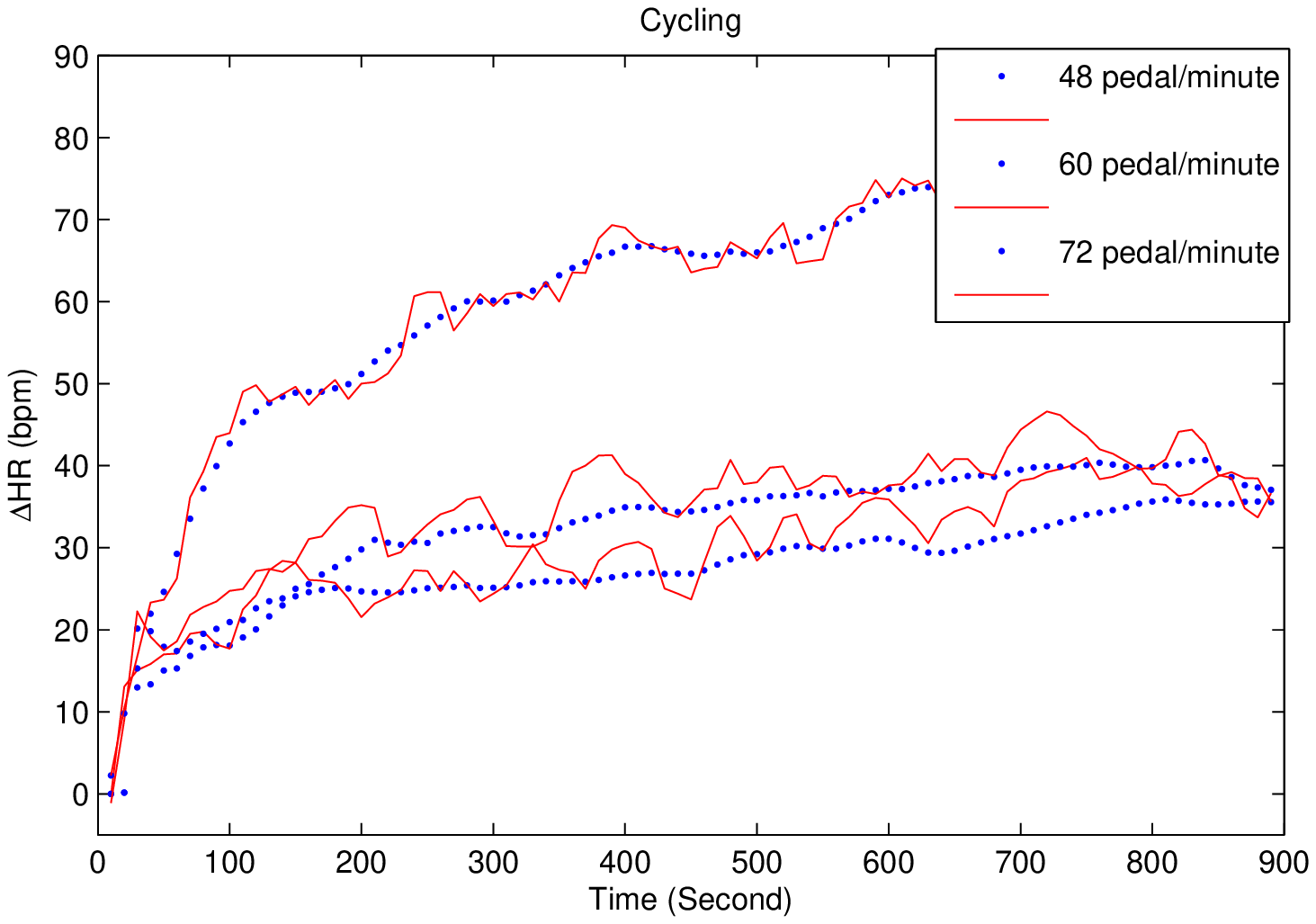}
\caption{Subject S2: Measured and estimated $\Delta$HR response for a cycling exercise.} \label{LTVcyclingted}
\end{figure}

\newpage
Figures \ref{tedcycLTVpa} show the time evolutions of the LTV model parameters at ER of 48, 60 and 72 pedal/min. for Subject S2. The simulated HR response using LTV models is shown in Fig.  \ref{LTVcyclingted} In all three cases of cycling exercise, the performance of the LTV model is better than LTI models as shown in Fig. \ref{tedcycmod}. The MSE between the measured and estimated values of $\Delta$HR using LTI model is 0.0894 bpm while Subject S2 engaged in cycling exercise at the ER of 72 pedal/min with $R =~0.9080$. Using the LTV model MSE gets reduced and its change to 0.0001 with $R =0.9762$. Similar findings were also observed at the ER of 48 and 60 pedal/min for Subject S2 in Fig.  \ref{LTVcyclingted} and Table \ref{statmodel}.
\newpage
\textbf{Rowing}

\begin{figure}
\centering
\includegraphics[scale=0.8]{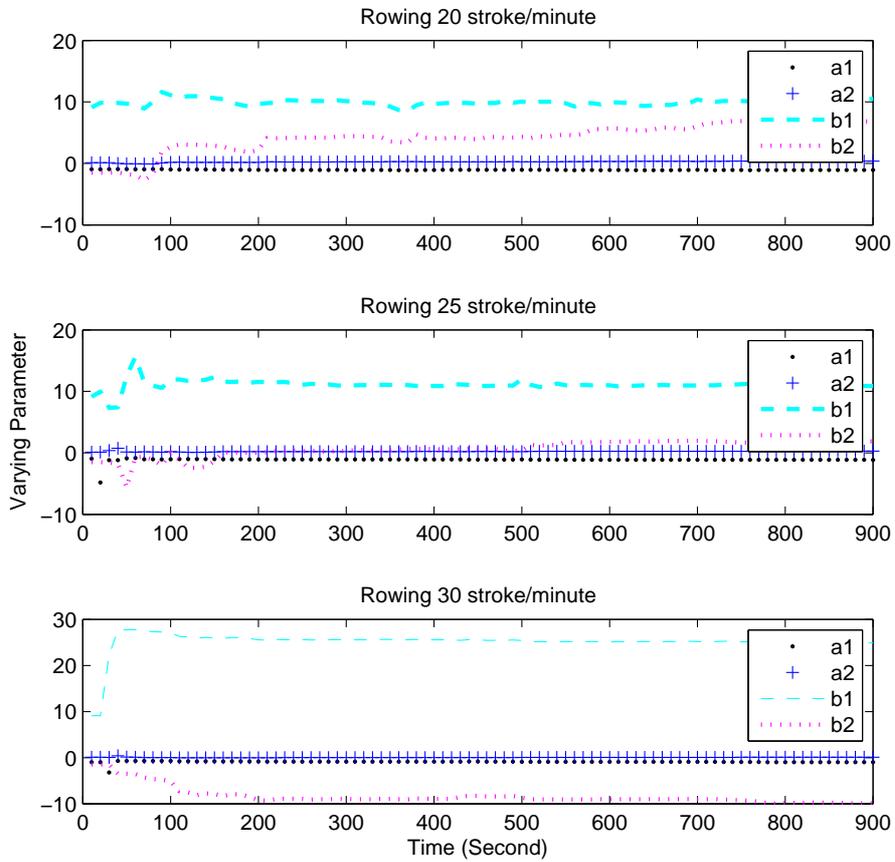}
\caption{Subject S1: Estimated LTV parameters for a rowing exercise.} \label{hoarowLTVpa}
\end{figure}

\begin{figure}
\centering
\includegraphics[scale=0.7]{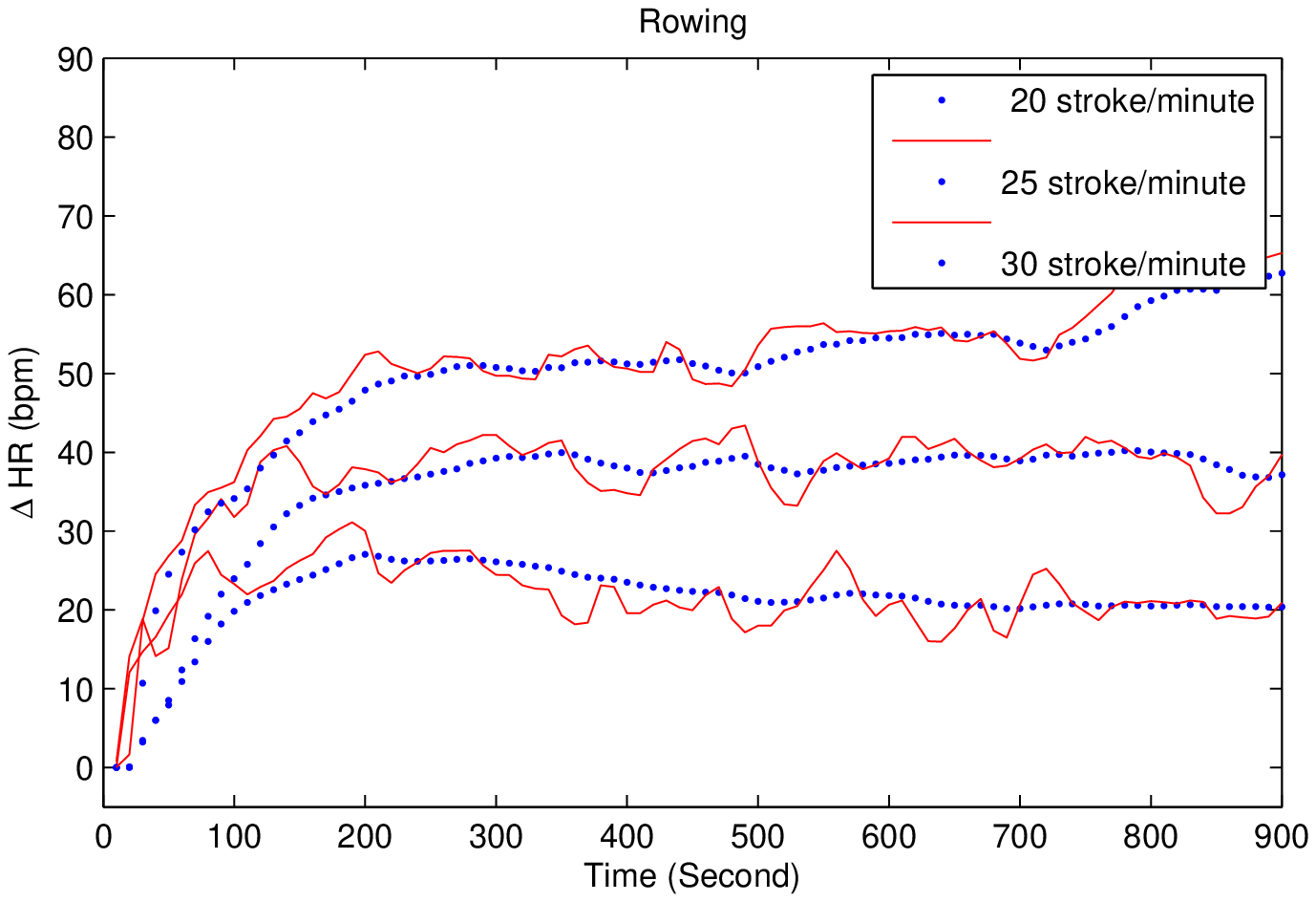}
\caption{Subject S1: Measured and estimated $\Delta$HR response for a rowing exercise.} \label{hoarowLTV}
\end{figure}

Figures \ref{hoarowLTVpa} show the time evolutions of the LTV model parameters at ER of 20, 25 and 30 strokes/min for Subject S1. The simulated HR response using LTV model at three intensity levels are shown in Fig. \ref{hoarowLTV}. In all three cases of rowing exercise, the simulated output of the $\Delta$HR response gets better using the LTV model as compared LTI models as shown in Fig. \ref{hoarowmod}. The LTI model gives MSE=0.5632 bpm while  Subject S1 engaged in rowing exercise at the ER of 30 stroke/min with $R =~0.8010$. Using the LTV model MSE gets reduced and its change to 0.0115 bpm with $R =0. 9058$. Similar findings were also observed at the ER of 20 and 24 stroke/min for  Subject S1 as shown in Fig.  \ref{hoarowLTV} and Table \ref{statmodel}.

\begin{figure}
\centering
\includegraphics[scale=0.8]{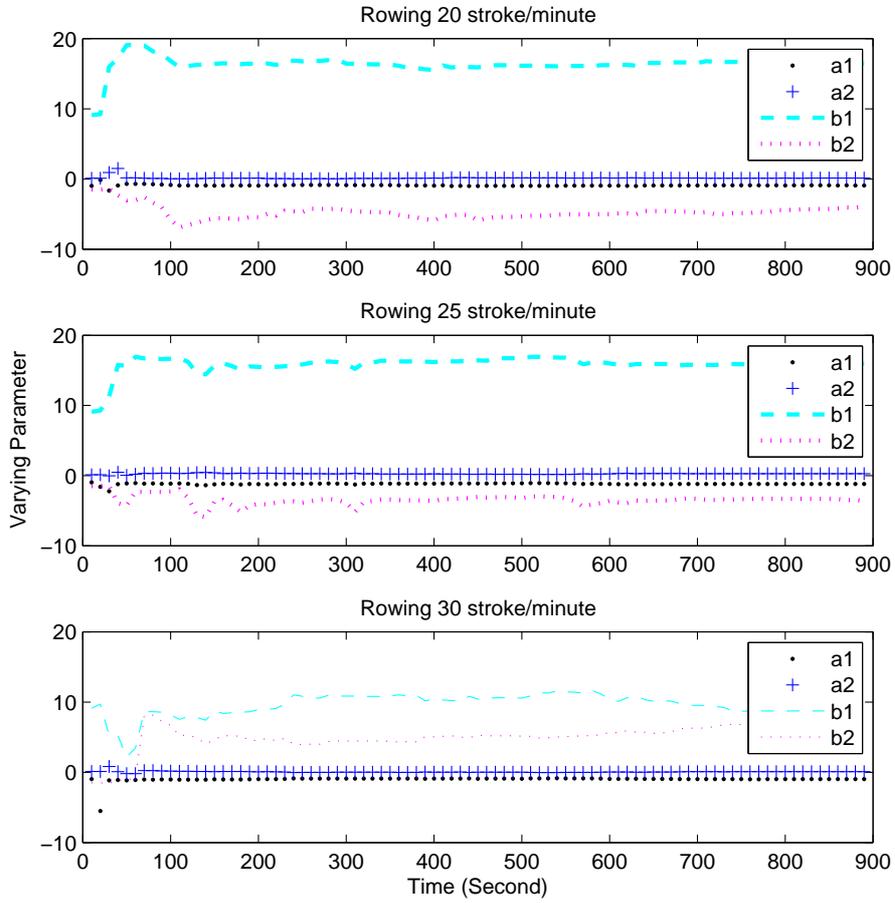}
\caption{Subject S2: Estimated LTV parameters for a rowing exercise.} \label{tedrowLTVpa}
\end{figure}

\begin{figure}
\centering
\includegraphics[scale=0.7]{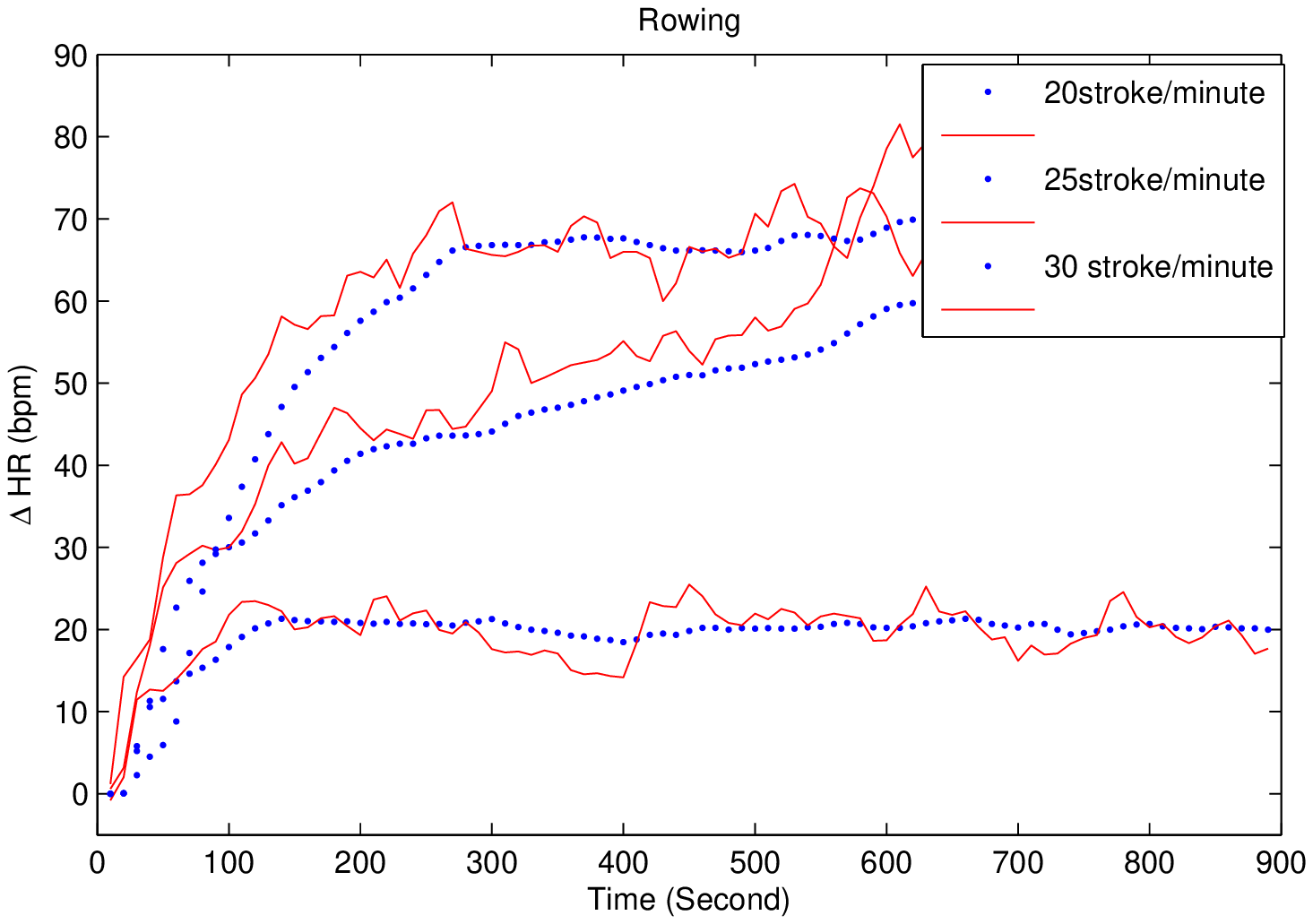}
\caption{Subject S2: Measured and estimated $\Delta$HR response for a rowing exercise.} \label{tedrowLTV}
\end{figure}

\newpage
Fig. \ref{tedrowLTVpa} shows the time evolutions of the LTV model parameters at ER of 20, 25 and 30 strokes/min for Subject S2. The simulated HR response using LTV models of Subject S2 is shown in Fig.  \ref{tedrowLTV}. In all three cases of rowing exercise for Subject S2, the simulated output of the $\Delta$HR response gets better using the LTV model as compared LTI models as shown in Fig.  \ref{tedrowmod}. The MSE between the measured and estimated values of $\Delta$HR using LTI model is 0.0062 bpm when Subject S2 is engaged in rowing exercise at the ER of 30 stroke/min with $R =~0.8829$. With LTV model,  MSE gets reduced to 0.0016 bpm with $R =0. 9589$. Similar findings were also observed at the ER of 20 and 25 stroke/min for Subject S2 as shown in Fig.  \ref{tedrowLTV} and Table \ref{statmodel}.

Finally the following conclusions were drawn from the simulation results:
\begin{enumerate}
\item{Estimation Case: the HR exhibits some degree of time variations irrespective of the type of rhythmic exercise and hence a model with time varying parameters may provide a better representation for the HR response than that of models with constant parameters.}
\item{Control Case: the adaptive control approach may be suitable to cater subject and exercise variations and hence leads to the improvement in control
    performance while controlling  the HR during these rhythmic exercises.}
    \end{enumerate}

Based on these observations, an indirect adaptive H$_\infty$ controller was designed and implemented in real-time to control the HR for a rhythmic exercise. The following section presents the design of indirect adaptive H$_\infty$ controller.
\newpage
\section{Adaptive H$_\infty$ Control Design for Unknown Type of Exercise}\label{section4.4}
The physiological control of HR reveals that HR was usually controlled by applying a conventional PI control \cite{Kawada1999a, Kawada1999b}, Biofeedback control, and Robust H$_\infty$ control techniques. Recently, advanced control techniques such as MPC has been widely used; see section \cite{Su2010}. Since the use of conventional PI control for regulation increases the workload beyond the safety limits; and if HR fails to track HR$_T$ and thus causes a safety concern, therefore this approach is considered to be unsafe for an exercising subject \cite{Kawada1999a,Kawada1999b}. The physiological control of HR using self biofeedback control was also reported in the papers \cite{Goldstein1977, Dardik1991a, Sada1999}. This biofeedback control of HR engages human brain to manipulate the exercise intensity using the HR$_{meas}$ along with the visual and audible alarming command to reach HR$_T$ during exercise. Therefore, the accuracy of controlling HR is limited to individual perception to achieve HR$_T$. This control method requires training to attain maximum performance for HR regulation. Moreover, the HR adapts the psychological factor in achieving the HR$_T$ \cite{Dardik1991a}, as brain knows the HR$_{meas}$ to control HR based on HR$_{meas}$. This leads inaccuracy in predicting the exercise intensity based on the actual physical exertion. Furthermore, the control performance of HR using biofeedback is limited for low and moderate intensity exercises \cite{Goldstein1977}. Modern control techniques are also used to control the HR response during machine dependent exercises such as Treadmill machine \cite{Su2007a, Cheng2008, Su2010}.  So far, the robust H$_\infty$ control technique emerges as most reliable and effective modern control technique for controlling the HR response during machine dependent exercises due to the fact  that it provides effective control for linear systems and has high robustness and stability in adverse operating condition such as parameter variation, high noise disturbance environment, actuator saturation and model uncertainty; see e.g., \cite{PUS00,VaUgr2000,Savkin2002}. Furthermore to impose the constraint on control input for HR regulation, the MPC technique was also employed in the literature \cite{Su2010}. The MPC approach requires accurate model to achieve desired performance, this model is unable to cater model uncertainty or parameter variations. These control approaches (PI, BIOFEEDBACK, H$_\infty$) were applied using the workload of an exercise machine. In the case of an unknown type of rhythmic exercise, the engagement of an unknown type of biomechanics varies the HR response and limits the ER in a specified bandwidth; see \cite{Cheng2009}. These variations are subject dependent and vary from subject to subject. Therefore, an effective control approach is required to cope with these variations. These variations in HR response was structured using a 2$^{nd}$-order LTI model with perturbed noise and can be estimated in real-time using RLS based on the KF approach; see Section \ref{section4.3}. In control point of view, such type of modelling problem is considered as known model structure and unknown time varying parameters and it can be controlled using indirect adaptive control approaches.  As we found that the H$_\infty$ control approach is most emerging technique for PCS \cite{Frederick2005}. Therefore, this H$_\infty$ control approach is integrated with the indirect adaptive control scheme to  manipulate ER as control input during an unknown type of rhythmic exercise.

In section \ref{section4.3}; we presented the analysis of HR response using LTI and  LTV models during walking, cycling and rowing exercises. It was observed that the HR response varies in terms of transient and steady state behaviours due to exercise and subject variations. Moreover, HR also varies with the change in the intensity level, and each rhythmic activity has its own bandwidth for manipulating the exercise intensity.  However, these variations in HR response are structured using a 2$^{nd}$-order LTI model with perturbed noise and can be estimated in real-time using RLS based on the KF approaches.

An indirect adaptive control approach has two design steps for the HR control system. The first step includes the estimation of the unknown model parameters of\ref{Eq12}. Second step includes the computation of the H$_\infty$ controller based on the identified system. The designed H$_\infty$ controller in second step will be used in the next interval of  $t$' for estimating the control input for exercising subject. These two steps require initialisation of KF for system identification, and a design of initial a H$_\infty$ controller for manipulating the ER.\\
\textbf{Initialisation of Kalman Filter for System Identification}

In this study, the unknown system is identified using RLS based on KF in a closed loop gain which uses the current values of the ER estimate (ER$_{est}$) and HR$_{meas}$ for estimating the LTV model parameters. The initialisation of the KF as an adaptation gain requires the initial values of the $\hat{\theta}_{0}$, $R_1$, $R_2$ and $P_t$ for estimating the LTV model parameters. These variables were initialised as described in section \ref{ltv}.\\
\textbf{Initialisation of Adaptive H$_\infty$ Controller}

\begin{figure}
\centering
\includegraphics[scale=0.8]{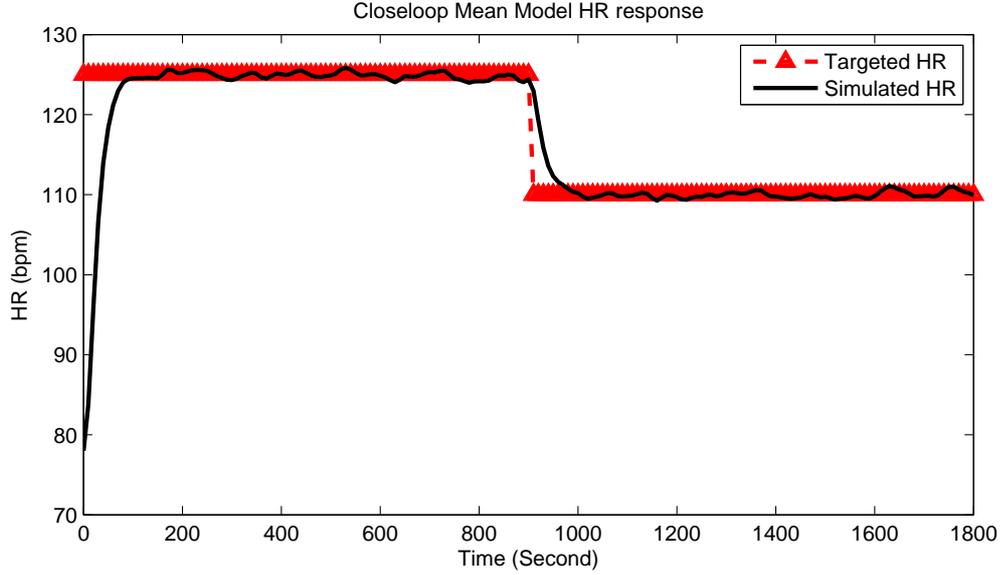}
\caption{Mean model HR response using  H$_\infty$ controller.} \label{modelcloop}
\end{figure}

An estimated mean model for walking, cycling and rowing exercises is given in \ref{Eq17} and Table \ref{statpara}. This mean model was used to design an initial discrete mixed H$_\infty$ controller.
\begin{equation}\label{Eq17}
y_t= -0.9698y_{t-1}+0.1119y_{t-2}+9.1032u_{t-1}-1.4981u_{t-2}
\end{equation}
The discrete time mixed sensitivity $H_\infty$ controller was not directly computed for the discrete time system. The selection of the weighting functions for this problem was implemented using its continuous counterpart. The discrete time model as given in \ref{Eq17} was transformed into the continues-time model (with a sampling period of 10 seconds) to design the weighing functions, i.e., W$_1$,W$_2$ and W$_3$. These weighing functions alter the closed loop system to track the desired time domain specifications. Our HR control system uses ER as control input with the desired ideal time specification such as  zero steady state error, zero overshoot, and settling time $\leq 200$ seconds. The weight functions were selected as  follows; see section \ref{controller} for further details.
\[W_1=\frac{10^{-005}s+0.004}{s+1.5\times10^{-006}},~~W_2=\frac{0.0003s+7.8\times10^{-5}}{s+0.25},~~
W_3=\frac{0.5s+0.262}{s + 0.5},\]
The continuous model was augmented with $W_1$, $W_2$ and $W_3$. This augmented model achieved the desired time domain performance in the closed loop. Furthermore the augmented plant is then converted to a discrete time model. Finally, a mixed sensitivity controller was also designed for the discrete-time system and its transfer function is given in \ref{Eq18}.
\begin{equation}\label{Eq18}
K_z(t=nk)=\frac{U(z)}{E(z)}= \frac{0.003205 z^{3}+0.0005673 z^{2}+ -0.002706z -1.293\times^{-005}} {z^{3}-2.091z^{2}+1.477z-0.3862}
\end{equation}
It is important to note that the order of the designed controller is higher than the order of the plant as in \ref{Eq18}. This increase is mainly because of the introduction of the weight functions. Since controller is designed for augmented plant and hence the order of controller is higher than that of the plant as in \ref{Eq12}. This is a well-known method for H$_\infty$ controller design; see e.g. \cite{SP96}. The simulated response of the discrete time feedback system using the mean model is given in Fig. \ref{modelcloop}, which shows that the closed loop system meets the desired time domain specifications.\\
\textbf{Indirect Adaptive Control Design}

So far, we discussed the initialisation of KF for system identification, and also presented the design of a initial controller using mean model of walking, cycling and rowing. The KF variables that were used for initialisation of KF and the designed initial mixed H$_\infty$ controller K$_z$(t=nk) was integrated in a closed loop system as the initial parameters for Indirect Adaptive Controller Scheme. The overall sampling time of the HR control system for an unknown type rhythmic activity was selected as 10 seconds. The obtained values of the ER$_{est}$ and HR$_{meas}$ is continuously used to estimate LTV model parameters in each interval of time $t$ and the predicted values of the P(t) and $\hat{\theta}$(t) will be assigned in the next interval of time ($t+1$) for the estimation of $\hat{\theta}(t+1)$.

The generalised transfer function of an adaptive H$_\infty$ controller is given in \ref{Eq18A}.
\begin{equation}\label{Eq18A}
K(z)=\frac{U(z)}{E(z)}= \frac{B_0 z^{-1}+B_1 z^{-2}+B_2 z^{-3}+B_3z^{-4}}{A_0z^{-1}+A_1z^{-2}+A_2z^{-3}+A_3z^{-4}}
\end{equation}

where, $E(t)=HR_{T}(t)-HR_{meas}(t)$. $B_0,~B_1,~B_2,~B_3$ are the feed-forward coefficients and $A_0,~A_1,~A_2,~A_3$ are the feedback coefficients of our adaptive controller. Based on the identified LTV model parameters, controller parameter \ref{Eq18A} is estimated satisfying a certain time based condition.
 This time based condition is selected because of the estimation of the 2$^{nd}$-order LTV model parameters  in the closed loop. The LTV model requires 2 samples of the measurements of the input (ER) and output (HR) at the time instant of $t-1$ and $t-2$. At N=3, the values of $\Delta$HR$_{meas}$ and ER$_{est}$ are available at the instant of $t-1$ and $t-2$. The estimated 2$^{nd}$-order LTV model using the RLS can easily adapts the accurate variations of the subjects and type of the exercise based on the measurements of ER and HR. Hence, the initial controller is designed at the N=3 at the start of the exercise.
 Now, the H$_\infty$ controller is required to continuously tune its coefficient based on the identified model parameters.  As, the HR$_T$ profile changes the controller demand i.e., ER$_T$, which results to switch the individual's Subject model parameters.  Thus, the previous estimated controller may not able to accommodate the variation of the model parameters because of the ER$_T$ level. Therefore, after adapting the exercise variations, the adaptive control algorithm adapts the model parameters variation at every 12 samples of the measurements of ER and HR. During the interval N= 3 at the start and N = 12 during the exercise, the controller is considered as invariant to achieve HR$_T$ profile.

The time based condition is defined as N$\%$delay $=~0$. In this expression, N$\%$delay denotes the remainder of the positive whole number N divided by delay. Previously, the H$_\infty$ controller was considered to be invariant and was used in closed loop for HR regulation until this condition gets satisfied. First computation of a controller $delay= 3$ is selected followed by delay $=12$.  In this way, indirect adaptive $H_\infty$ controller is re-designed firstly after 30 seconds and then every 120 seconds of an exercise.
\subsection{Simulation Study}\label{subsection4.4a}
\begin{figure}
\includegraphics[scale=0.5]{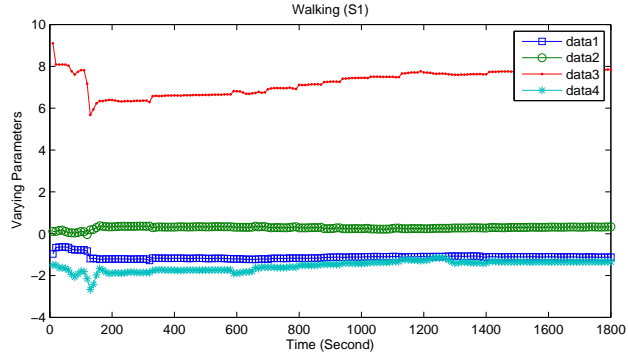}
\centering
\caption{Estimated 2$^{nd}$-order LTV model parameters for Subject S1 using his switch model for a walking exercise.}\label{hoawalkconpar}
\end{figure}

In order to validate the designed controller before it is actually implemented on human's platform, computer simulations were performed. For simulation purposes, subject HR response was assumed to be behaving like a switch model based on the intensity levels for each type of the exercise. This switch model was obtained using the LTI models as in Table \ref{arxparameter}  for  Subject S1 and Subject S2. The switching of the model parameters were therefore decided by the output of a controller, i.e., target ER (ER$_T$). To check the parameter variations in the presence of noise, the output of the switched model,   i.e., subject HR was disturbed with the sinusoidal noise. The simulations are based on the assumption that subject exactly follows the ER$_T$ and hence, ER$_T$ = ER$_{est}$. However, in real-time implementation we require an actuating system to achieve ER$_T$, and this system is known as Human Actuating System (HAS).\\
\textbf{Walking}
In the case of walking exercise, the LTI model parameters were identified at the speed of 4.5, 5.5 and 6.5 km/hr and the mean ER$_{est}$ are obtained at these intensity levels of 0.9596, 1.050 and 1.2 Hz, respectively. Based on these assumptions, the switch model for Subject S1 and Subject S2 has been  developed for walking exercise and is tabulated in Table \ref{Walking}.
\begin{ourtable}
\centering
\caption{Switch model for a walking exercise.}\label{Walking}
\begin{center}
\tabcolsep 0.5pt
\begin{tabular}{c c c}
\hline
\hline
Subject S1\\
\hline
\hline
\\
&1.03$\Delta$HR(t-1)-0.16$\Delta$HR(t-2)+0.12ER(t-1)+4.20ER(t-2) &\mbox{if}~~ER(t)$\leq$0.95Hz\\
$\hat{\Delta HR}(t)$=&0.80$\Delta$HR(t-1)+0.07$\Delta$HR(t-2)+15.16ER(t-2)-8.73ER(t-1)&\mbox{if} 0.96Hz$\leq$ER(t)$\leq$1.05Hz\\
&0.90$\Delta$HR(t-1)-0.014$\Delta$HR(t-2)+6.03ER(t-1)-0.45ER(t-2) &\mbox{if}~~ER(t)$>$1.050 Hz\\
\\
\hline
\hline
Subject S2\\
\hline
\hline
\\
&0.47$\Delta$HR(t-1)+0.07$\Delta$HR(t-2)+8.17ER(t-1)+6.52ER(t-2) &\mbox{if}~~ER(t)$\leq$0.95Hz\\
$\hat{\Delta HR}(t)$=&0.71$\Delta$HR(t-1)-0.01$\Delta$HR(t-2)+13.31ER(t-1)-1.32ER(t-2)&\mbox{if} 0.96Hz$\leq$ER(t)$\leq$1.05Hz\\
&0.50$\Delta$HR(t-1)+0.30$\Delta$HR(t-2)+3.00ER(t-1)+7.33ER(t-2) &\mbox{if}~~ER(t)$>$1.05Hz\\
\\
\hline
\hline
\end{tabular}
\end{center}
\end{ourtable}
\newpage
\textbf{Subject S1}\\The closed loop indirect adaptive H$_\infty$ controller was simulated for Subject S1 using his switch model (given in Table \ref{Walking}) for walking exercise. The estimated  2$^{nd}$-order LTV model structure parameters (a$_{1,t}$, a$_{2,t}$, b$_{1,t}$ and b$_{2,t}$) in the presence of noise are shown in Fig. \ref{hoawalkconpar}, which shows the the time evolution of these parameters. The simulated response of HR has been shown in Fig.  \ref{hoawalkHR}, where the Subject S1 achieves the desired HR$_T$ profile. Whereas in the case of exercising phase, the switch model brings the system to the desired HR$_T$ profile with an overshoot of 1.6\% which seems to be a noise disturbance. Similarly the  Subject S1 experiences an approximate under-shoot of 3\% during recovery phase before actually tracking the HR$_T$ profile. Hence, the developed control approach is capable of adapting the variations in model parameters due to change in the controller demand i.e., ER$_T$. It can be also observed from the Fig. \ref{hoacycconER} that the indirect adaptive H$_\infty$ controller tracks the HR$_T$ in a desired bandwidth of the ER$_T$ and also adapts the model parameters variations due to intensity levels of walking exercise.\\
\begin{figure}
\centering
\includegraphics[scale=0.5]{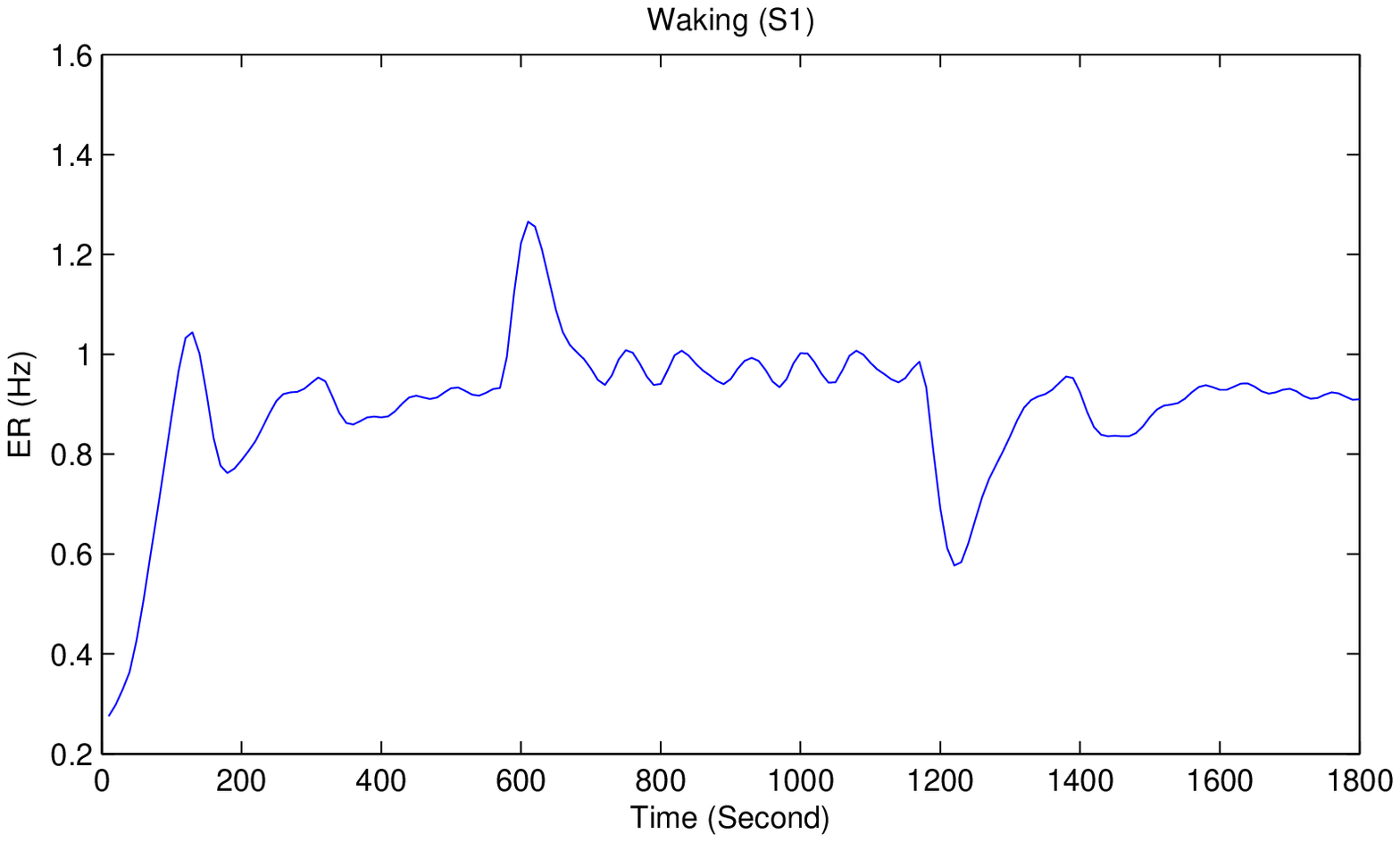}
\caption{ER$_T$ for Subject S1 switch model for a walking exercise.} \label{hoawalkER}
\end{figure}
\begin{figure}
\centering
\includegraphics[scale=0.5]{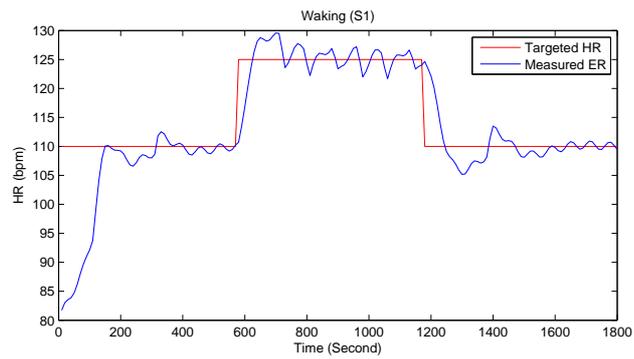}
\caption{HR$_T$ and simulated HR response of the switch model for Subject S1 using an adaptive H$_\infty$ controller for a walking exercise.} \label{hoawalkHR}
\end{figure}
\textbf{Subject S2}\\Similar observations were observed for Subject S2 using his switch model developed for walking exercise. The estimated model parameters of Subject S1 using 2$^{nd}$-order LTV model structure in the presence of noise are shown in Fig. \ref{tedwalkconpar}. The simulated response of HR is shown in Fig. \ref{tedwalkconHR}. The result indicates that the Subject S2 achieves the HR$_T$ profile. It is worth noting form the Fig. \ref{tedwalkconinpER} that the indirect adaptive H$_\infty$ controller adapts the ER$_T$ in a desired bandwidth and also adapts the model parameters variations due to intensity levels of walking exercise for the Subject S2 as well.\\
\begin{figure}
\includegraphics[scale=0.5]{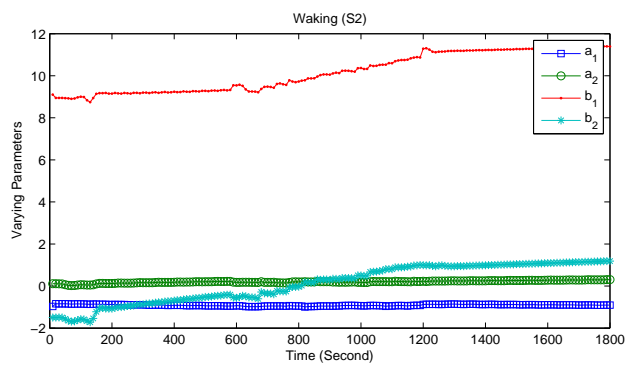}
\centering
\caption{Estimated 2$^{nd}$-order LTV model parameters for Subject S2 using his switch model for a walking exercise.} \label{tedwalkconpar}
\end{figure}
\begin{figure}
\centering
\includegraphics[scale=0.5]{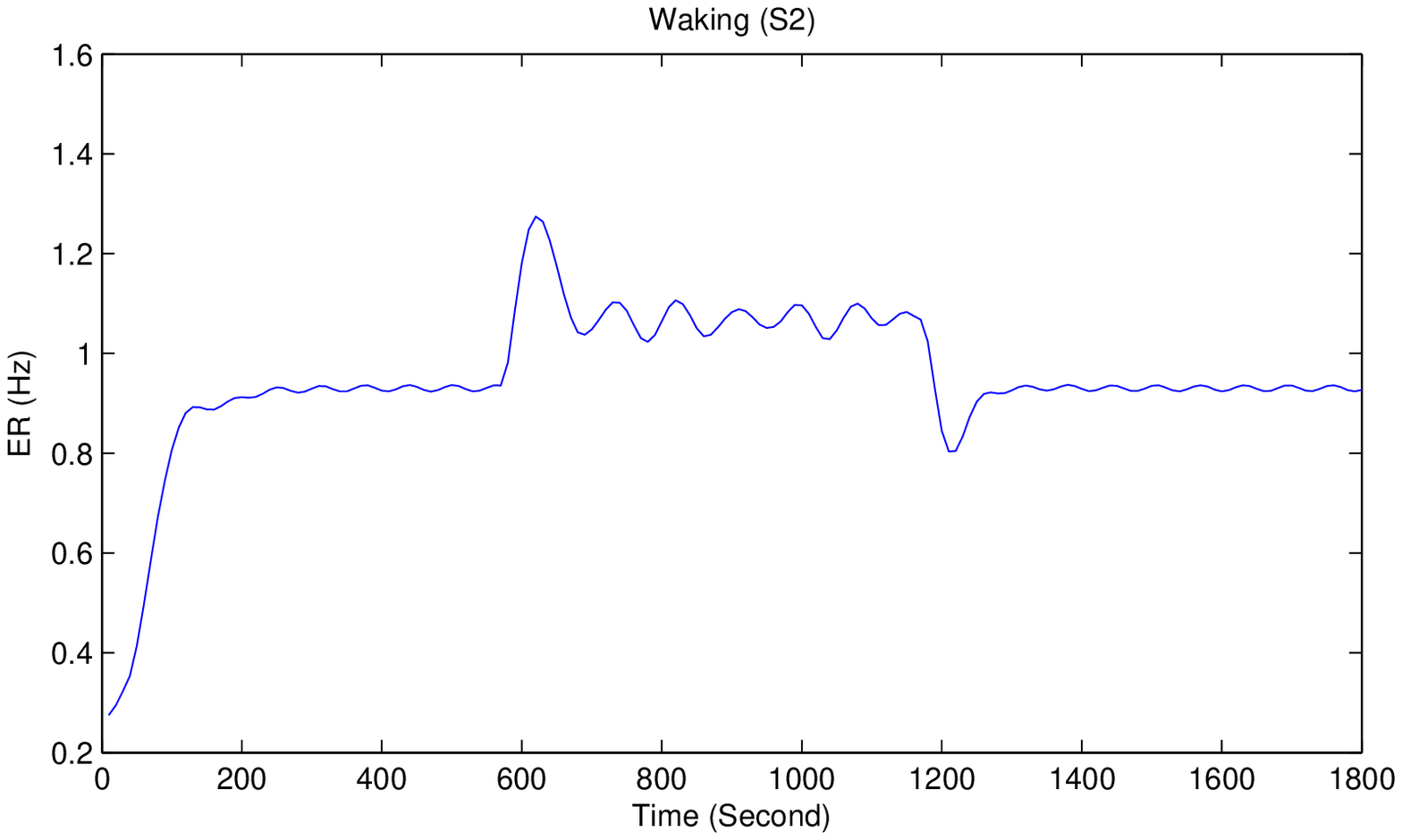}
\caption{ER$_T$ for Subject S2 switch model for a walking exercise.} \label{tedwalkconinpER}
\end{figure}
\begin{figure}
\centering
\includegraphics[scale=0.5]{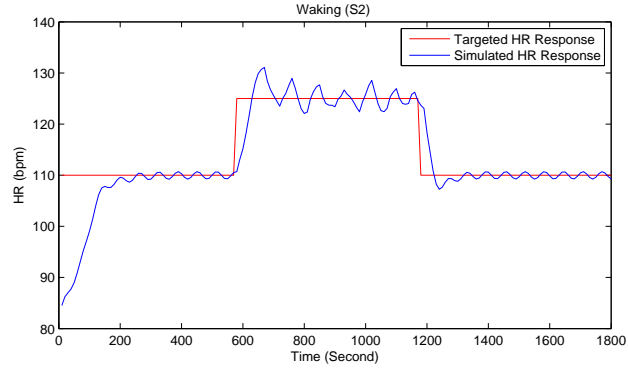}
\caption{HR$_T$ and simulated HR response of the switch model for Subject S2 using an adaptive H$_\infty$ controller for a walking exercise.} \label{tedwalkconHR}
\end{figure}
\newpage
\textbf{Cycling}

\begin{figure}
\includegraphics[scale=0.5]{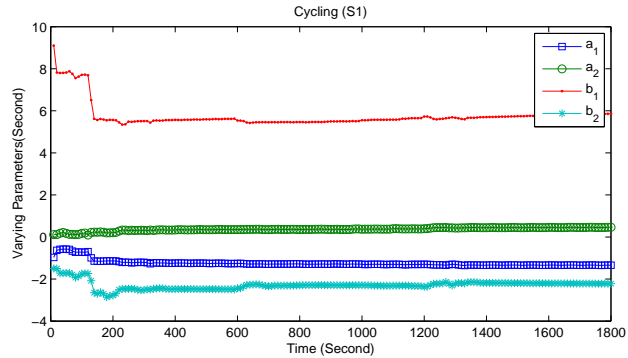}
\centering
\caption{Estimated 2$^{nd}$-order LTV model parameters of Subject S1 using his switch model for a cycling exercise.} \label{hoacycconpar}
\end{figure}

\begin{figure}
\centering
\includegraphics[scale=0.5]{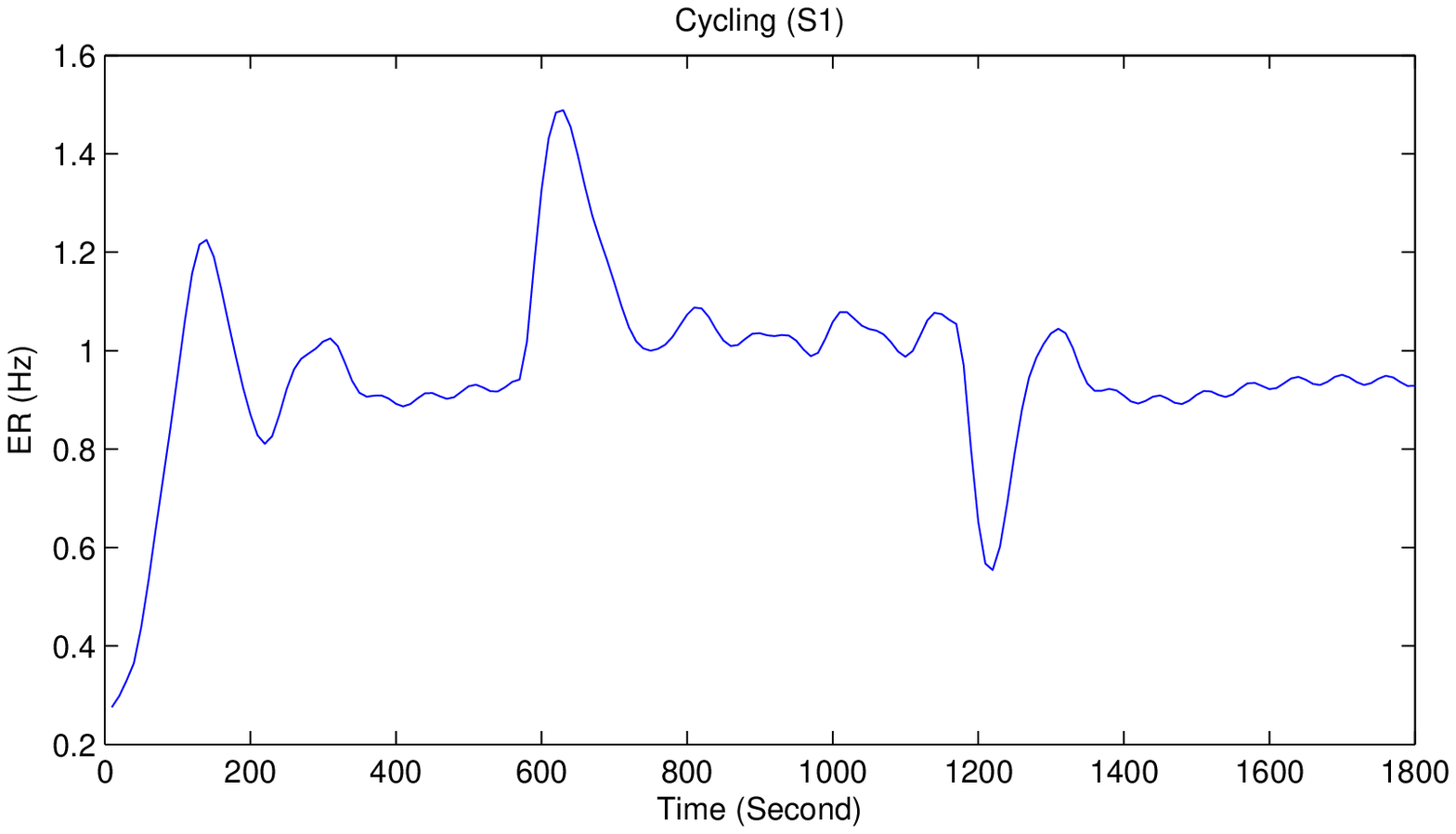}
\caption{ER$_T$ for Subject S1 switch model for a cycling exercise.} \label{hoacycconER}
\end{figure}

\begin{figure}
\centering
\includegraphics[scale=0.52]{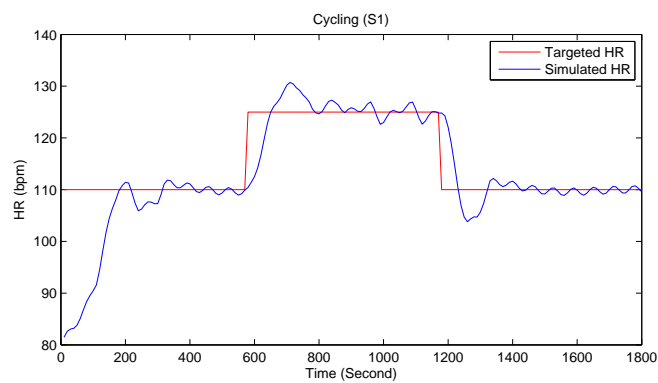}
\caption{HR$_T$ and simulated HR response of the switch model for Subject S1 using an adaptive H$_\infty$ controller for a cycling exercise.} \label{hoacycconHR}
\end{figure}

\begin{figure}
\includegraphics[scale=0.5]{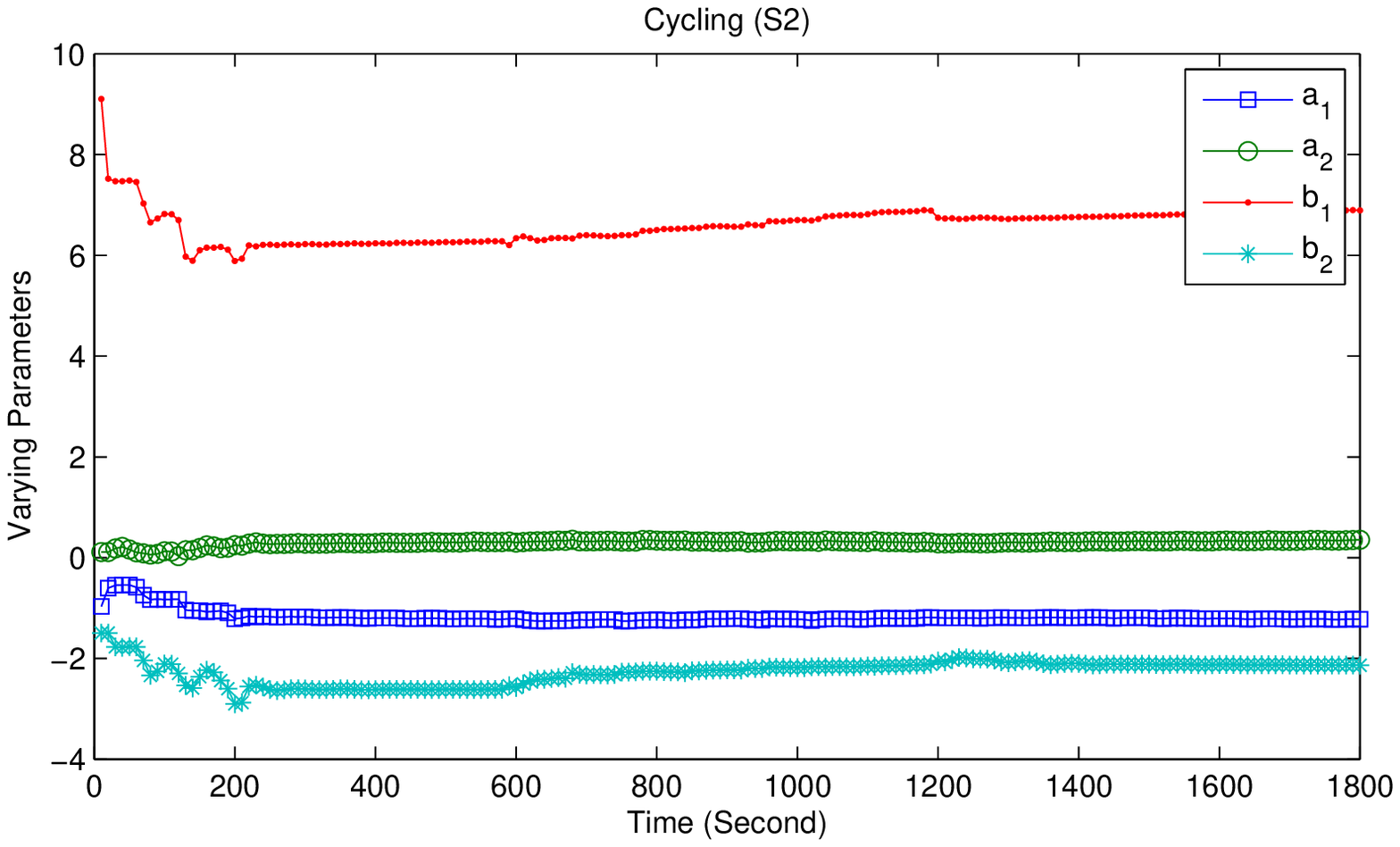}
\centering
\caption{Estimated 2$^{nd}$-order LTV model parameters for Subject S2 using his switch model for a cycling exercise.} \label{tedcycconpar}
\end{figure}

\begin{figure}
\centering
\includegraphics[scale=0.5]{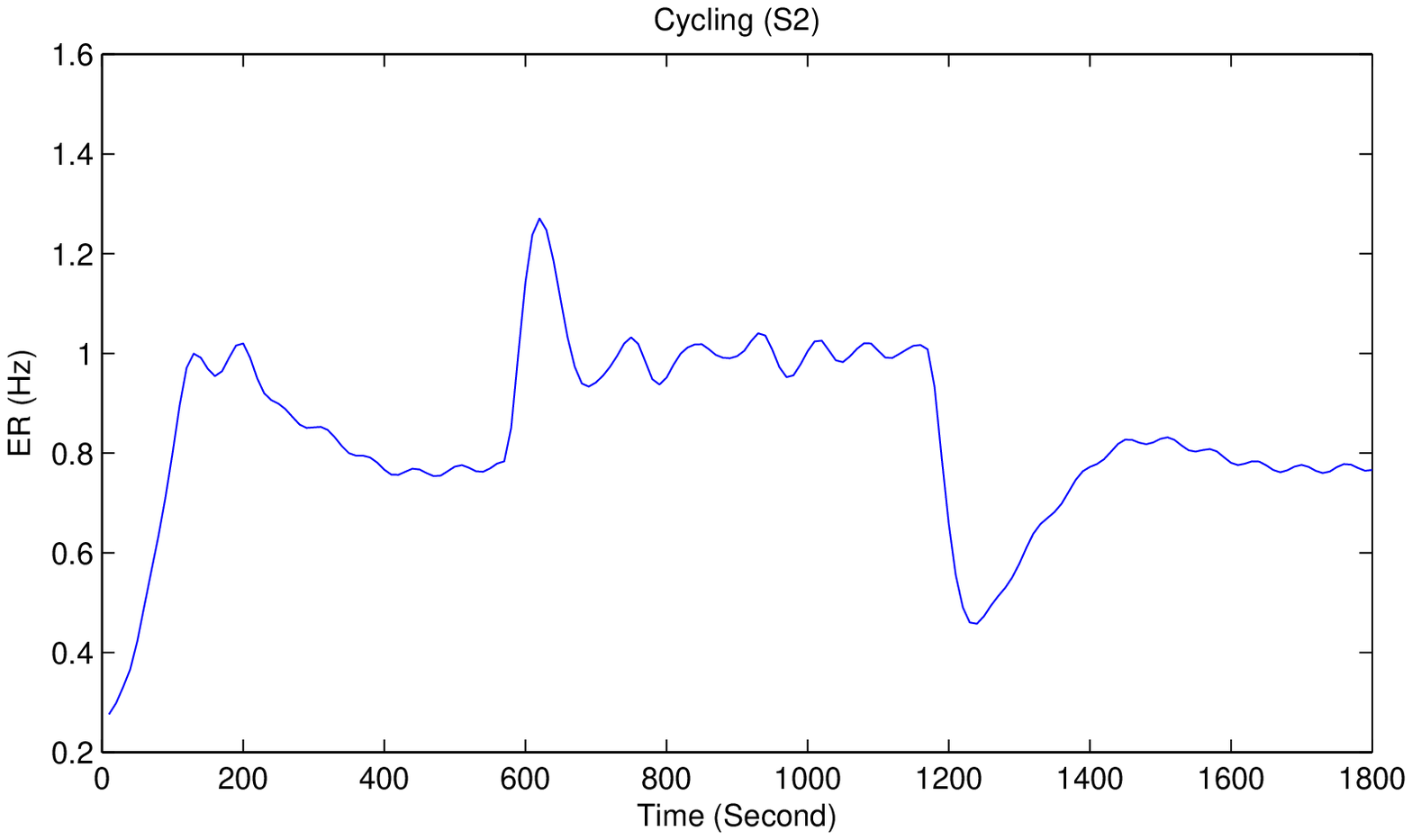}
\caption{ER$_T$ for Subject S2 switch model for a cycling exercise.}\label{tedcycconER}
\end{figure}

\begin{figure}
\centering
\includegraphics[scale=0.5]{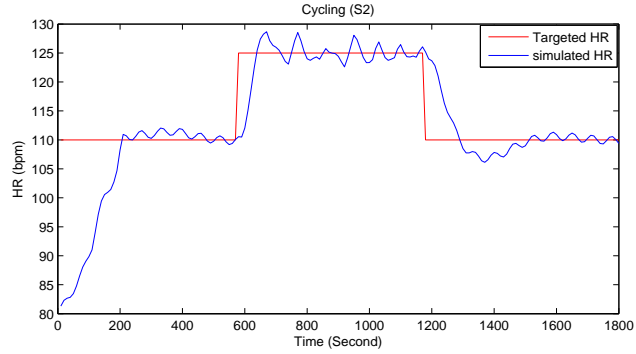}
\caption{HR$_T$ and simulated HR response of the switch model for Subject S2 using an adaptive H$_\infty$ controller for a cycling exercise.} \label{tedcycconHR}
\end{figure}

In the case of cycling exercise the LTI model parameters were identified at the cadence rate of 48, 60 and 72 Pedals/min and the mean estimated values of ER at these intensity levels are 0.8520, 1.013 and 1.2 Hz. Based on these assumptions the switch model for Subject S1 and Subject S2 was developed for cycling exercise and is tabulated in Table \ref{Cycling}.\\
\begin{ourtable}
\centering
\caption{Switch model for a cycling exercise.}\label{Cycling}
\begin{center}
\tabcolsep 0.5pt
\begin{tabular}{c c c}
\hline
\hline
Subject S1\\
\hline
\hline
\\
&1.10$\Delta$HR(t-1)-0.24$\Delta$HR(t-2)+0.10ER(t-1)+3.35ER(t-2) &\mbox{if}~~ER(t)$\leq$0.85Hz\\
$\hat{\Delta HR}(t)$=&1.11$\Delta$HR(t-1)-0.18$\Delta$HR(t-2)+2.00ER(t-1)+0.094ER(t-2)&\mbox{if} 0.85Hz$\leq$ER(t)$\leq$1.013Hz\\
&1.10$\Delta$HR(t-1)-0.13$\Delta$HR(t-2)+6.10ER(t-1)-4.69ER(t-2) &\mbox{if}~~ER(t)$>$1.013 Hz\\
\\
\hline
\hline
Subject S2\\
\hline
\hline
\\
&1.11$\Delta$HR(t-1)-0.20$\Delta$HR(t-2)+0.73ER(t-1)+2.11ER(t-2) &\mbox{if}~~ER(t)$\leq$0.85Hz\\
$\hat{\Delta HR}(t)$=&0.96$\Delta$HR(t-1)-0.03$\Delta$HR(t-2)+13.62ER(t-1)-10.87ER(t-2)&\mbox{if} 0.852Hz$\leq$ER(t)$\leq$1.013Hz\\
&0.89$\Delta$HR(t-1)+0.30$\Delta$HR(t-2)+7.89ER(t-1)-2.86ER(t-2) &\mbox{if}~~ER(t)$>$1.013 Hz\\
\\
\hline
\hline
\end{tabular}
\end{center}
\end{ourtable}
\newpage
\textbf{Subject S1}\\The closed loop indirect adaptive H$_\infty$ controller was simulated for Subject S1 using his switch model obtained for cycling exercise. The estimated model parameters of Subject S1 using 2$^{nd}$-order LTV model structure in the presence of noise are shown in Fig.  \ref{hoacycconpar}. The simulated response of HR is shown in Fig. \ref{hoacycconHR} indicate that Subject S1 achieves the HR$_T$ profile. However, during exercising phase the initial HR response got the  approx. overshoot of 4\%. The system settles to the desired HR$_T$ profile after a small interval of time. During recovery phase of the cycling exercise, switch model for Subject S1 closely tracks the HR$_T$ profile after experiencing  approx. undershoot of 3.2\%. Fig. \ref{hoacycconER} indicates that the HR$_\infty$ controller adapts the ER$_T$ in a desired bandwidth and also adapts the model parameters variations due to intensity levels of cycling exercise.\\
\textbf{Subject S2}\\Similar observations were observed for Subject S2 using his switch model for cycling exercise. The estimated model parameters of Subject S2 using 2$^{nd}$-order LTV model structure in the presence of noise are shown in Fig. \ref{tedcycconpar}. The developed control approach adapts the model parameter variations. The simulated response of HR is shown in Fig.  \ref{tedcycconHR}, result indicates that the Subject S2 achieves the HR$_T$ profile.  However, during recovery phase of an exercise the Subject S2 experience the approx. undershoot of 2.7\% and after a certain interval of time H$_\infty$ controller tracks the HR$_T$ profile. Fig.  \ref{tedcycconER} indicates the indirect adaptive H$_\infty$ controller adapts the ER$_T$ in a desired bandwidth and also adapts the model parameters variations due to intensity levels of cycling exercise.\\
\newpage
\textbf{Rowing}

\begin{figure}
\includegraphics[scale=0.5]{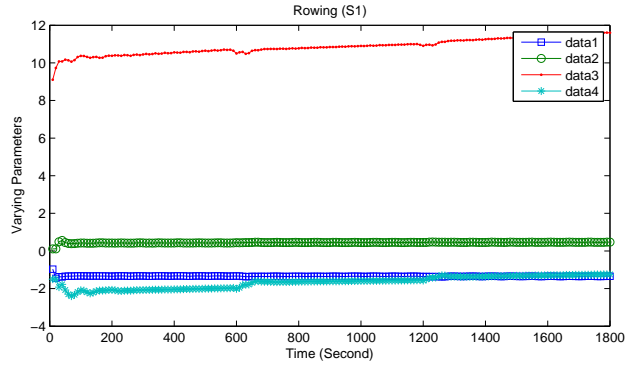}
\centering
\caption{Estimated 2$^{nd}$-order LTV model parameters for Subject S1 using his switch model for a rowing exercise.} \label{hoarowconcpara}
\end{figure}

\begin{figure}
\centering
\includegraphics[scale=0.5]{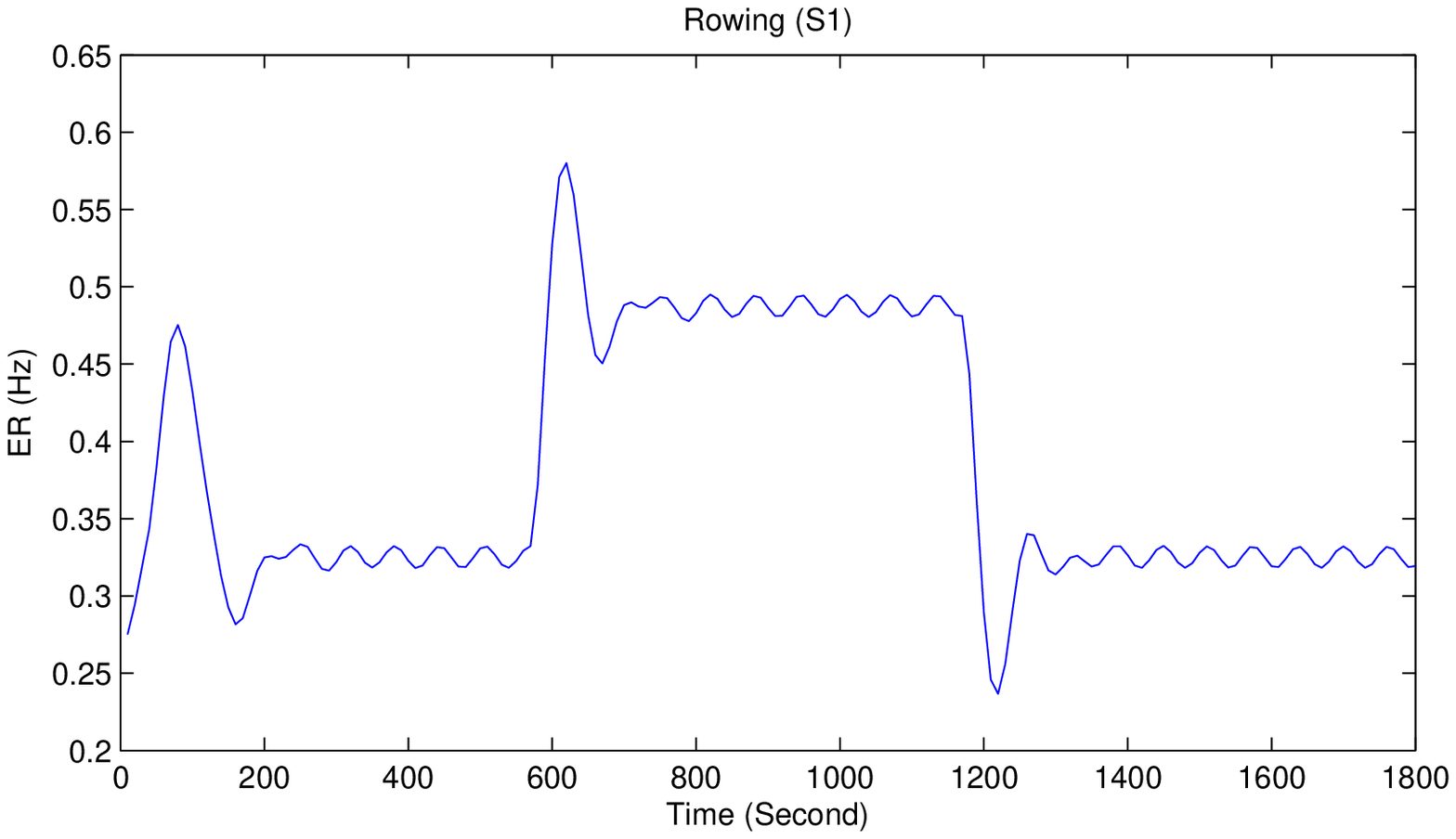}
\caption{ER$_T$ for Subject S1 switch model for a rowing exercise.} \label{hoarowconER}
\end{figure}

\begin{figure}
\centering
\includegraphics[scale=0.5]{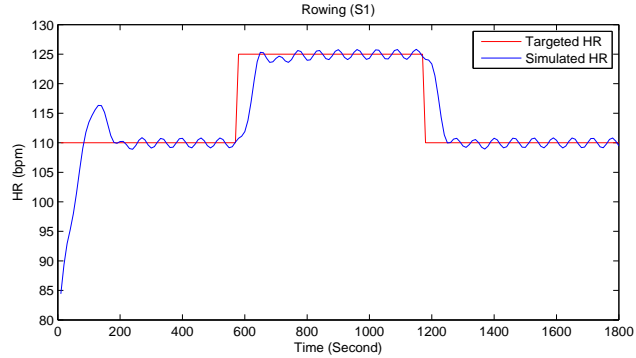}
\caption{HR$_T$ and simulated HR response of the switch model for Subject S1 using an adaptive H$_\infty$ controller for a rowing exercise.} \label{hoarowconHR}
\end{figure}

In the case of rowing exercise, LTI model parameters were identified at the speed of 20, 25 and 30 strokes/min and the mean estimated values of ER at these intensity levels are 0.333, 0.42 and 0.5 Hz, respectively. Based on these assumptions, the switch model for Subject S1 and Subject S2 was developed for rowing exercise and is tabulated in Table \ref{Rowing}.
\begin{ourtable}
\centering
\caption{Switch model for a rowing exercise.}\label{Rowing}
\begin{center}
\tabcolsep 0.5pt
\begin{tabular}{c c c}
\hline
\hline
Subject S1\\
\hline
\hline
\\
&1.22$\Delta$HR(t-1)-0.44$\Delta$HR(t-2)-0.46ER(t-1)+14.78ER(t-2) &\mbox{if}~~ER(t)$\leq$0.33Hz\\
$\hat{\Delta HR}(t)$=&1.14$\Delta$HR(t-1)-0.28$\Delta$HR(t-2)+9.70ER(t-1)+3.22ER(t-2)&\mbox{if} 0.331Hz$\leq$ER(t)$\leq$0.42Hz\\
&1.20$\Delta$HR(t-1)-0.22$\Delta$HR(t-2)+33.74ER(t-1)-27.38ER(t-2) &\mbox{if}~~ER(t)$>$0.42Hz\\
\\
\hline
\hline
Subject S2\\
\hline
\hline
\\
&0.92$\Delta$HR(t-1)-0.13$\Delta$HR(t-2)+23ER(t-1)-11.24ER(t-2) &\mbox{if}~~ER(t)$\leq$0.33Hz\\
$\hat{\Delta HR}(t)$=&1.30$\Delta$HR(t-1)-0.32$\Delta$HR(t-2)+19.69ER(t-1)-13.20ER(t-2)&\mbox{if} 0.331Hz$\leq$ER(t)$\leq$0.42Hz\\
&1.06$\Delta$HR(t-1)-0.15$\Delta$HR(t-2)+2.07ER(t-1)+11.25ER(t-2) &\mbox{if}~~ER(t)$>$0.42Hz\\
\\
\hline
\hline
\end{tabular}
\end{center}
\end{ourtable}
\newpage
\textbf{Subject S1}\\The closed loop indirect adaptive H$_\infty$ controller was simulated for Subject S1 using his switch model obtained for rowing exercise. The estimated model parameters of Subject S1 using 2$^{nd}$-order LTV model structure in the presence of noise are shown in Fig.  \ref{hoarowconcpara}. The developed control approach adapts the model parameter variation. The simulated response of HR is shown in Fig.  \ref{hoarowconHR}, this result indicates that the Subject S1 achieves the HR$_T$ profile. However, the initial HR response has an approx. overshoot of 4.32\% during the warm up phase before achieving the desired HR$_T$ profile. More importantly, Fig. \ref{hoarowconER} indicates that the indirect adaptive HR$_\infty$ controller adapts the ER$_T$ in a desired bandwidth and the model parameters variations due to intensity levels of a rowing exercise.\\

\begin{figure}
\includegraphics[scale=0.5]{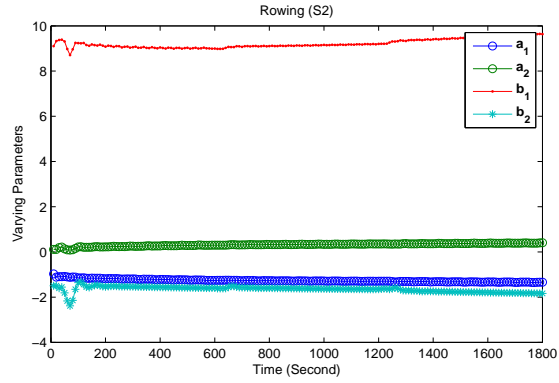}
\centering
\caption{Estimated 2$^{nd}$-order LTV model parameters for Subject S2 using his switch model for a rowing exercise.} \label{tedrowconpara}
\end{figure}

\begin{figure}
\centering
\includegraphics[scale=0.5]{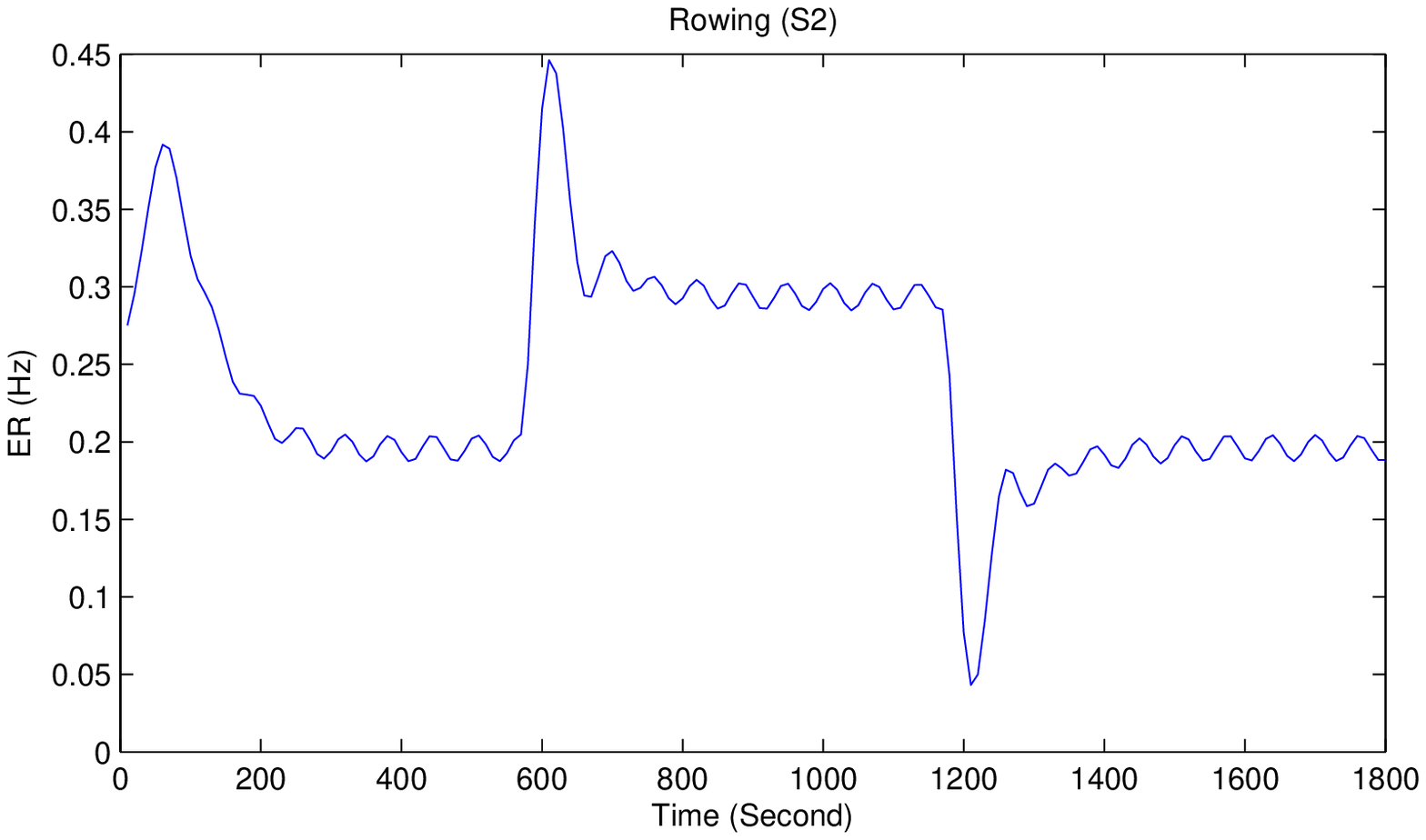}
\caption{ER$_T$ for Subject S2 switch model for a rowing exercise.} \label{tedrowconER}
\end{figure}

\begin{figure}
\centering
\includegraphics[scale=0.5]{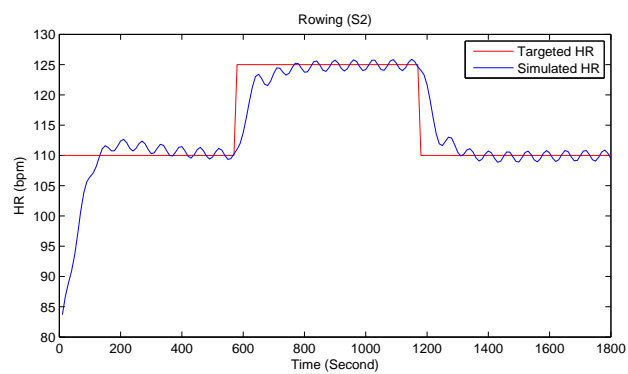}
\caption{HR$_T$ and simulated HR response of the switch model for Subject S2 using an adaptive H$_\infty$ controller for a rowing exercise.} \label{tedrowconHR}
\end{figure}

\textbf{Subject S2}\\Similar observations were observed for Subject S2 using the switch model obtains for rowing exercise. The estimated model parameters of Subject S2 using 2$^{nd}$-order LTV model structure in the presence of noise are shown in Fig.  \ref{tedrowconpara}. The developed control approach adapts the model parameter variations. The simulated response of HR is shown in Fig. \ref{tedrowconHR} which indicates that the Subject S2 achieves the HR$_T$ profile in all phases of an exercise. Fig. \ref{tedrowconER} indicates that the indirect adaptive H$_\infty$ controller adapts the ER$_T$ in a desired bandwidth during warmup and exercising phase of a rowing exercise.

The simulation results during walking, cycling and rowing exercises clearly indicate that the developed control methodology is capable of tracking the exercise and subject variations in order to achieve the desired HR$_T$ profile.  As a result, the transient response varies subject to subject in order to achieve the HR$_T$ profile.  The average RMSE between the HR$_T$ and HR simulated of two subjects is calculated 5.58 bpm, 6.52 bpm  and 3.5 bpm for a walking, a cycling and a rowing, respectively. Some of the cases of walking, cycling and rowing experienced the average approx. overshoot of 2.52\% and 3.12\% while tracking the HR$_T$ profile. Hence, the overshoot within the range of 0 to 4.5\%  and RMSE between 0 to 6.52 bpm as a steady state error is considered as the acceptable range for the purpose of the validation for the real-time HR regulation system.
\section{Real-time Implementation of Heart Rate Regulation System Using H$_\infty$ Controller}\label{section4.5}
\begin{figure}
\begin{centering}
\includegraphics[scale=0.6]{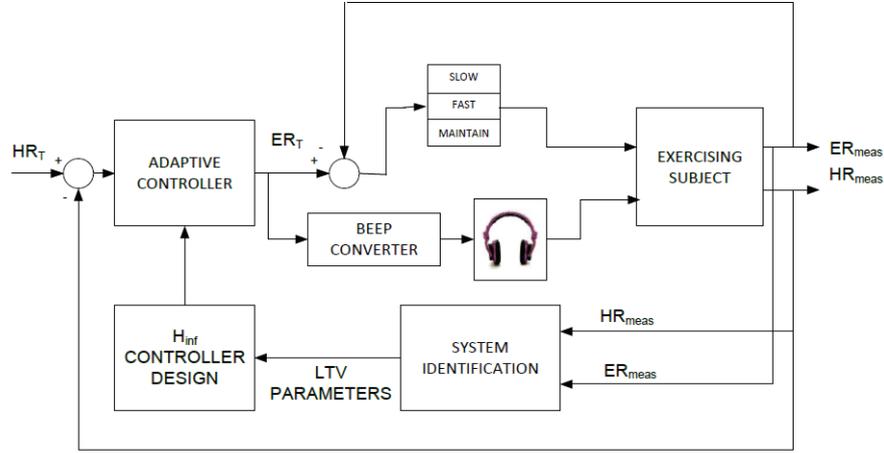}
\caption{Block diagram of the real-time implementation of HR regulation system.}\label{blkdiaga}
\end{centering}
\end{figure}
\begin{figure}
\begin{centering}
\includegraphics[scale=0.55]{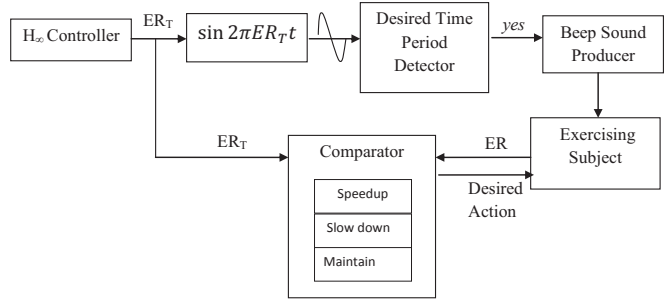}
\caption{Block diagram of the human actuating system.}\label{blkHAS}
\end{centering}
\end{figure}

An indirect adaptive H$_\infty$ control design and the simulations for HR regulation system during unknown type of activity has been presented in section \ref{section4.4}. This section presents the real-time implementation of the HR regulation system similar to one demonstrated in section \ref{subsection4.4a}. The developed controller was implemented in real-time using the LabVIEW software from National Instruments (NI) as shown in the system block diagram \ref{blkdiaga}. A human actuating system (HAS) was also developed in real-time in order to establish the communication between the controller and exercising subject using the block diagram of the HAS as shown in Fig. \ref{blkHAS}.
The role of HAS is to actuate the exercising subject to perform a desired exercising task in terms of an audible beep and alarming signal to achieve the targeted HR. The beep signal represents the desired time period of the physical activity and the alarming signal indicates the desired action that is required by the subject to achieve HR$_T$. The desired time period to achieve ER$_T$ is extracted by detecting its zero crossing. Moreover, the desired actions are speed up, slow down or maintain the rate of current activity, it has been generated using the comparison of the current values of ER$_T$ and ER$_{est}$.

\subsection{System Description}
The implementation of the designed HR regulation system requires the measurement of HR and ER, which are abbreviated as HR$_{meas}$ and ER$_{meas}$ in Fig.  \ref{blkdiaga}. These quantities need to be monitored continuously in real time. The HR$_{meas}$ was achieved using HR measuring system with sampling rate  of 10 seconds. We refer to section \ref{HRmeasured} for further details about HR measurement.

The ER$_{meas}$ of the exercising subject is achieved using its estimate, i.e., ER$_{est}$. This estimate is achieved using TA measurement, which detects the acceleration of the body along $x$, $y$ and $z$ axes and represents periodicity of rhythmic movement.

For further details; see section \ref{TAmeasured}. 
At the start of a particular exercise, initial mean controller computes the ER${_T}$ and gives this input to HAS for the first 30 seconds (N=3, with a sampling interval of 10 seconds). After the period of 30 seconds, LTV model parameters are calculated at every sampling interval, whereas the $H_\infty$ controller is adaptively re-designed after every 120 seconds; see section \ref{section4.5}.

The working principle of the developed system can be explained in the following sequential steps:
\begin{enumerate}
\item{LTV model parameters are identified using the current measurements of HR and ER.}
\item{The final identified LTV model is used to estimate the coefficients of H$\infty$ controller and consequently the estimated controller will be applied in the next sampling interval.}
\item{The H$\infty$ controller estimates the ER$_{T}$ based on the values HR$_{T}$ and HR$_{meas}$.}
\item{ An audio beep signal is generated and its frequency is determined by ER$_{T}$.}
\item{A visual command signal is also generated below a certain threshold through HAS  which guides the exerciser to do a particular action during exercise to achieve ER$_T$ such as slow-down, speed up in order to maintain the exercising pace.}
\end{enumerate}

\subsection{Subjects}
To examine the performance of the developed system, 6 healthy subjects were required to participate in this exercising activity. The physical characteristics i.e., age, weight and height of these participating subjects are given in Table \ref{Phychar}. Written informed consent was obtained from all the participants of this study.
\begin{table}
\centering
\caption{Subjects: physical characteristics.}\label{Phychar}
\begin{center}
\tabcolsep 2.0pt
\begin{tabular}{c c c c c}
\hline
Subject & Age (yrs) & Mass (kg) & Height (cm) & BMI\\
\hline
1   &	  24  &	78	&  184 &	23.3\\
2	&     33  &	77	&  167 &	27.6\\
3	&     34  &	76	&  174 &	25.1\\
4	&     35  &	71.6&  165 &	26.3\\
5	&     39  &	95	&  175.3  &	30.9\\
6	&     32  &	72	&  154.3  &  30.2\\
Mean$\pm$std	& 32.8$\pm$4.9	& 78.3$\pm$8.6	& 169.9$\pm$10.2 & 27.2$\pm$2.9 \\
\hline
\end{tabular}
\end{center}
\end{table}
\subsection{Exercising Protocol}
The subjects were required to familiarise themselves with the experimental setup. Before beginning the exercise, an initial 5 minute  recording of the HR during the resting period was used to obtain the HR$_{rest}$. This HR$_{rest}$ was used in the process of system identification in the form of closed loop. The reference HR profile was selected in three phases of an exercise. The first phase of exercise is considered as warm up period of 10 minutes duration. During the warm-up period, HR$_{T}$ is selected as 100 bpm. The second phase of exercise is considered as an exercising period again with the duration of 10 minutes. During the exercising period HR$_{T}$ is selected as 115 bpm which is 65\% of HR$_{max}$.  The third phase is considered as the cool-down period with duration of 10 minutes as well. During the cool down period HR$_{T}$ is selected as 100 bpm. The total duration of an exercise with this protocol is 30 minutes.
\subsection{Performance Measure}
To evaluate the performance of the closed loop system, three quantitative performance measures are defined as follows:
\begin{enumerate}
\item{ First is the root mean square (RMS) tracking error. This gives the quality of the tracking between the $HR_{meas}$ against HR$_{T}$(t) and is given by:
\begin{equation}\label{Eq19}
e(t)_{rms}=\sqrt{\frac{1}{N}\sum_{t=1}^{N}(HR_{T}-HR_{meas})^{2}}
\end{equation}
}
\item{The second criterion is based on the changes in the controller output signal ER$_ {T}$(t) . The RMS value of this signal and is given by:
\begin{equation}\label{Eq20}
\Delta ER_{T}(t)_{rms}= \sqrt{\frac{1}{N-1}\sum_{t=2}^{N}(ER_{T}(t)-ER_{T}(t-1))^{2}}
\end{equation}}
\item{Third is the fitness of quality measure between the ER$_{T}$(t) (i.e., input to the HAS) and the output of the exerciser (ER$_{est}$). This is calculated by finding the correlation coefficient $R$ and is given by:
\begin{equation}\label{Eq21}
R=\frac{\sqrt{E(ER_{T}ER_{est})-E(ER_{T})E(ER_{est})}}{\sqrt{E(ER_{T}^{2})-E^{2}(ER_{T})}\sqrt{E(ER_{est}^{2})-E^{2}(ER_{est})}}\\
\end{equation}
Where $E(\cdot)$ is the expected value and is computed by summing over the entire time period.}
\end{enumerate}
\subsection{Experimental Results}
As mentioned earlier, the developed HR regulation system is based on 6 healthy subjects performing cycling and rowing exercises. The participating subjects have different physical characteristics as given in Table \ref{Phychar}.\\
\textbf{Subject S1}
\begin{figure}
\centering
\includegraphics[scale=0.55]{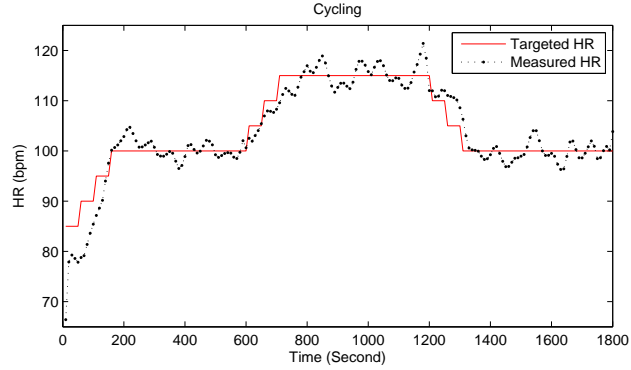}
\caption{HR profile tracking of Subject S1 during  cycling.} \label{S1HRcyc}
\end{figure}
\begin{figure}
\centering
\includegraphics[scale=0.55]{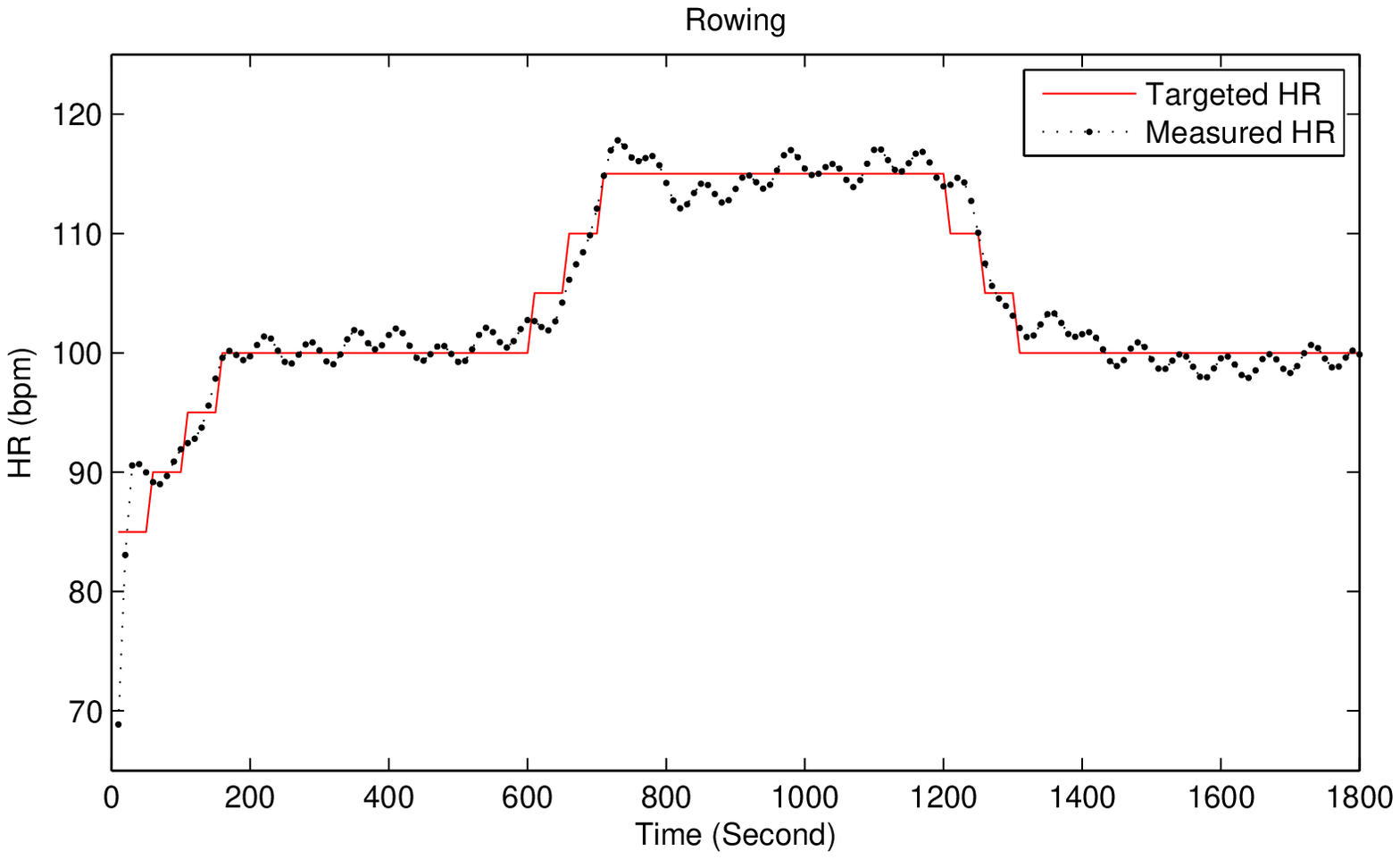}
\caption{HR profile tracking of Subject S1 during rowing.} \label{S1HRrow}
\end{figure}
\begin{figure}
\centering
\includegraphics[scale=0.55]{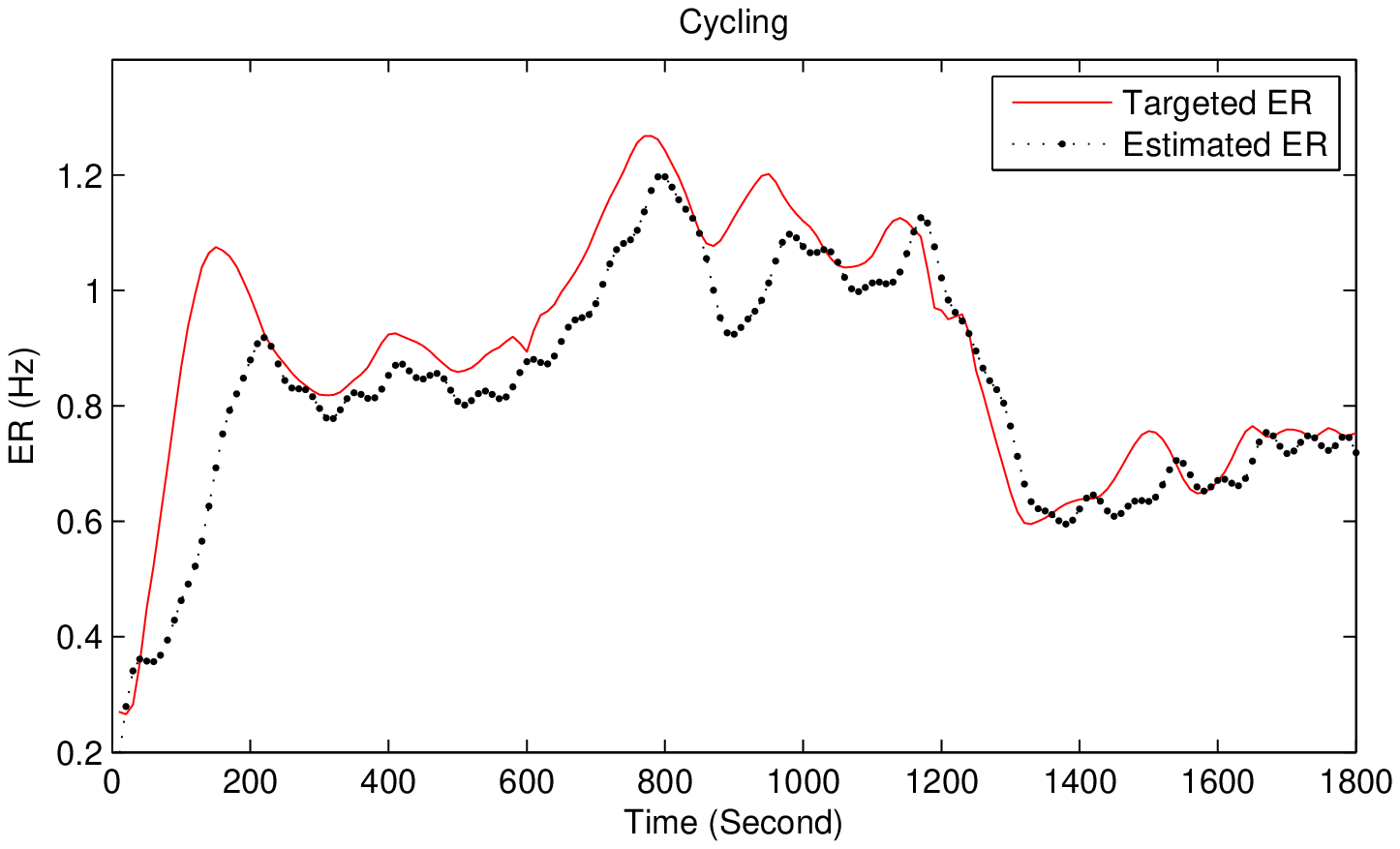}
\caption{Tracking of Subject S1 between ER$_{T}$ (solid line) and ER$_{est}$ (dotted line) during cycling.} \label{S1ERcyc}
\end{figure}
\begin{figure}
\centering
\includegraphics[scale=0.55]{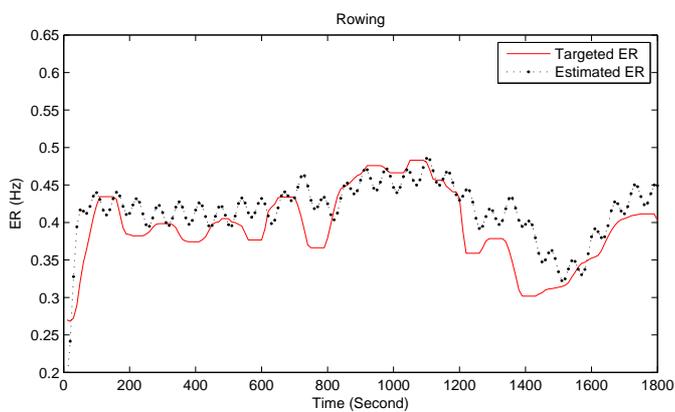}
\caption{Tracking of Subject S1  between ER$_{T}$ (solid line) and ER$_{est}$ (dotted line) during rowing.} \label{S1ERrow}
\end{figure}

Figures \ref{S1HRcyc} and \ref{S1HRrow} show the HR$_T$ profile tracking of Subject S1 using the indirect adaptive H$_\infty$ controller and HAS during cycling and rowing exercises.  In the case of cycling, the controller adapts the ER$_T$ in the range of 0.27 to 1.2 Hz from rest to the exercising phase for subject S1. Similarly in the case of rowing, the controller adapts the ER$_T$ in the range of 0.27 to 0.45 Hz. These results show the controller adapts the exercise variations and manipulates the control input, i.e., ER$_T$ within the required bandwidth of a cycling and a rowing exercises to achieve the desired HR$_T$ profile. The HAS interprets the ER$_T$ and the tracking performance of Subject S1 using HAS is represented in the Fig. \ref{S1ERcyc} and Fig. \ref{S1ERrow} during cycling and rowing, respectively. It can be observed from these figures that the ER$_{T}$ was closely followed by the Subject S1. As a result, HR response of the Subject S1 was achieved the desired HR$_{T}$ profile using the developed control approach and HAS, irrespective of the type of rhythmic activity.\\
\textbf{Subject S2}
\begin{figure}
\centering
\includegraphics[scale=0.55]{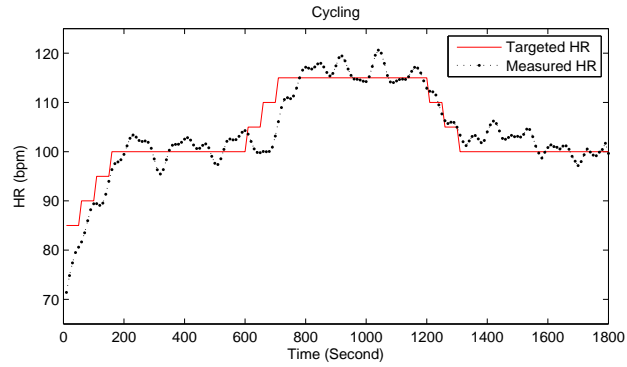}
\caption{HR profile tracking of Subject S2 during  cycling.} \label{S2HRcyc}
\end{figure}
\begin{figure}
\centering
\includegraphics[scale=0.55]{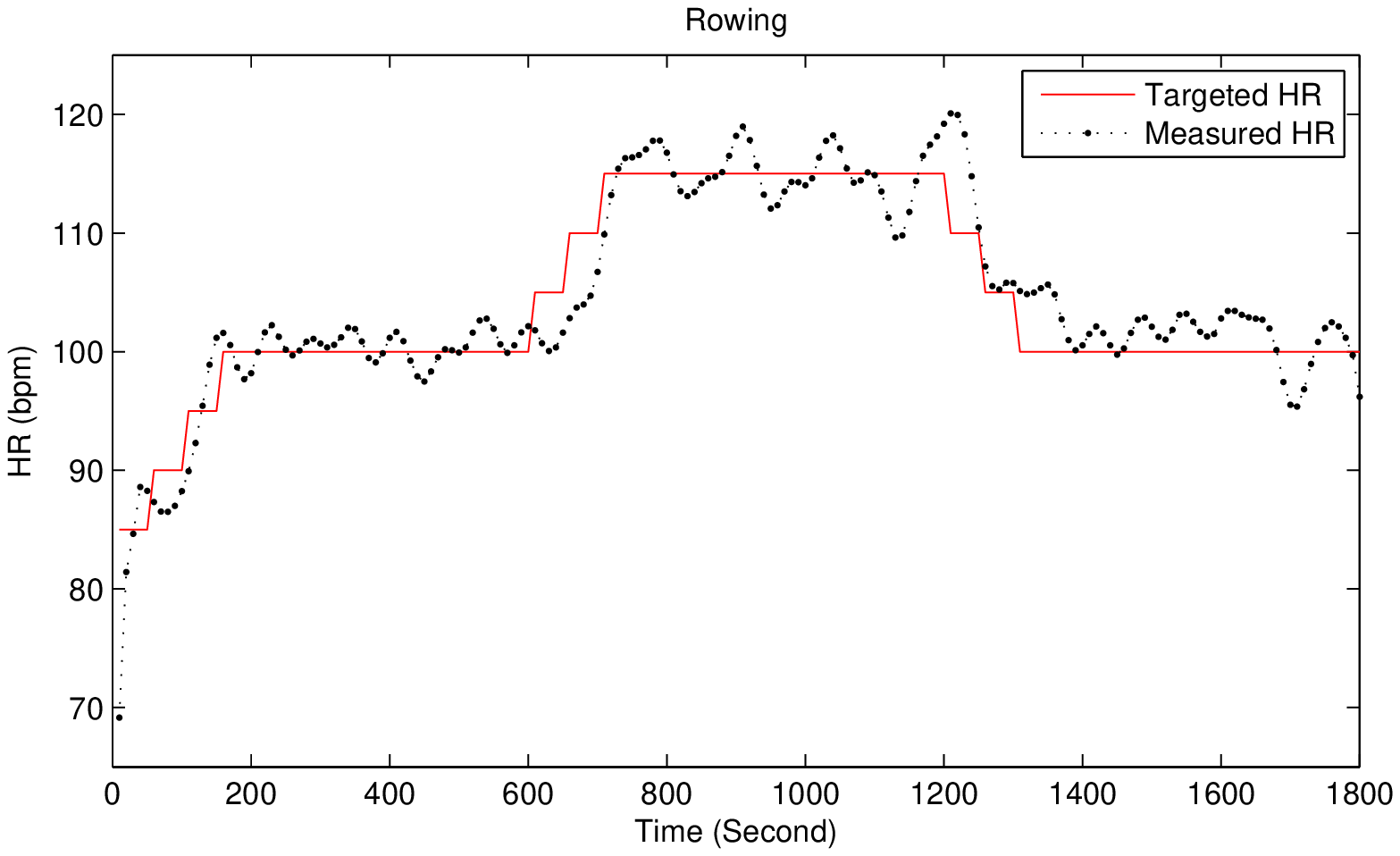}
\caption{HR profile tracking of Subject S2 during rowing.} \label{S2HRrow}
\end{figure}
\begin{figure}
\centering
\includegraphics[scale=0.55]{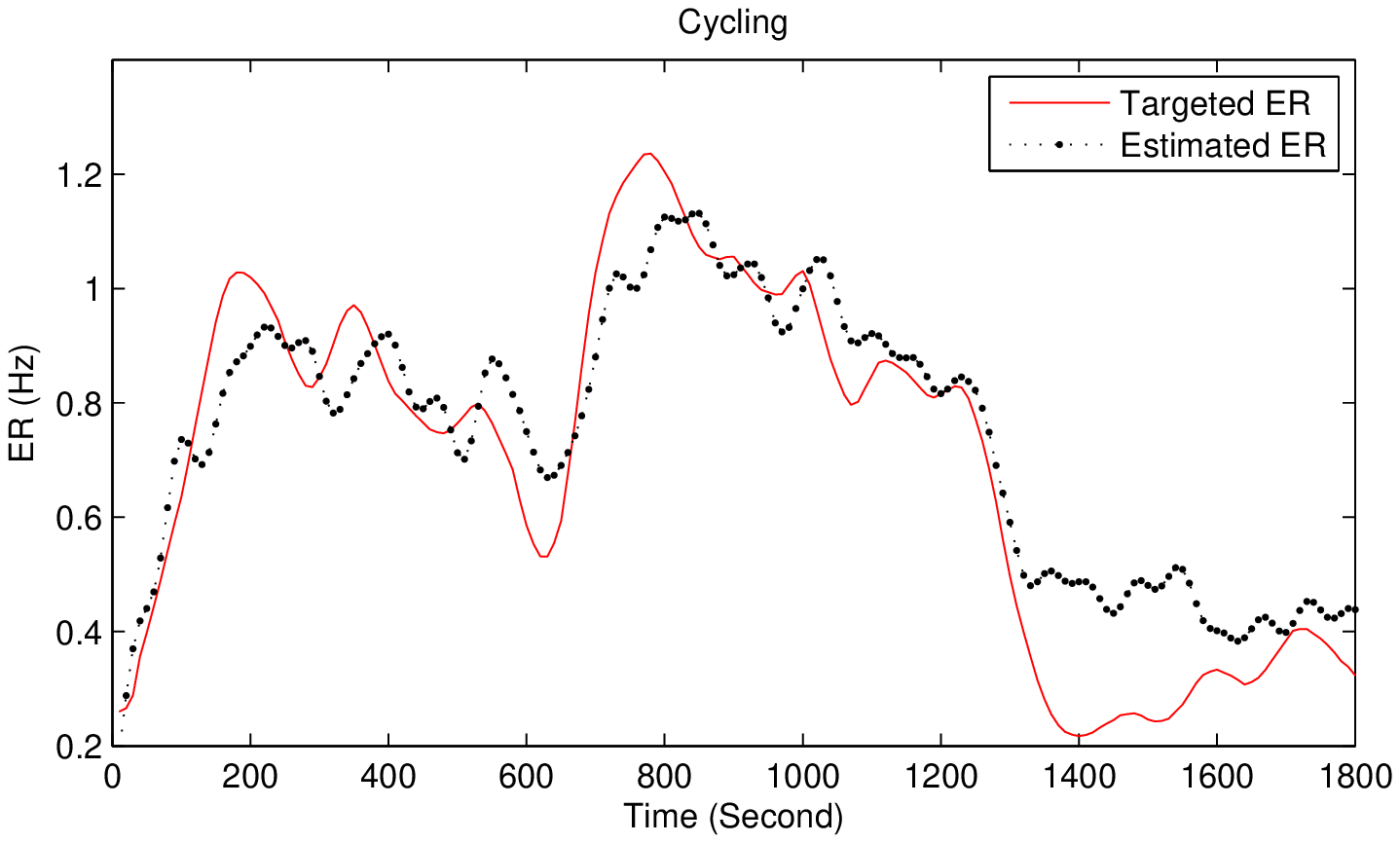}
\caption{Tracking of Subject S2 between ER$_{T}$ (solid line) and ER$_{est}$ (dotted line) during cycling.} \label{S2ERcyc}
\end{figure}
\begin{figure}
\centering
\includegraphics[scale=0.55]{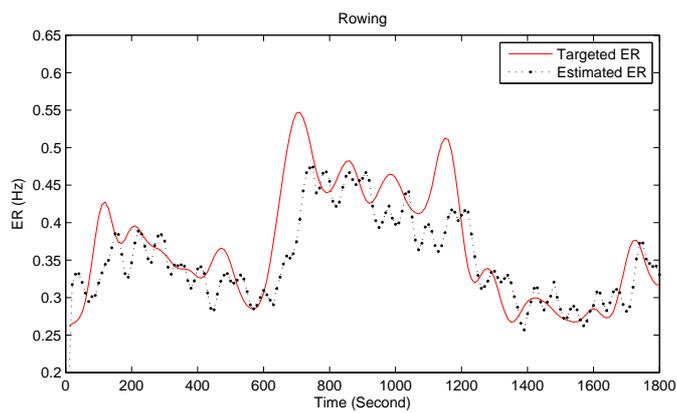}
\caption{Tracking of Subject S2 between ER$_{T}$ (solid line) and ER$_{est}$ (dotted line) during cycling.} \label{S2ERrow}
\end{figure}

Figures \ref{S2HRcyc} and \ref{S2HRrow} represent the HR$_T$ profile tracking of Subject S2 using the indirect adaptive H$_\infty$ controller and HAS during cycling and rowing exercises. In the case of cycling, the controller adapts the ER$_T$ in the range of 0.27 to 1.12 Hz from rest to the exercising phase for Subject S2. Similarly in the case of rowing, the controller adapts the ER$_T$ in the range of 0.27 to 0.533 Hz. These results show the controller adapts the exercise variations and manipulates the control input, i.e., ER$_T$ within the required bandwidth of a cycling and a rowing exercises to achieve the desired HR$_T$ profile.
The tracking performance of the Subject S2 using HAS is represented in Fig.  \ref{S2ERcyc} and Fig. \ref{S2ERrow} during cycling and rowing, respectively. These figures indicate that the ER$_{T}$ was closely followed by the Subject S2. As a result,  HR response of the Subject S2 was also achieved the desired HR$_T$ profile, irrespective of the type of rhythmic activity and subject's physical variations.\\
\textbf{Subject S3}
\begin{figure}
\centering
\includegraphics[scale=0.55]{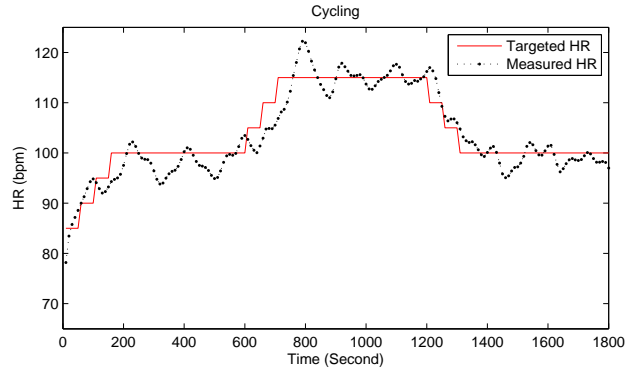}
\caption{HR profile tracking of Subject S3 during cycling.} \label{S3HRcyc}
\end{figure}
\begin{figure}
\centering
\includegraphics[scale=0.55]{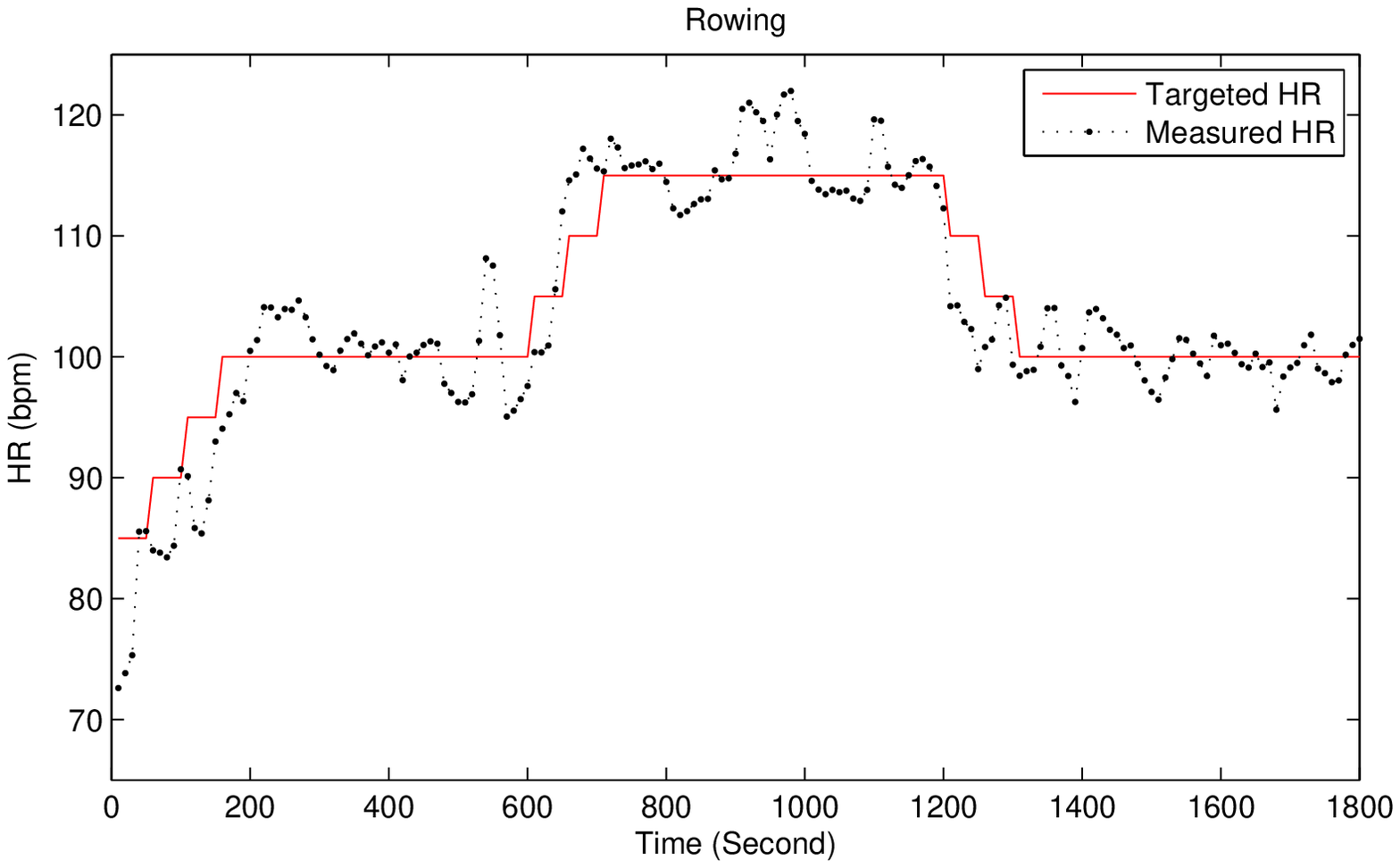}
\caption{HR profile tracking of Subject S3 during rowing.} \label{S3HRrow}
\end{figure}
\begin{figure}
\centering
\includegraphics[scale=0.55]{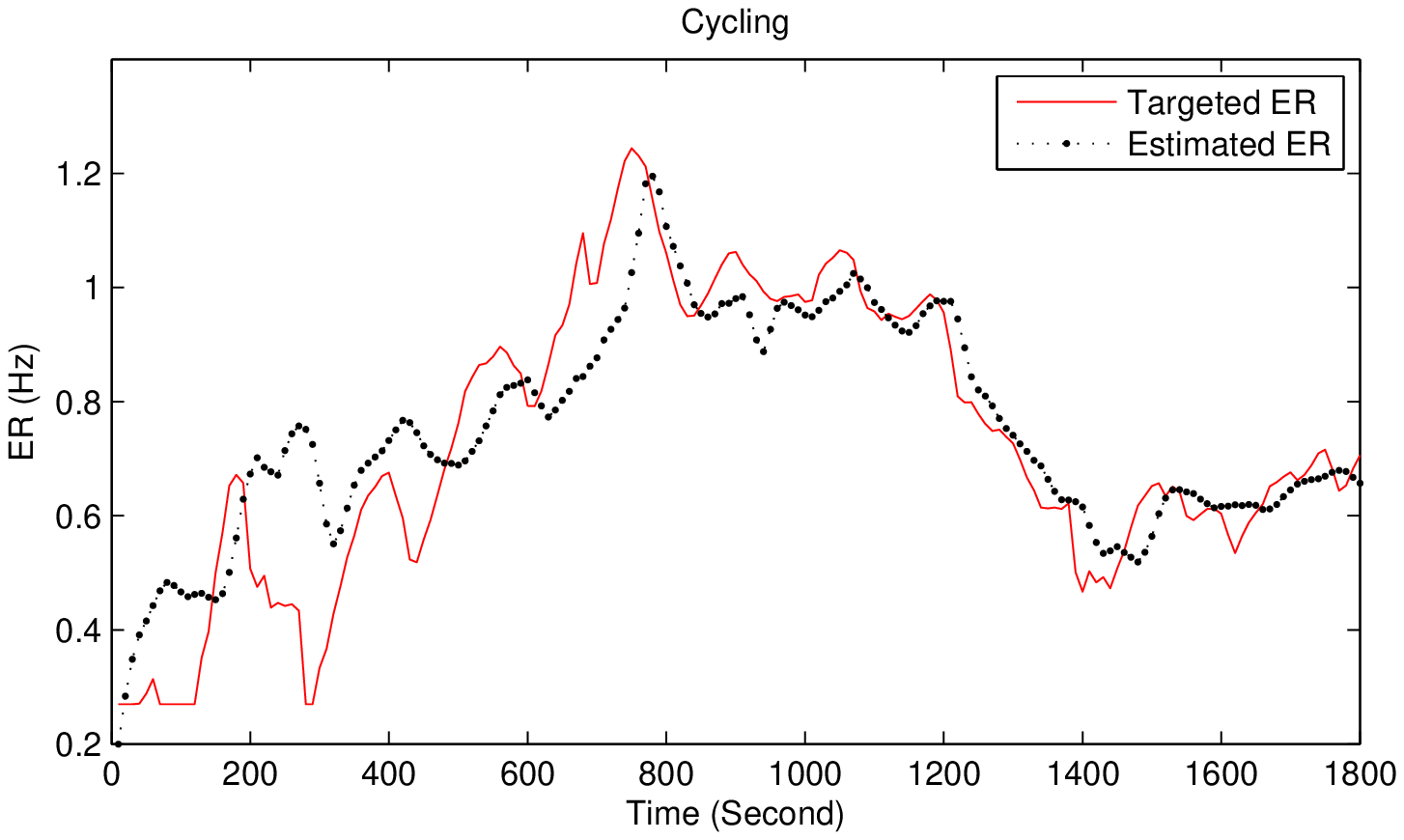}
\caption{Tracking of Subject S3 between ER$_{T}$ (solid line) and ER$_{est}$ (dotted line) during cycling.} \label{S3ERcyc}
\end{figure}
\begin{figure}
\centering
\includegraphics[scale=0.55]{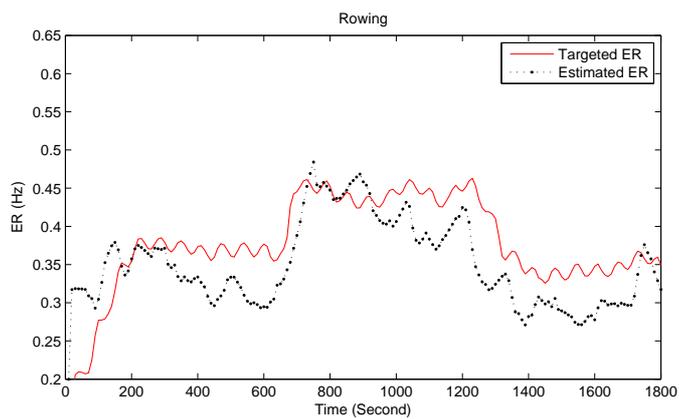}
\caption{Tracking of Subject S3 between ER$_{T}$ (solid line) and ER$_{est}$ (dotted line) during rowing.} \label{S3ERrow}
\end{figure}

Figures \ref{S3HRcyc} and \ref{S3HRrow} show the HR$_T$ profile tracking of Subject S3 using indirect adaptive H$_\infty$ controller and HAS during cycling and rowing exercises. The tracking performance of the Subject S3 using HAS is represented in Fig. \ref{S3ERcyc} and Fig. \ref{S3ERrow} during cycling and rowing, respectively. In the case of cycling, the controller adapts the ER$_T$ in the range of 0.27 to 1.21 Hz from rest to the exercising phase for Subject S3. Similarly in the case of rowing, the controller adapts the ER$_T$ in the range of 0.27 to 0.46 Hz. These results show the controller adapts the exercise variations and manipulates the control input, i.e., ER$_T$ within the required bandwidth of a cycling and a rowing exercises to achieve the desired HR$_T$ profile.
These figures indicate that the ER$_{T}$ was closely followed by the Subject S3. As a result, HR response of the Subject S3 has also been achieved at the targeted HR$_T$ profile irrespective of the type of rhythmic activity and subject's physical variations.\\
\newpage
\textbf{Subject S4}\\
\begin{figure}
\centering
\includegraphics[scale=0.55]{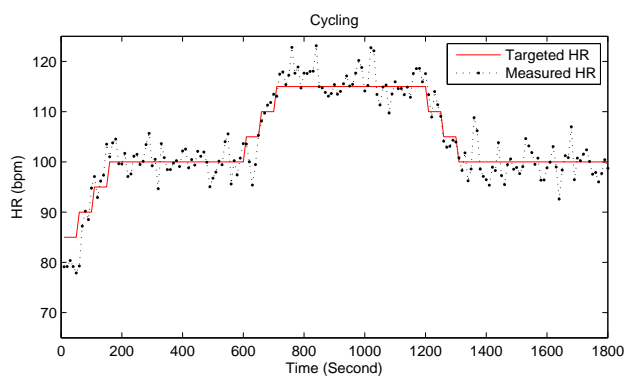}
\caption{HR profile tracking of Subject S4 during cycling.} \label{S4HRcyc}
\end{figure}
\begin{figure}
\centering
\includegraphics[scale=0.55]{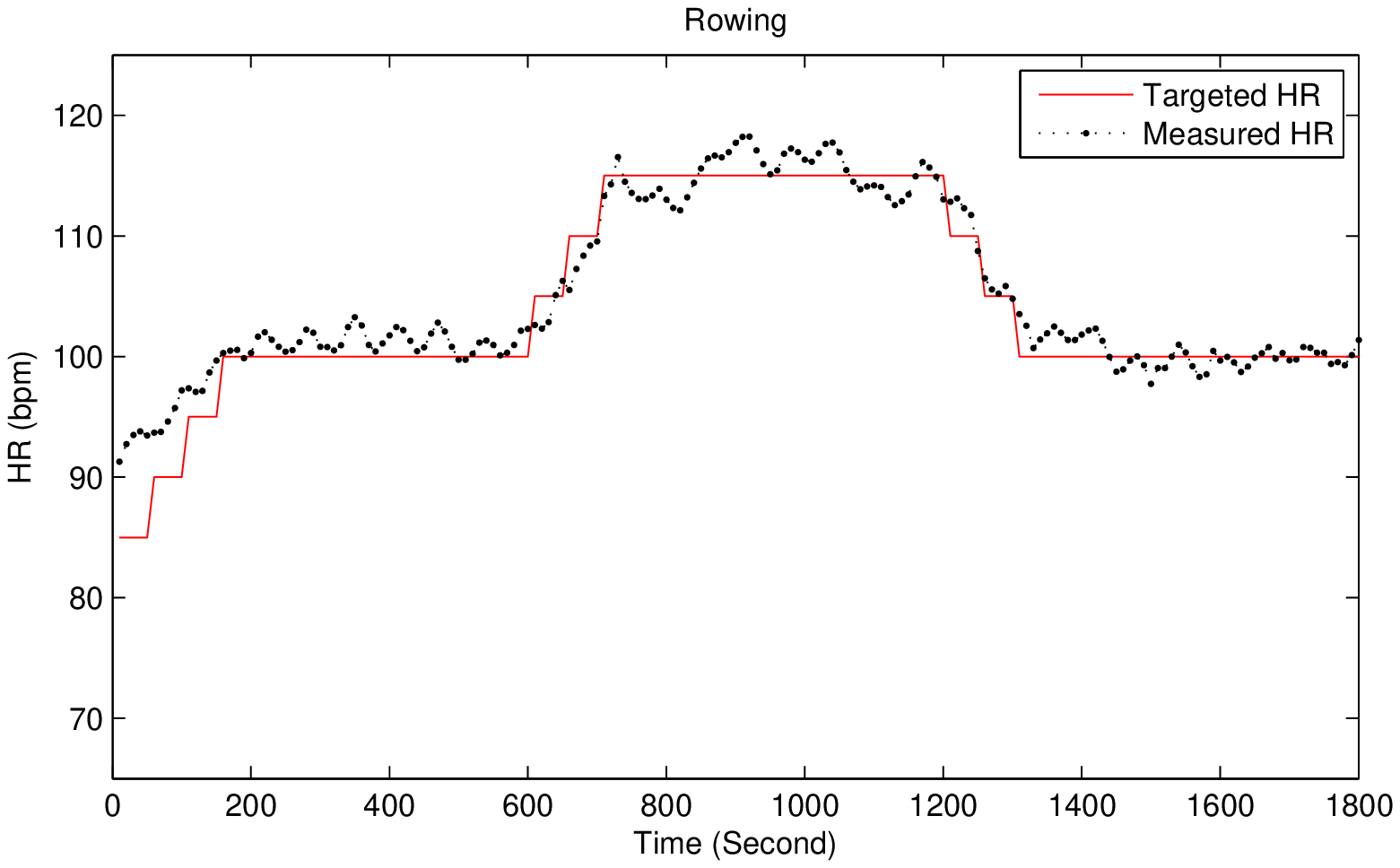}
\caption{HR profile tracking of Subject S4 during rowing.} \label{S4HRrow}
\end{figure}
\begin{figure}
\centering
\includegraphics[scale=0.55]{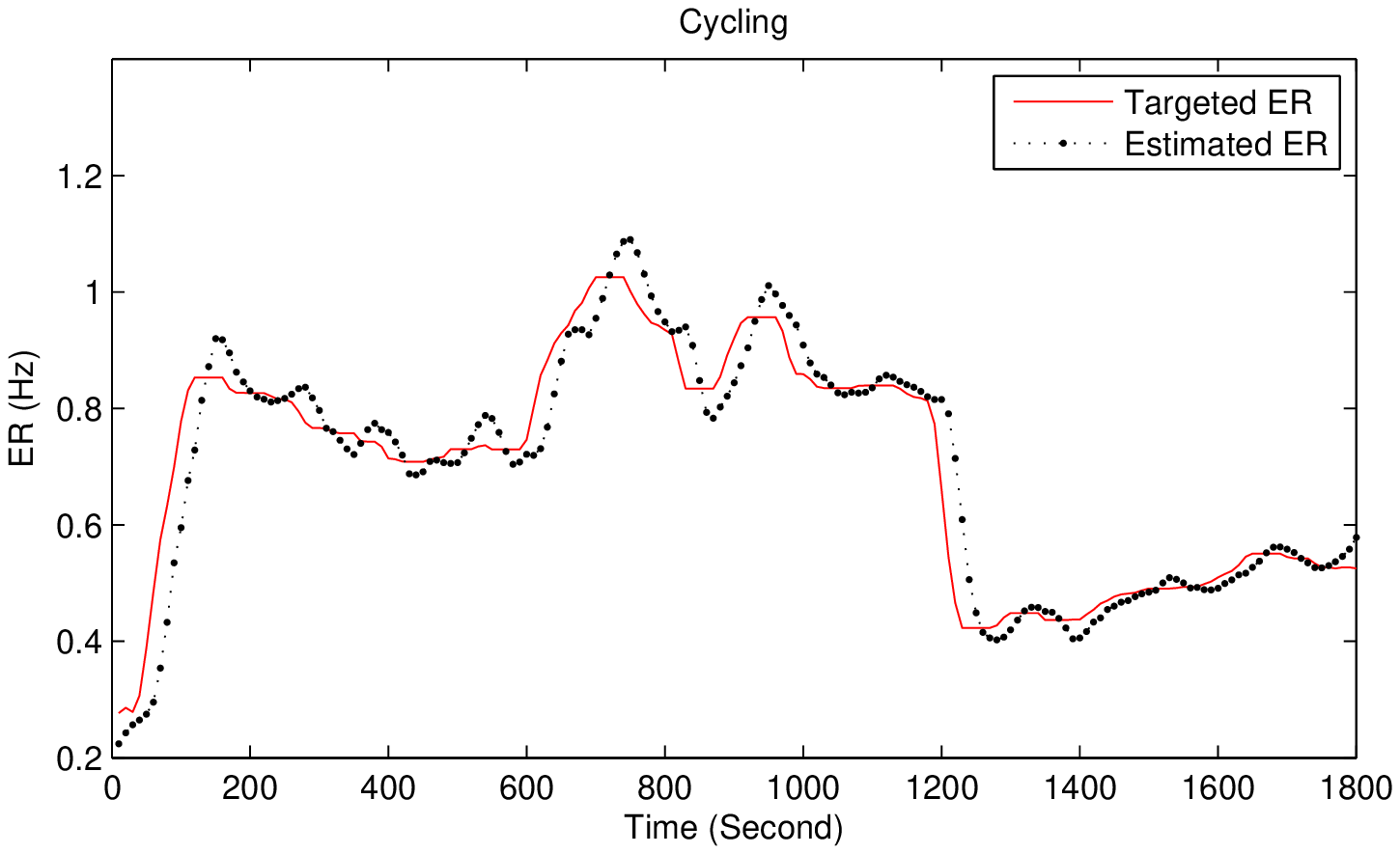}
\caption{Tracking of Subject S4 between ER$_{T}$ (solid line) and ER$_{est}$ (dotted line) during cycling.} \label{S4ERcyc}
\end{figure}
\begin{figure}
\centering
\includegraphics[scale=0.55]{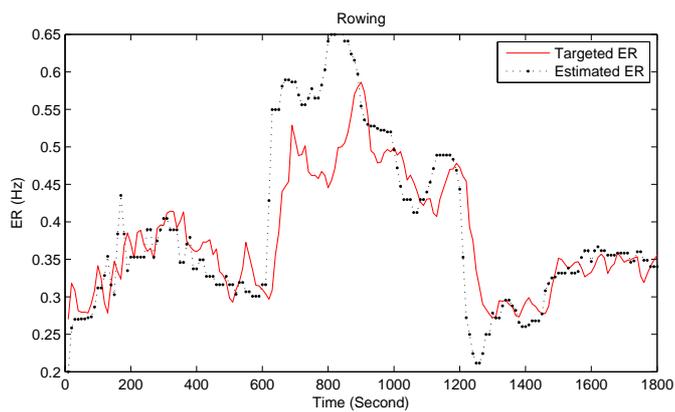}
\caption{Tracking of Subject S4 between ER$_{T}$ (solid line) and ER$_{est}$ (dotted line) during rowing.} \label{S4ERrow}
\end{figure}

Figures \ref{S4HRcyc} and \ref{S4HRrow} show the HR$_T$ profile tracking of Subject S4 using indirect adaptive H$_\infty$ controller and HAS during cycling and rowing exercises. The tracking performance of the Subject S4 using HAS is represented in Figs. \ref{S4ERcyc} and \ref{S4ERrow} during cycling and rowing, respectively. In the case of cycling, the controller adapts the ER$_T$ in the range of 0.27 to 1.1 Hz from rest to the exercising phase for Subject S4. Similarly in the case of rowing, the controller adapts the ER$_T$ in the range of 0.27 to 0.55 Hz. These results show the controller adapts the exercise variations and manipulates the control input, i.e., ER$_T$ within the required bandwidth of a cycling and a rowing exercises to achieve the desired HR$_T$ profile.
These figures indicate that the ER$_{T}$ was closely followed by the Subject S4. As a result, HR response of the Subject S4 achieved the HR$_T$ profile also, irrespective of the type of rhythmic activity and subject's physical variations.\\
\textbf{Subject S5}
\begin{figure}
\centering
\includegraphics[scale=0.55]{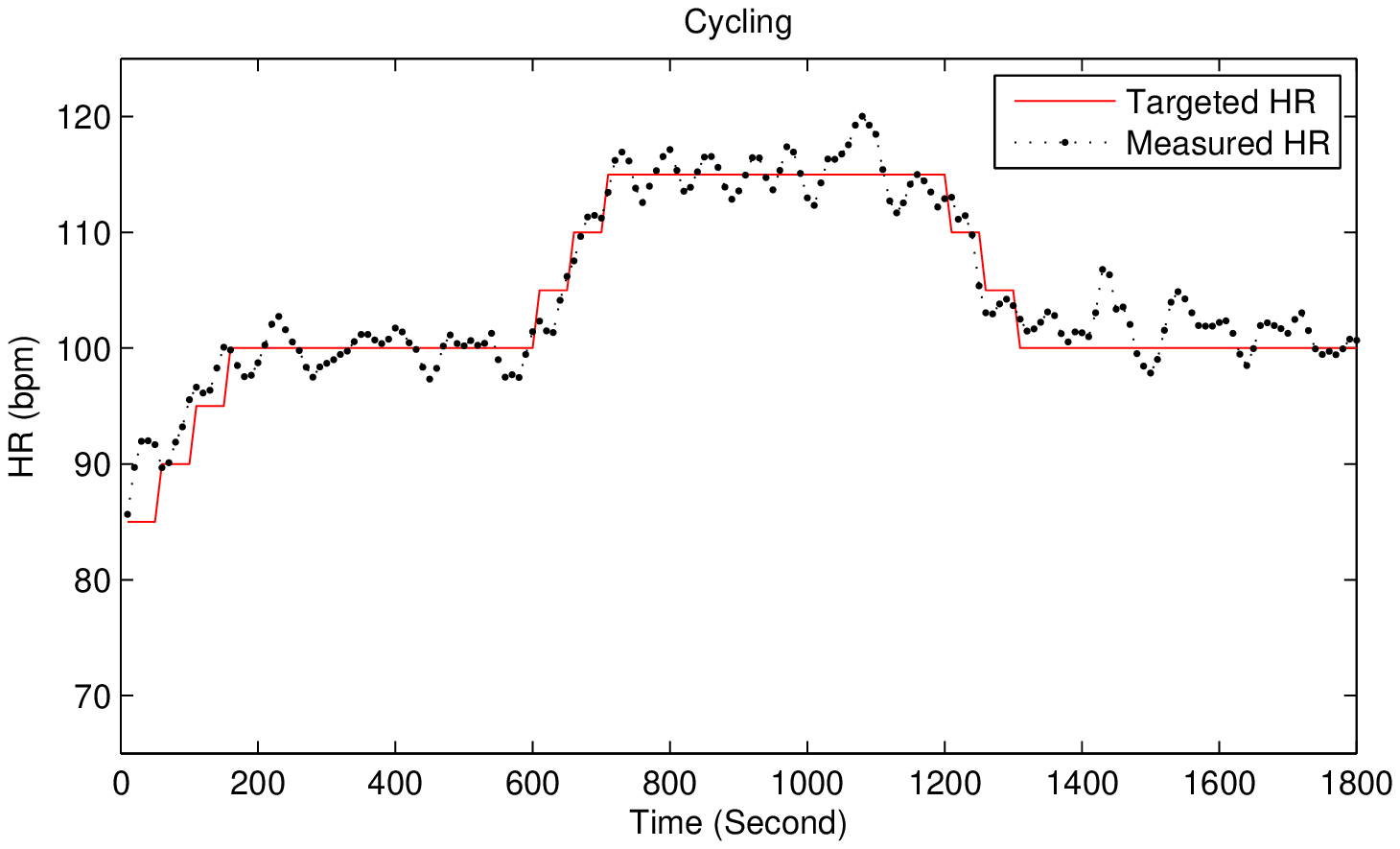}
\caption{HR profile tracking of Subject S5 during cycling.} \label{S5HRcyc}
\end{figure}
\begin{figure}
\centering
\includegraphics[scale=0.55]{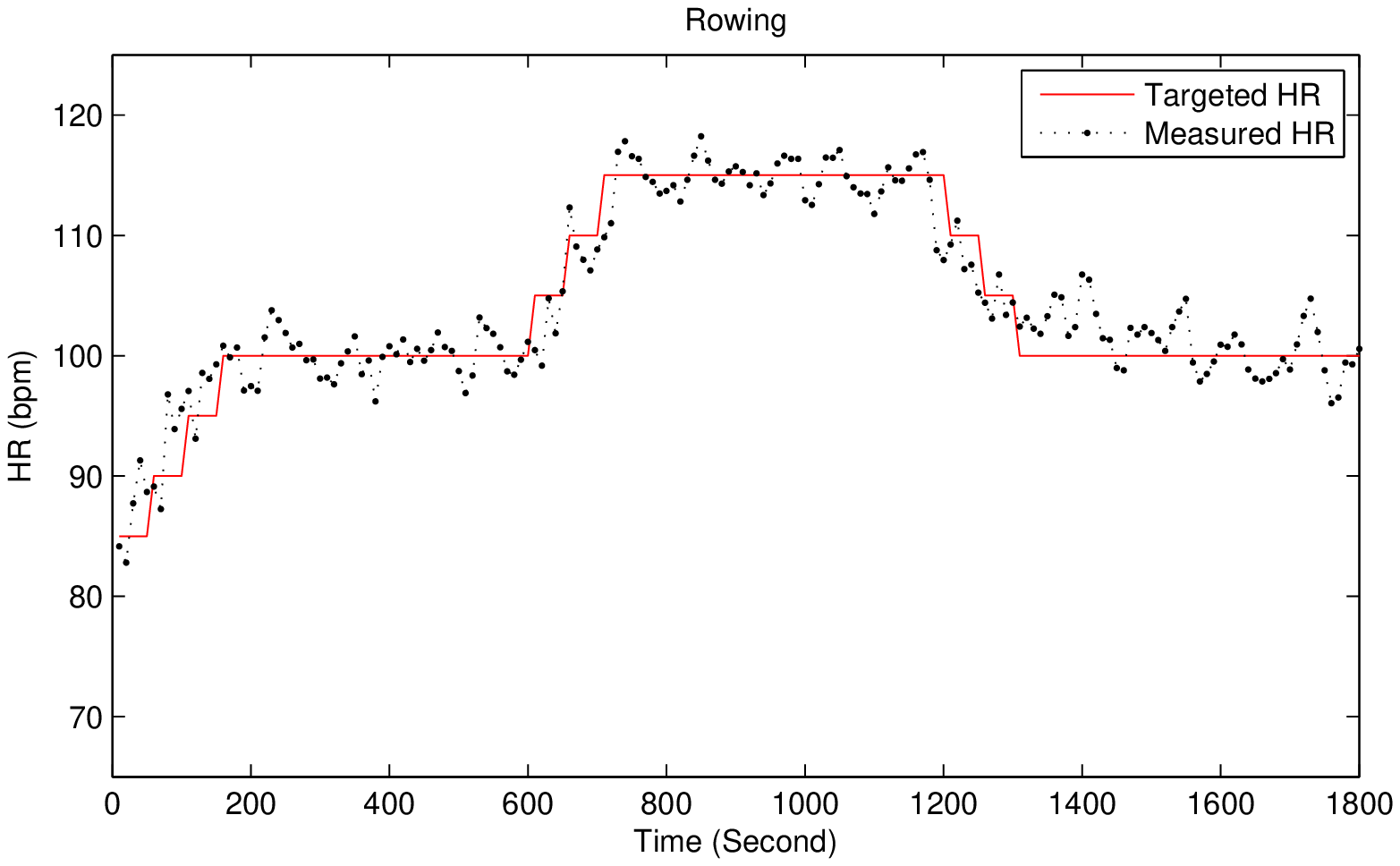}
\caption{HR profile tracking of Subject S5 during rowing.} \label{S5HRrow}
\end{figure}
\begin{figure}
\centering
\includegraphics[scale=0.55]{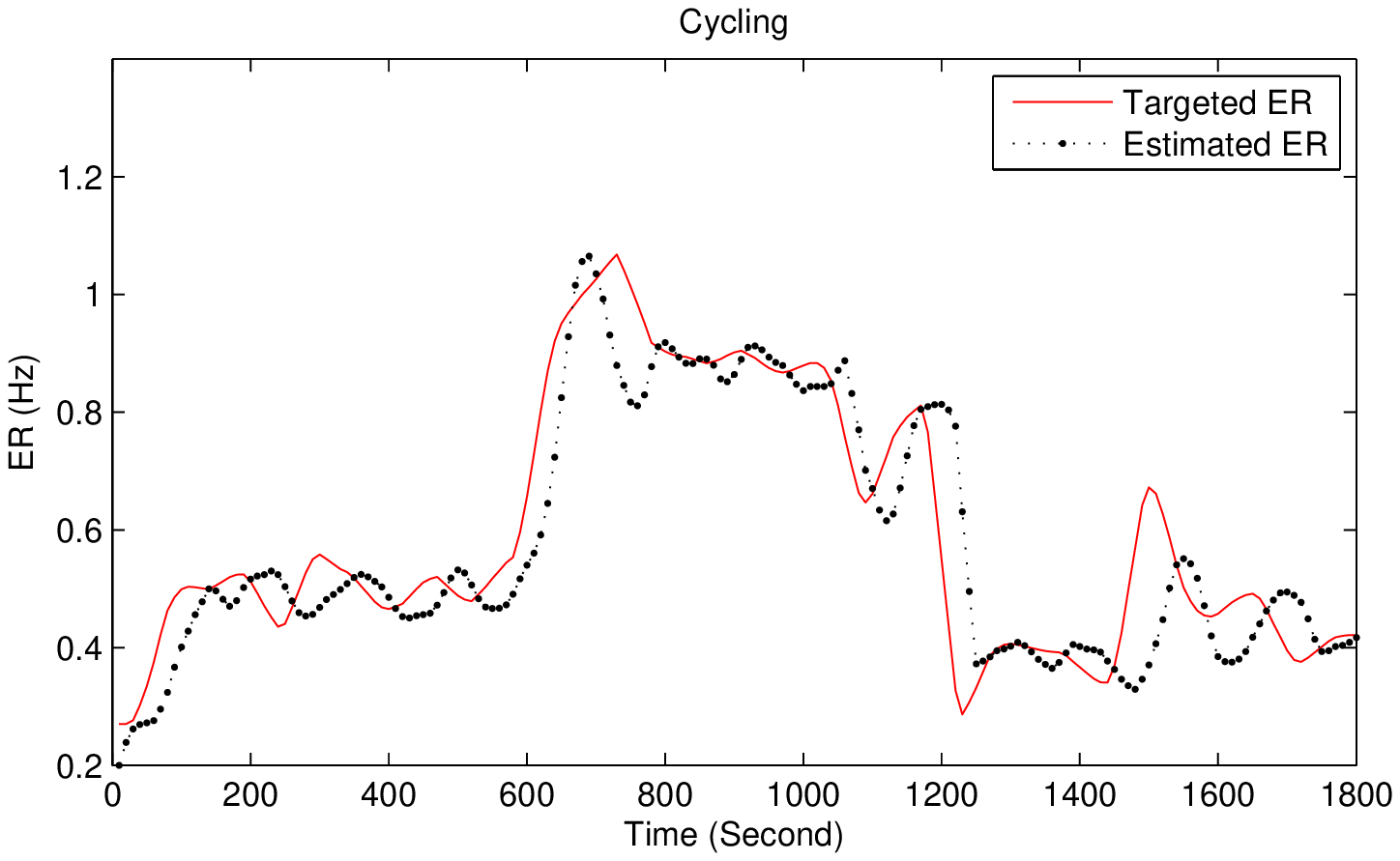}
\caption{Tracking of Subject S5 between ER$_{T}$ (solid line) and ER$_{est}$ (dotted line) during cycling.} \label{S5ERcyc}
\end{figure}
\begin{figure}
\centering
\includegraphics[scale=0.55]{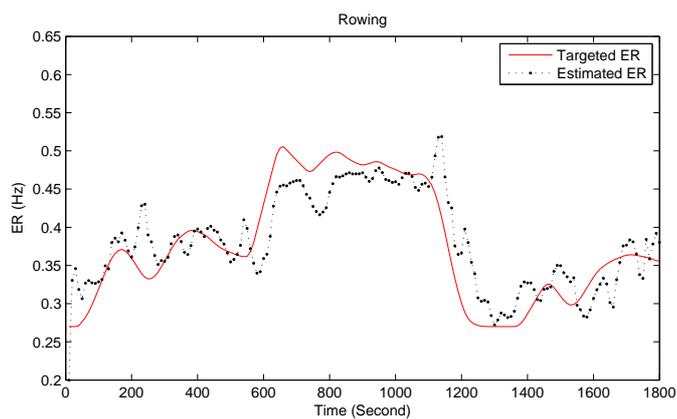}
\caption{Tracking of Subject S5 between ER$_{T}$ (solid line) and ER$_{est}$ (dotted line) during rowing.} \label{S5ERrow}
\end{figure}

Figures \ref{S5HRcyc} and \ref{S5HRrow} show the HR$_T$ profile tracking of Subject S5 using indirect adaptive H$_\infty$ controller and HAS during cycling and rowing exercises. The tracking performance of the Subject S5 using HAS is represented in Figs. \ref{S5ERcyc} and \ref{S5ERrow} during cycling and rowing, respectively.  These figures indicate that the ER$_{T}$ was closely followed by the Subject S5. In the case of cycling, the controller adapts the ER$_T$ in the range of 0.27 to 1.1 Hz from rest to the exercising phase for Subject S5. Similarly in the case of rowing, the controller adapts the ER$_T$ in the range of 0.27 to 0.50 Hz. These results show the controller adapts the exercise variations and manipulates the control input, i.e., ER$_T$ within the required bandwidth of a cycling and a rowing exercises to achieve the desired HR$_T$ profile. As a result, HR response of the Subject S5 also achieved the HR$_T$ profile, irrespective of the type of rhythmic activity and subject's physical variations.\\
\textbf{Subject S6}
\begin{figure}
\centering
\includegraphics[scale=0.55]{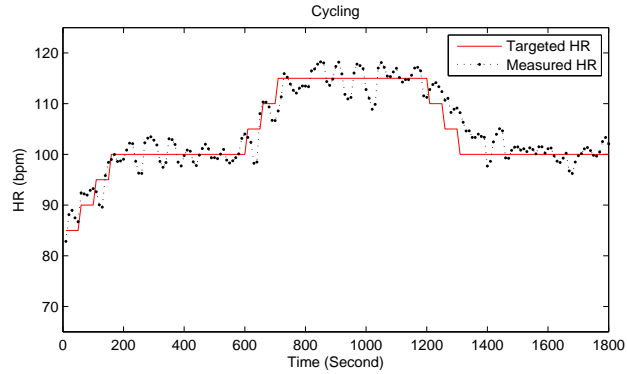}
\caption{HR profile tracking of Subject S6 during cycling.} \label{S6HRcyc}
\end{figure}
\begin{figure}
\centering
\includegraphics[scale=0.55]{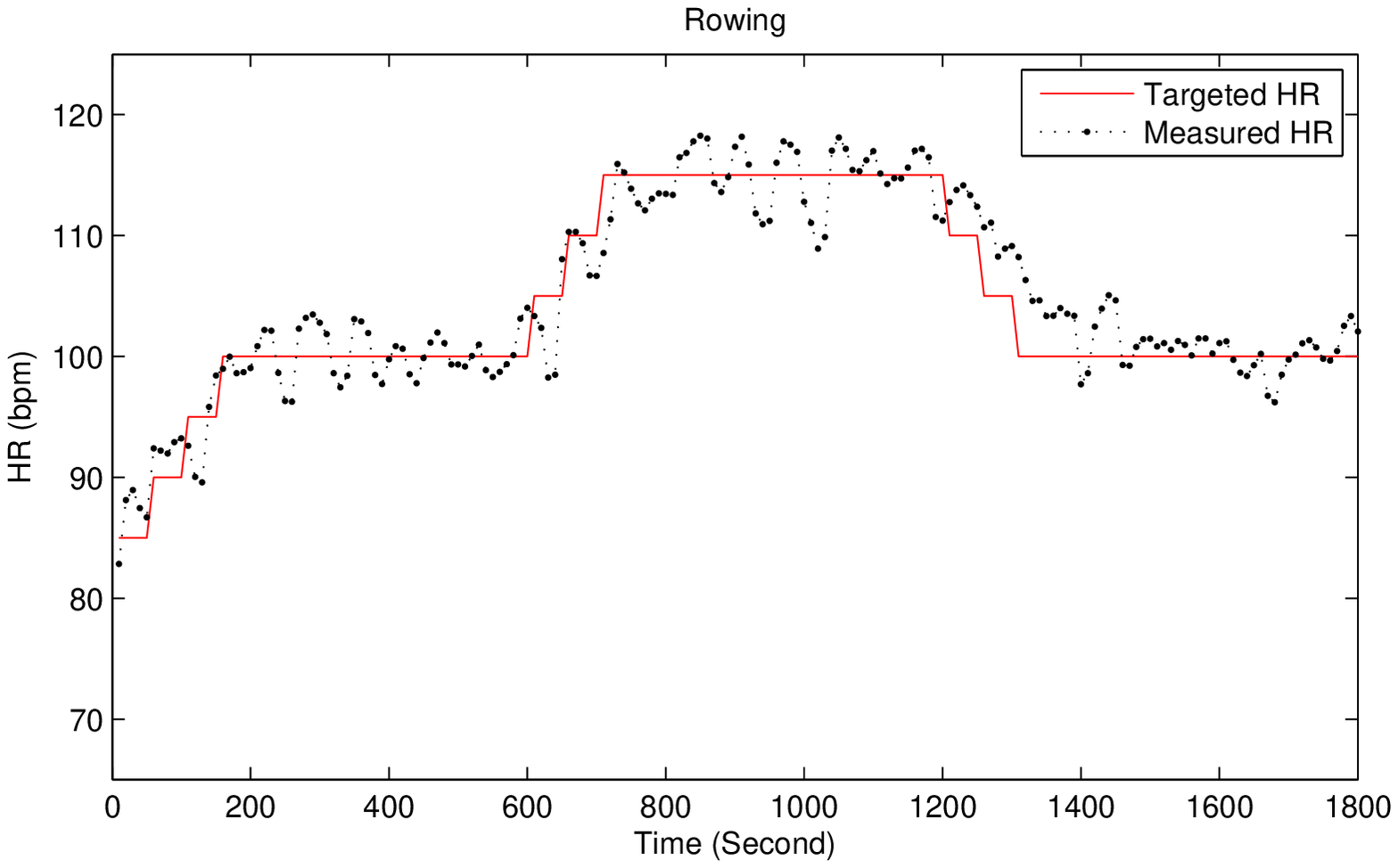}
\caption{HR profile tracking of Subject S6 during rowing.} \label{S6HRrow}
\end{figure}
\begin{figure}
\centering
\includegraphics[scale=0.55]{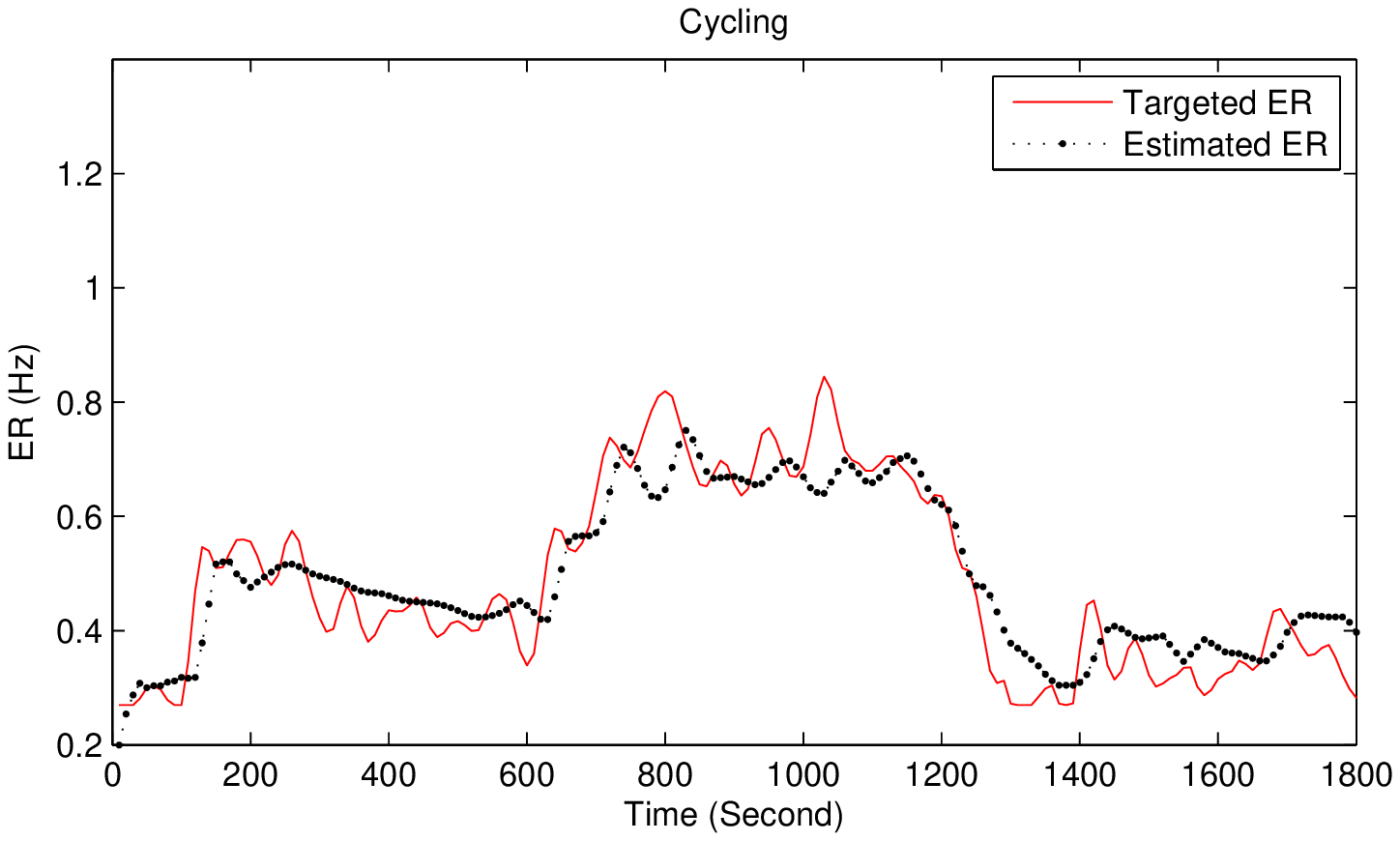}
\caption{Tracking of Subject S6 between ER$_{T}$ (solid line) and ER$_{est}$ (dotted line) during cycling.} \label{S6ERcyc}
\end{figure}
\begin{figure}
\centering
\includegraphics[scale=0.55]{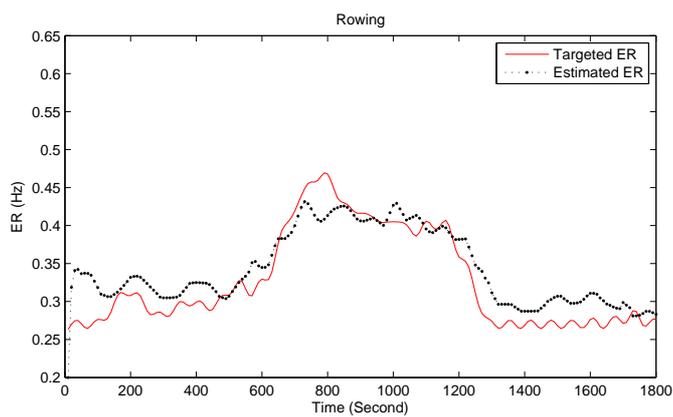}
\caption{Tracking of Subject S6 between ER$_{T}$ (solid line) and ER$_{est}$ (dotted line) during rowing.} \label{S6ERrow}
\end{figure}

Figures \ref{S6HRcyc} and \ref{S6HRrow} show the HR$_T$ profile tracking of Subject S6 using indirect adaptive H$_\infty$ controller and HAS during cycling and rowing exercises. The tracking performance of the Subject S6 using HAS is represented in Figs. \ref{S6ERcyc} and \ref{S6ERrow} during cycling and rowing exercises respectively. These figures indicate that the ER$_{T}$ was closely followed by the Subject S6. In the case of cycling, the controller adapts the ER$_T$ in the range of 0.27 to 0.45 Hz from rest to the exercising phase for Subject S5. Similarly in the case of rowing, the controller adapts the ER$_T$ in the range of 0.27 to 0.50 Hz. These results show the controller adapts the exercise variations and manipulates the control input, i.e., ER$_T$ within the required  bandwidth of a cycling and rowing exercises to achieve the desired HR$_T$ profile. As a result, HR response of the Subject S6 also achieved the HR$_T$ profile, irrespective of the type of rhythmic activity and subject's physical variations.

The developed control system was also validated using statistical analyses. Table \ref{stateanalysis} presents the RMS tracking error between $HR_{T}$ and $HR_{meas}$ for all six subjects during cycling and rowing exercises. These RMS values were calculated using \ref{Eq19}. The average RMS tracking error for cycling and rowing is 3.1857 bpm and 2.9396 bpm, respectively.
 These tracking errors  were found to be within acceptable range of the simulation study.
  The experimental results illustrate that the developed system can successfully track the variations in HR response while achieving the HR$_T$ profile during any type of the rhythmic exercising activity.

The RMS value of the change in controller output, i.e., $\Delta ER_{T}(t)$ during cycling and rowing exercises for all subjects is also tabulated in table \ref{stateanalysis}. The results prove that the controller adapts the changes in the HR${_T}$ very smoothly. An assessment of HAS is also presented in Table \ref{stateanalysis} in terms of the correlation co-efficient $R$ calculated between ER$_{T}$(t) and ER$_{est}$(t). A high value of $R$ indicates that all subjects responded well to follow the ER$_{T}$(t), with an average correlation co-efficient of 0.914 and 0.8425 for cycling and rowing exercises, respectively.

\begin{ourtable}
\caption{Statistical analysis of 6 subjects during cycling and rowing.}
\label{stateanalysis}
\begin{center}
\tabcolsep 2.0pt
\begin{tabular}{c c c c c c c}
\hline
~\pounds~~& \multicolumn{2}{c}{~~$e(t)_{rms}$Hz~~} & \multicolumn{2}{c}{~$\Delta ER_{T}(t)_{rms}$BPM~~} & 
\multicolumn{2}{c}{~$R(ER_{T},ER_{meas})$~~} \\
\hline
S.No&~~~Cycling&~~Rowing&~~~Cycling&~~Rowing~~~&Cycling~~&Rowing~~~\\
\hline
1&  3.8460&	3.6500&	0.0027&  0.00080&	0.8750&	0.7660\\
2&	4.0140&	3.2410&	0.0004&  0.00060&	0.9490&	0.8780\\
3&	3.3540&	3.4850&	0.0024&  0.00031&	0.8900&	0.8210\\
4&	3.1160&	2.7990&	0.0014&  0.00041&	0.9540&	0.8240\\
5&	2.1390&	1.9420&  0.0008&  0.00052&	0.8940&	0.8890\\
6&	2.6440&	2.5190&  0.0001&  0.00001&	0.9230&	0.8770\\
\hline
Mean &3.1857 & 2.9396 & 0.0013 & 0.00031  &	0.9140  &  0.8425\\
\hline
Std &   0.7132 & 0.6459 &   0.0011 &0.00040 &  0.0003 & 0.0785\\	
\hline
\end{tabular}
\end{center}%
\end{ourtable}
\begin{figure}
\centering
\includegraphics[scale=0.45]{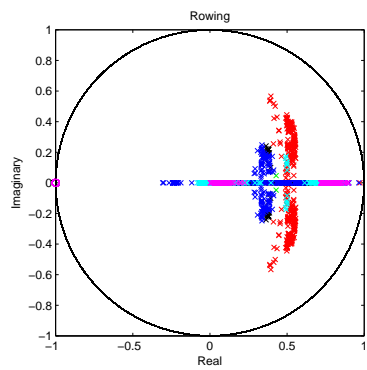}
\caption{Pole zero map during a rowing exercise.} \label{PZmap1}
\end{figure}
\begin{figure}
\centering
\includegraphics[scale=0.45]{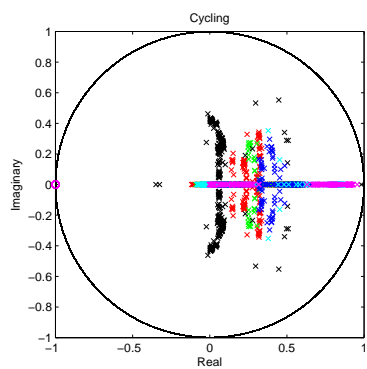}
\caption{Pole zero map during a cycling exercise.} \label{PZmap2}
\end{figure}
The pole zero map of the identified system in closed loop for all six subjects during rowing and cycling exercises is
shown in Fig.  \ref{PZmap1} and Fig. \ref{PZmap2} respectively to prove the stability of the adaptive control system. These results indicate that the poles and zeros are inside the unit circle indicating that the overall system is stable throughout the exercise for all six subjects. This is not a mathematically rigorous analysis. From a mathematical viewpoint, the proposed control system is a switched system, and its mathematically rigorous stability analysis can be conducted by recently developed
techniques for stability analysis of switched and networked systems; see e.g., \cite{Skafidas1999,Matveev2000,Matveev2001,Matveev2003,Savkin2006a,Savkin2007,Matveev2009}. However, such mathematically rigorous stability analysis is outside of the scope of this report.

In conclusion, the experimental results prove that the proposed HR regulation system is capable enough to regulate the human HR even for unknown subjects and type of exercises.
\section{Discussions}\label{section4.6}
 The experimental data analyses showed that HR response is varied in terms of the steady state and transient response due to interpersonal variability. It also varies its transient and steady behaviour with the type of rhythmic activity. Based on these observations, the relationship between ER and HR during walking, cycling and rowing exercises were analysed at various intensity levels using the LTI model with the variable p order. The results of the model identification tests show that a simple 2$^{nd}$-order LTI model is able to accurately model the HR dynamics during low and moderate intensity of exercise.  This model order was selected based on AIC. Except four cases are suggested the model (p=3)-order.  These four types are linked primarily with the high intensity exercises.  Since, this study required a universal model structure that can modelled the HR response during low, moderate and high intensity exercise. Therefore, the model structure p=2 was selected for all LTI models that were estimated from input-output data sets at various intensity levels during walking, cycling and rowing exercises. The performance analysis of LTI models at various intensity during walking, cycling and rowing exercise show that the simulated response of $\Delta$HR closely represents the actual measurement of $\Delta$HR during low and moderate intensity levels. However, the performance of the LTI at high intensity gets worst as exercise intensity increases, the measurement $\Delta$HR shows delay to reach the steady state. This delay in HR response represents the cardic vascular drift \cite{Cheng2008}. Therefore, LTV gives better description of the HR response than LTI models. Hence, an adaptive control approach is suitable for controlling the HR response in which type of activity is unknown.

 Based on these facts, an indirect adaptive H$_\infty$ controller was designed and simulated in Matlab. These simulations for walking, cycling and rowing were made based on the assumption that the exercising subject exactly follows the controller output i.e., ER$_{T}$.  The simulation study indicates that the designed control approach is able to manipulate ER within its required exercise bandwidth (see section \ref{section4.3}) to achieve the HR$_T$, irrespective of the type exercise, and subject physical variations.  The average RMSE between the HR$_T$ and HR simulated of two subjects is calculated 5.58 bpm, 6.52 bpm  and 3.5 bpm for a walking, a cycling and a rowing, respectively. Some of the cases of walking, cycling and rowing experienced the average approx. overshoot of 2.52\%, 3.12\% and 4.5\% while tracking the HR$_T$ profile. Hence, the overshoot within the range of 0 to 4.5\%  and RMSE between 0 to 6.52 bpm as a steady state error have been considered as an acceptable range for the purpose of the validation in the real-time for HR regulation system. Moreover, the simulated response of the HR in closed-loop for subjects S1 and S2 did not give similar behaviour during onset and offset of an exercise.

 The developed HR regulation system was tested on 6 healthy subjects performing cycling and rowing exercises. The participating subjects had different physical characteristics which vary the controller demand of ER to achieve the offset free HR profile tracking and a controller acquires an ability to track these changes in order to achieve the  HR$_T$ profile for an individual. The real-time experimental study showed that  the targeted exercise rate (ER$_{T}$) was generated by the adaptive H$_\infty$ within the desired bandwidth of the exercise. This ER$_T$  was closely followed by the exerciser using HAS and able to achieve the targeted HR (HR$_ {T}$) irrespective of the type of exercise was being done. The subjects achieved the HR$_T$ with average root mean square (RMS) tracking error of 3.1857 bpm and 2.9396 bpm, respectively, during cycling and rowing exercises. These results illustrate that the developed system can successfully track the changes in HR while tracking the reference HR profile during any type of the rhythmic activities. The controller has ability to smoothly adapt the changes in the set point of HR. To assess how well the human actuation system (HAS) worked by using the beep signal, the correlation co-efficient $R$ was calculated between the ER$_{T}$(t) and ER$_{est}$(t). A high value of $R$ indicated that all subjects had responded well to follow the ER$_{T}$(t), with an average correlation co-efficient of 0.914 and 0.8425 for cycling and rowing exercises. Moreover, some cases of the human subject during these exercises, experienced the overshoot within defined range of the simulation study. However, steady state was maintained through the course of exercise, irrespective of the type of exercise. Therefore, we can conclude that the developed system for HR control during unknown type of rhythmic activities achieves the desired performance in real-time. The obtained performance of the developed system is in the acceptable range in terms of overshoot and steady state error. As, it was found in the simulation study. To be statistically significant, these assertion need to be tested on larger group of subjects. \\
Furthermore, the performance of the developed system is based on an assumption that HR depends on the ER. However, the emotional factors distort the correlation between ER and HR. In this scenario, the developed adaptive robust control approach may cater the variations in the HR response by considering it as a noise disturbance. Therefore, the further experimental study will be carried out in the presence of the non-metabolic factors.\\

 Another future research will be focussed to improve the functioning of the existing control approach using the advance robust control technique; see \cite{Savkin1995a,Savkin1999}. \\
Moreover, the existing control techniques will be applied to regulate HR at the constant level of HR$_T$, while subject engaged in multiple exercising activities. As, the ACSM suggested that the total physical condition is possible while subject engages in the multiple activities \cite{Balday2000}. Moreover, the developed control methodology opens the door for gaited patient and arthritic patients to perform their exercise effectively and allow the body joints to enhance its movement's capability without doing any fatigue \cite{Khunt2004}.
 Lastly, this developed system can be used for online regulation of HR during rhythmic activities. As, the developed control approach does not require information about the type of rhythmic activity. It only requires the current values of ER and HR which can easily communicate through internet to achieve ER$_T$.
\section{Concluding Remarks}\label{section4.7}
In this study, an adaptive HR regulation system has been developed to regulate the HR of exercising subjects at the desired level during various unknown types of rhythmic activities. The system tracks the desired HR profile faithfully using an indirect adaptive H$_\infty$ controller. An important feature of the designed control system is the generation of the rhythmic movement required to achieve the HR$_{T}$. To author's knowledge to date, the proposed methodology of controlling HR in real-time has never been developed and implemented. The unique essence of the proposed methodology is that system has no information about the type of rhythmic exercise and the only information available from the subjects is the HR and ER. The system's performance is evaluated using real-time experiments carried out on 6 healthy subjects during cycling and rowing exercises. Such a system can be further enhanced to design an optimal exercising protocol for a gaited patient, cardiovascular and athletes, and would be really useful for cardiac rehabilitation programs.
\chapter{Estimation and Control of Oxygen Consumption During Cycling and Rowing Exercises}\label{ch:alt}

\section{Introduction}\label{intchap5}
The control of HR during rhythmic exercises was presented in Chapter \ref{ch:hrc}. The results concluded that HR could be controlled via manipulating the ER as a control input. However, HR response during exercise does not only represent the metabolic demands, it also includes non-metabolic factors such as individual emotions, air temperature and body posture. These non-metabolic factors distort the correlation between ER and HR \cite{Pulkkinen2003}. Therefore, the physiological control of HR during rhythmic activities is not a good solution to regulate the actual workload. On the other hand, measurements of VO$_2$ has been found as an actual indicator of the energetic demands during a physical activity \cite{ACSM2001}. It is considered as the most accurate physiological variable to quantify the intensity of aerobic activities. As the direct measurement of VO$_{2}$ during a physical activity is unwieldy, and this difficulty poses the limitations for a control design in developing an effective VO$_2$ regulation system.
To date, various methods have been developed to estimate VO$_2$ based on non-invasive measurements of an exercise intensity. The existing literature reveals that HR is a good estimator of VO$_2$, and various techniques had been developed to estimate VO$_2$ based on HR \cite{Spurr1988,Ceesay1989,Ainsile2003, Wester2009}. In particular, traditional estimation of VO$_2$ using HR is through the individual's HR-VO$_2$ calibration curve i.e., TRAD(HR-VO$_2$) \cite{Spurr1988,Ceesay1989}. The calibration curve is based on steady state values of the HR only excluding any transients and dynamic variations especially during exercises.  Paper \cite{Smolander2007} proposed a new methodology known as HR-HR-variability (HR-HRV) method for VO$_2$ estimation. This methodology uses ON/OFF dynamics and derives RespR from HRV. In addition, it uses the individual physical parameters such as age, weight and height to predict VO$_2$ in real-time. Furthermore, this method distinguishes between non-metabolic and metabolic demands by using the RespR as an input of the estimator. The results depict that the HR-HRV method is useful for predicting average VO$_2$ in the field and does not give the actual steady state values of an individual in contrast to the TRAD (HR-VO$_2$) curve calibration method. Moreover, this method does not require any calibration of an estimator due to individual's physical variations. Thus, these methods are not very useful in the measurement of VO$_2$ while controlling VO$_2$ during an exercise. The relationship between exercise intensity and VO$_2$ was modelled by \cite{Su2007C} using the Hammerstein approach during treadmill walking exercise. This model gives the precise estimation of the transient and steady state behaviour of the average VO$_2$ for six subjects. Therefore this model is only useful to design a control system for VO$_2$ regulation during treadmill walking exercise.\\
A new methodology for VO$_2$ estimation is presented in this Chapter.  Several inexpensive and easy-to-use sensors have been used to measure HR, RespR and ER non-invasively. From among these measured variables, ER is quantified as a fundamental variable of exercise intensity and is responsible for producing the dynamic change in HR, RespR and VO$_2$. These deviations in HR, RespR and VO$_2$ have been achieved through cycling ergo-meter and rowing machine by keeping a constant resistance of the machine against the body weight. That's how variations in these physiological variables (HR, RespR and VO$_2$) are exclusively dependent on ER. The developed estimating approach uses a concept that the VO$_{2max}$ and HR$_{max}$ are the indicators of the individual's physical fitness. Based on this concept, a new method of VO$_2$ estimation is presented that can cater subject to subject variations without going through the expensive calibration of an estimator.

\section{Proposed VO$_2$ Estimation Methodology}\label{Mthchap5}
\begin{figure}
\begin{centering}
\includegraphics[scale=1.45]{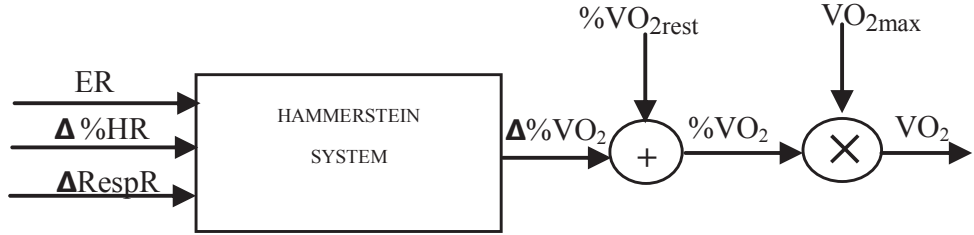}
\caption{A block diagram of VO$_2$ estimator.}\label{Fig1}
\end{centering}
\end{figure}
During rhythmic exercise, the repetitive body movements are responsible for producing the change in HR, VO$_2$ and RespR from their respective baseline measurements. The main goal of the proposed VO$_2$ estimator is to accurately predict the dynamic and steady state behaviour of VO$_2$ during rhythmic exercises. The proposed VO$_2$ estimator uses ER, $\Delta$RespR and the deviation between the percentage change in HR$_{max}$ from resting to the exercising phase, i.e., $\Delta$\%HR$_{max}$ as the inputs of the estimator. The block diagram of the proposed estimator is given in Fig. \ref{Fig1}. \\
The VO$_2$ varies with incremental exercise intensity, and maximum variation is possible at full aerobic capacity \cite{10,110}. Therefore, the deviation between the percentage change in VO$_{2max}$ from the resting to the exercising phase,  i.e., $\Delta$\%VO$_2$ is found to be a key element in determining VO$_2$ estimate. This quantity is computed using \ref{Eq3} and is responsible for determining the variations among subjects based on individual physical fitness.\\
\begin{equation}\label{Eq3}
\%VO_{2}= \frac{VO_{2}}{VO_{2max}},\\
\Delta\% (VO_{2}) = \%VO_{2meas}-\%VO_{2rest}
\end{equation}
The $\Delta$\%VO$_{2}$ can be estimated via Hammerstein model; see section \ref{hammodel}. The inputs to this estimator are obtained using mathematical manipulation of HR$_{max}$, VO$_{2rest}$, VO$_{2max}$ and $\Delta$RespR. Among these values, HR$_{max}$, VO$_{2rest}$ and VO$_{2max}$ are computed by using the physical characteristics of the exercising individual and resting value of HR. $\Delta$RespR is determined by using resting value of RespR.\\
\subsection{Estimating Rest and Maximum VO$_2$}\textbf{Estimating VO$_{2rest}$}\\
Resting VO$_2$ (VO$_{2rest}$) estimate can be obtained through several different equations. The most common estimate for VO$_{2rest}$ in exercise physiology is to use one metabolic unit (1 MET) which assumes 3.5 ml/kg/min  of  VO$_{2rest}$ in a seated position \cite{Brooks2005}. However, this conversion has been questioned because it is based on one 70kg, 40 years old man \cite{Byrne2005}. A US government group, examining international dietary norms, derived the estimates of resting energy expenditure using gender, height and weight for various age groups \cite{Brooks2004}.
One of the most frequently cited set of equations for daily resting expenditure is the Harris-Benedict equations \cite{Pellet1990}. Generating absolute VO$_{2rest}$ values from the Harris-Benedict equations requires a series of transformations to adjust for rate (i.e., per day to per minute) and caloric equivalent (i.e., convert calories to litres of oxygen) and defined in \ref{A} and in \ref{B} for men and women, respectively and are expressed in l/min.
\begin{equation}\label{A}
VO_{2rest}= 66.5+ 13.75\times wt+5.03\times ht-6.75\times age
\end{equation}
\begin{equation}\label{B}
VO_{2rest}= 655.1 +9.56\times wt + 1.85\times ht - 4.68\times age
\end{equation}
The VO$_{2rest}$ is estimated using the Harris Benedict Equation. This formulation can cater to any individual and is subject to variations based on physical fitness. Based on this formulation, VO$_{2rest}$ was determined for all male subjects from their physical parameters, i.e., weight (wt), height (ht) and age. The calculated values of VO$_{2rest}$ of all six subjects using  \ref{A} were converted into ml/min/Kg via dividing by the individual weight. The obtained estimate of VO$_{2rest}$ for all subjects are given in Table \ref{table1} and were used  to compute initial conditions of the VO$_2$ estimator for both types of exercise. We assume that this quantity does not vary until the subjects achieve new physical parameters, i.e., age, weight and height.\\
\textbf{Estimating VO$_{2max}$}\\
\begin{equation}\label{C}
VO_{2max}=15.33\frac{HR_{max}}{HR_{rest}}
\end{equation}

Sports researchers and physiologists have developed different formulae for estimating VO$_{2max}$ directly \cite{Plowman2008,Uth2005}. A simple formula as in \ref{C} was developed \cite{Uth2005}  for the calculation of  VO$_{2max}$ based on the resting HR (HR$_{rest}$). In this equation, HR$_{max}=220-age$ and HR$_{rest}$ is the 20 seconds resting HR. The VO$_{2max}$ is dependent on HR$_{rest}$, which  is  normally varied due to non-metabolic factors.  We assumed that both the mean value HR$_{rest}$ and VO$_{2max}$ remain constant. Hence the relation between inputs (ER, $\Delta$\%HR$_{max}$ and $\Delta$RespR) and output ($\Delta$\%$VO_{2}$) can easily be estimated by using the Hammerstein model.

\subsection{Estimating $\Delta$\%VO$_{2}$}
The Hammerstein model is used as an estimator of $\Delta$\%VO$_{2}$, which is  identified separately for cycling and rowing exercises. These estimators require a merge dataset of the six healthy male subjects during each exercise consisting of $nu$ inputs (nu=3), i.e., $\Delta$\%HR$_{max}$, $\Delta$RespR, ER, and $ny$ output (ny=1), i.e., $\Delta$\%VO$_{2}$.  The datasets were obtained experimentally for both cycling and rowing exercises.

\section{Experimental Procedure}\label{expchap5}
A cycling ergo-meter and a rowing machine were used in the experiments where resistance was kept constant against body weights. This constant machine resistance is responsible for any deviations in HR, RespR and VO$_2$ as well as ER.  Our experimental protocol is based on these facts.
\subsection{Subjects}\label{Subjects}
\begin{ourtable}
\centering
\caption{Subjects: physical characteristics.}
\label{table1}
\begin{center}
\begin{tabular}{c c c c c c c}
\hline
Sub & Age & Mass & Height  &  HR$_{max}$ & VO$_{2max}$ & VO$_{2rest}$\\
    &$(yrs)$& $(Kg)$ & $(cm)$& $(bpm)$& $(ml/min/Kg)$&$(ml/min/Kg)$\\
\hline
S1  &	 28  &	60	&  160 &	192  &  39.77  & 3.5413\\
S2	&    29  &	60	&  164 &	191  &  39.57  & 3.662\\
S3	&    31  &	78	&  177 &	189  &  35.55  & 3.536\\
S4	&    25  &	63  &  167 &	195  &  31.42  & 2.765\\
S5	&    24  &	61	&  172 &	196  &  39.5   & 3.0816\\
S6  &     23  &	95	&  177 &    197  &  35.16  & 3.2408\\
\hline
Mean & 26.3  & 69.5	& 169.5&  193.33 & 36.86   & 2.1\\
Std &  2.65&14.263& 7.01 & 3.14&4.0151&0.3413\\

\hline
\hline
\end{tabular}
\end{center}
\end{ourtable}
Six healthy male subjects participated in the data collection activity. As mentioned earlier, the subjects were free of cardiac and pulmonary diseases and had no medical history of Cardio-respiratory, neurological disease and acute upper respiratory tract infection. Subjects aged between 18 and 35 years.   The subjects were clearly familiarised with the experimental procedure and possible risks involved before their participation in the activity and the consent form was duly signed by each of them.  Physical characteristics of the subjects are tabulated in Table \ref{table1}.

\subsection{Experimental Equipment}
The equipment used in the experiments consists of the following:
\begin{enumerate}
\item{A cycling ergo-meter}
\item{A rowing machine}
\item{Traix Accelerometer (TA) }
\item{Polar Belt }
\item{Respiratory Belt}
\item{Cosmed Kb4 gas analyser}
\item{Computer Station (CS)}
\end{enumerate}

The detailed description of the apparatus has already been given in Section \ref{equip}. It is important to mention here that the devices were calibrated carefully and the data were carefully monitored in real-time using LabVEIW for HR, RespR and ER, and VO$_2$ using Kb4 gas analyser software.
\subsection{Experimental Protocols}
Cycling and rowing exercises were performed by all participating individuals in three phases, i.e., resting, exercising and recovery. During the resting phase, subjects were requested to be seated comfortably at rest for 5 minutes.  The measurements thus obtained were used to determine the respective resting value of HR and RespR. The exercising phase lasted for 10 minutes immediately after the resting phase. Finally, the subjects were required to recover completely in about 5 minutes and this phase is known as the recovery phase. Each type of exercise was performed at four different levels of intensity. These four intensity levels are tabulated in Table \ref{table1} where the units of ER are pedals/min and strokes/min for cycling and rowing, respectively. These units were converted into Hz for system identification.
Another dataset for each type of exercise was obtained using the pseudo random binary sequence (PRBS) signal at two different levels of intensity. This input PRBS signal was implemented on a computer station in the form of beeps. These beeps reflect the exercise period or a desired time period to achieve a rhythmic movement.  In case of cycling, this ER was switched between 48 pedals/min and 60 pedals/min. In the case of rowing exercise, this ER value was switched between 28 strokes/min and 30 strokes/min. The switching of these frequencies was specified in the form of binary code, i.e., $0$ and $1$; $0$ represents the low intensity value of ER and $1$ represents the highest intensity value of ER. The binary code sequence for PRBS input during cycling and rowing exercise was selected as described in \cite{Su2007C} and is given here.
\begin{center}
 $1~0~0~1~0~1~1~1~0~1~1~0~0~0~1~1~1~1~1~0$
 \end{center}

The active bit was displayed to the exercising individual with the help of LED indicator on the computer station. This LED indicator turned ON if binary bit is $1$ and turned OFF if binary bit is $0$ which reflected the switching of high to low value of ER. Secondly, the unit of ER was converted into HZ and transformed it into the desired time period. This time period reflected the time to achieve the desired rhythmic movement in order to achieve the desired ER. This time period was communicated to the exercising individual in terms of a beep signal.
\begin{ourfigure}
\begin{centering}
\includegraphics[scale=0.75]{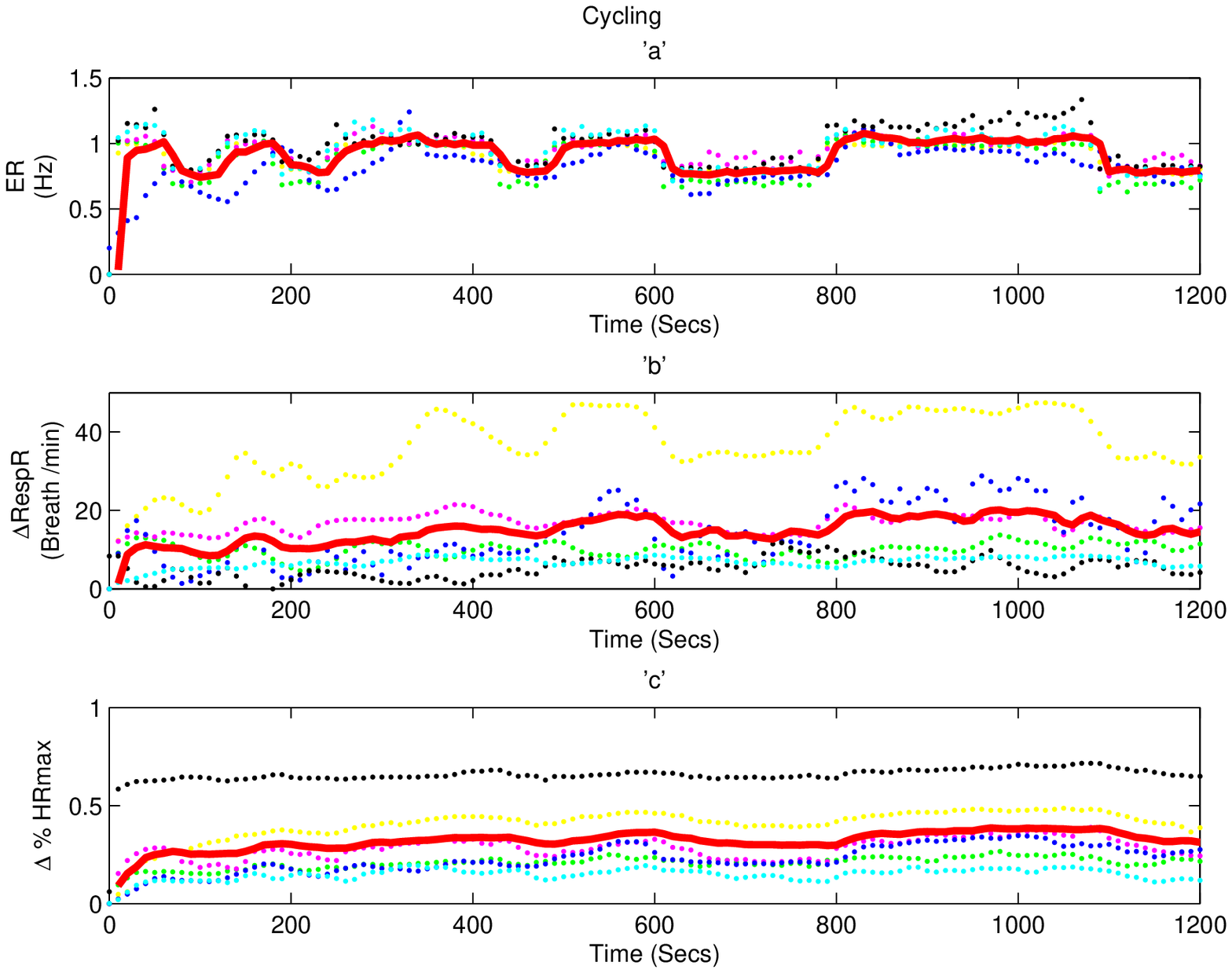}
\caption{$a) ER$ , b) $\Delta$RespR, and c) $\Delta$\%HR$_{max}$ during a cycling exercise.}\label{prbc2}
\end{centering}
\end{ourfigure}

\begin{ourfigure}
\begin{centering}
\includegraphics[scale=0.75]{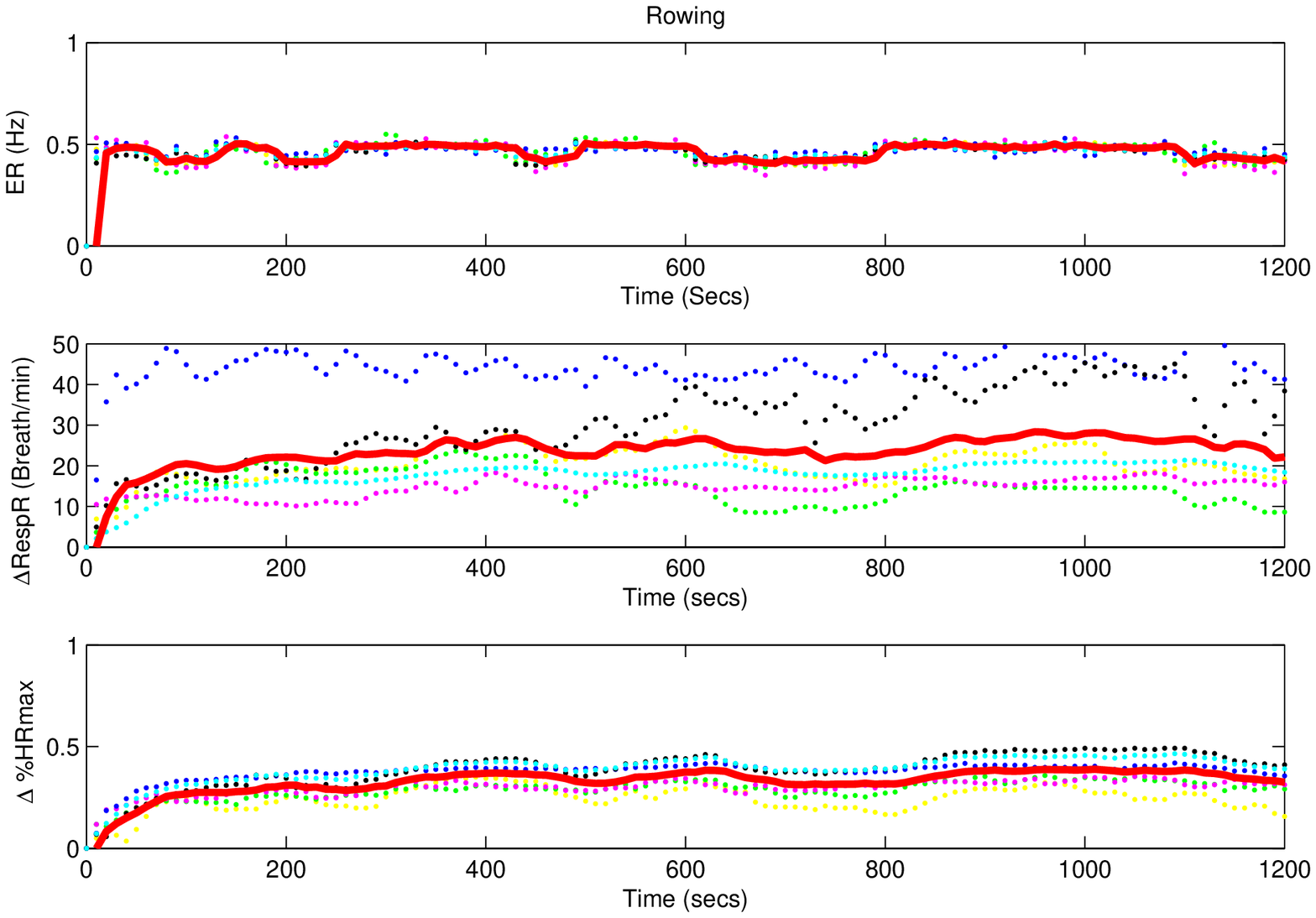}
\caption{$a) ER$ , b) $\Delta$RespR, and c) $\Delta$\%HR$_{max}$ during a rowing exercise.}\label{prbr2}
\end{centering}
\end{ourfigure}
Subjects were requested to precisely follow the beep signals and also closely monitor the switching of the LED signals in front of them. Meanwhile, the measurements of the desired ER (generated from beep signal) and estimated ER (obtained from TA measurements) were recorded and used for data processing lately
\subsection{Data Processing}
The obtained measurements of TA (ER estimation), HR, RespR and VO$_2$ were used to obtain the input and output merged datasets of cycling and rowing exercises. All the measurements of the HR, RespR and VO$_2$, were initially get filtered and interpolated at the sampling time of 10 seconds. These merged datasets were used for the identification of the Hammerstein models, and the individual datasets for all six subjects were used to validate the performance of the Hammerstein models obtained for cycling and rowing exercises.\\
\textbf{ER Estimate}\\
The ER estimate was achieved by using the measurement of TA; see \ref{TAmeasured} for further details.\\
\textbf{HR Measurements}\\
Measurement of HR was achieved using polar belt; sec e.g., \ref{HRmeasured}. Measurements of HR of an individual were subtracted from his HR$_{rest}$ (calculated as the mean value of the HR in resting phase). This deviation is used to calculate $\Delta$\%HR$_{max}$ by dividing it with the individual HR$_{max}$.\\
 \textbf{ $\Delta$RespR}\\
Respiratory belt was used to measure the RespR;  as in  \ref{RespRmeasured}.  Measurements of the RespR of an individual were subtracted from his/her RespR$_{rest}$ (calculated as the mean value of the RespR measurement obtained in resting phase) to obtain $\Delta$RespR.
 \textbf{$\Delta$\%$VO_{2}$}\\
 Measurements of the VO$_2$ of an individual were subtracted from his VO$_ {2rest}$ (calculated using the Harris Benedict Equation). This deviation is used to calculate $\Delta$\%VO$_{2}$ by dividing it with the individual VO$_{2max}$.\\
 The data processing for ER, HR, RespR and VO$_2$ was carried out for all six subjects and used to get the individual datasets and merged datasets for both cycling and rowing exercises.\\
All exercising individuals achieved the desired ER generated by the PRBS input as shown in Fig. \ref{prbc2}-a and Fig. \ref{prbr2}-a during cycling and rowing exercises, respectively. This ER produced the dynamic deviations in $\Delta$RespR, $\Delta$\%HR$_{max}$ and  $\Delta$\%$VO_2$ as shown in Fig. \ref{prbc2}b-c and Fig. \ref {VO22} for cycling exercise. Similar finding were observed in $\Delta$RespR, $\Delta$\%HR$_{max}$ and  $\Delta\%VO_2$  for rowing exercise, as shown in  Fig. \ref{prbc2}b-c and  Fig. \ref{VO22}. These variations ($\Delta$RespR, $\Delta$\%HR$_{max}$ and $\Delta$\%VO$_2$) were subject dependent and varied due to type of the exercise and its intensity levels.  Based on these observations, a $\Delta$\%VO$_2$ estimator was developed with inputs ER, $\Delta$RespR, $\Delta$\%HR$_{max}$ and the output $\Delta$\%VO$_2$  irrespective of the type of exercises.
\begin{ourfigure}
\begin{centering}
\includegraphics[scale=0.75]{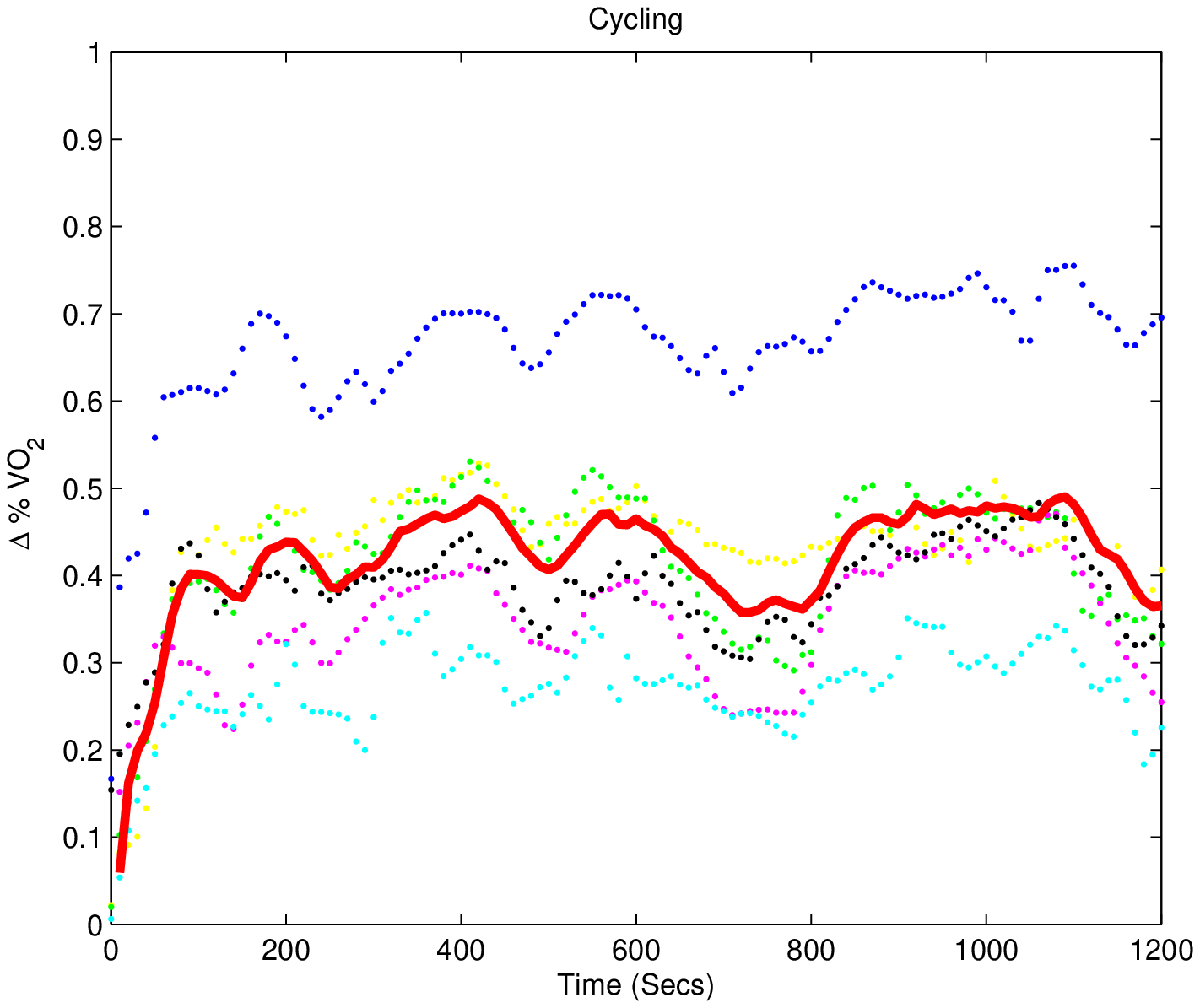}
\caption{$\Delta$\%VO$_2$ during cycling.}\label{VO21}
\end{centering}
\end{ourfigure}

\begin{ourfigure}
\begin{centering}
\includegraphics[scale=0.75]{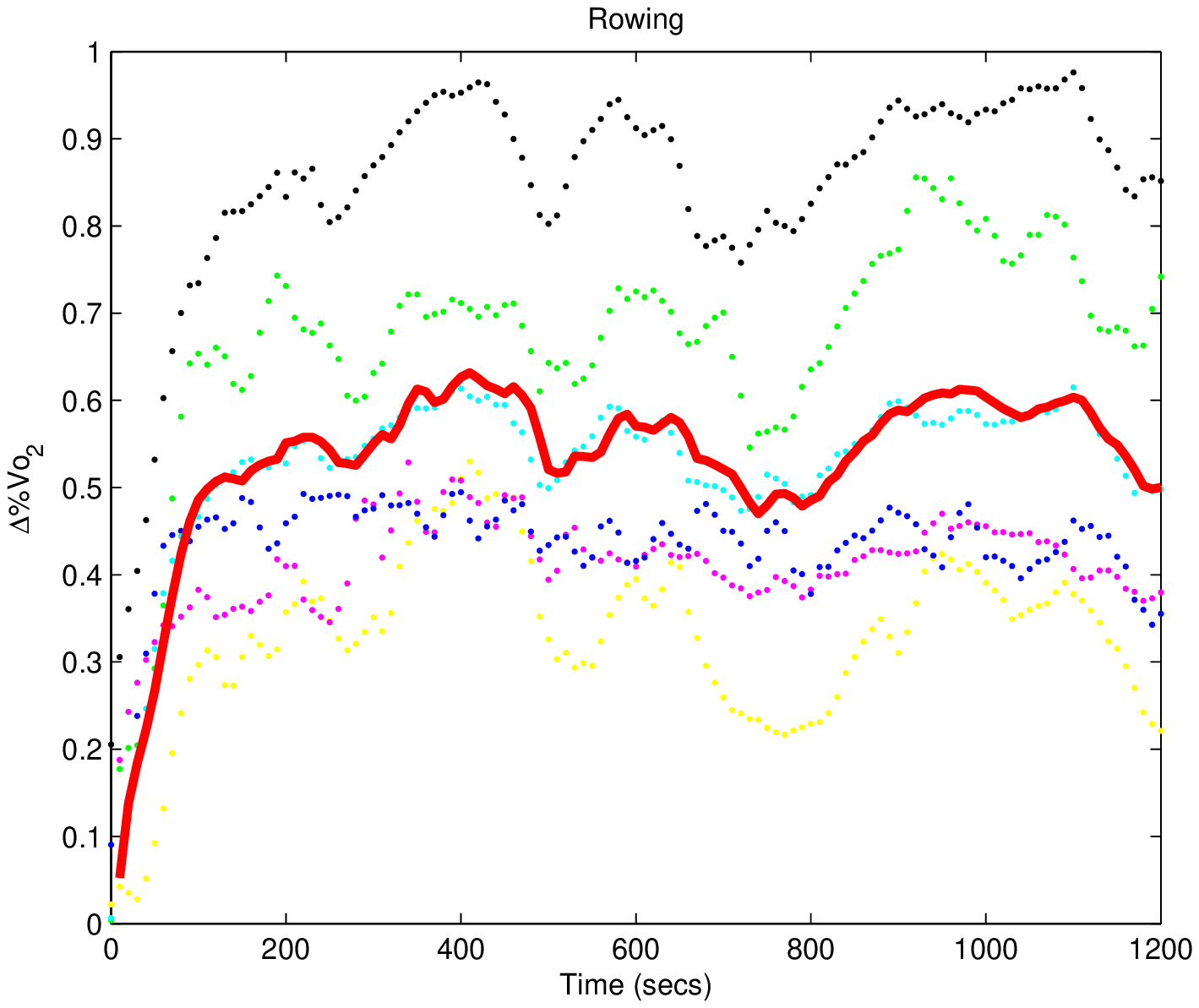}
\caption{$\Delta$\%VO$_2$ during rowing.}\label{VO22}
\end{centering}
\end{ourfigure}

\section{System Identification for $\Delta\% VO_2$}\label{sysiden}

The merged datasets obtained from a cycling, and a rowing exercise at various intensity levels were used to identify a Hammerstein model for each type of an exercise. It consists of two blocks, a static input nonlinearity functions $f(u)$  and a linear dynamic $G(z)$, as explained in  Section \ref{hammodel}. The dimensions of the $f(u)$ are $(3\times1)$, which is the transpose of an input vector $u(n)$. The input vector consists of the inputs ER, RespR, and $\Delta$\%HR$_{max}$.  $G(z)$ is a linear model with $nu$ inputs (nu=3) coming out from the functions $f(u)$ and $ny$ output (ny=1). The function $f(u)$ and $G(z)$ were identified  by minimising the error signal between measured output and estimated output ($\Delta \%VO_2$) through recursive model identification approach.
\subsection{Initialisation of Linear Block G(z) Hammerstein Model}
\begin{ourtable}
\caption{Estimated ARX model during cycling and rowing exercises.}
\label{table4}
\begin{tabular}{c c}
\hline
\hline
Cycling&\\
\hline
\hline
\\
&$A(z) = 1 - 1.349 z^{-1} + 0.7379 z^{-2} - 0.2106 z^{-3}$\\
&$B_1(z) = 0.1084 z^{-2} - 0.006307 z^{-3} + 0.01486 z^{-4}$\\
&$B_2(z) = 0.001481 q^{-1} - 0.001687 q^{-2}$\\
&$B_3(z) = 0.02519 q^{-3}$\\
\hline
\hline
Rowing&\\
\hline
\hline
\\
&$A(z) = 1 - 1.4 z^{-1} + 0.6581 z^{-2} - 0.1394 z^{-3} $\\
&$B_1(z) = 0.01638 z^{-2} + 0.07326 z^{-3} + 0.01223 z^{-4}$\\
&$B_2(z) = -0.001884 z^{-1} + 0.001405 z^{-2}$\\
&$B_3(z) = -0.002912 z^{-3}$\\
\hline
\hline
\end{tabular}
\end{ourtable}

The averaged datasets (inputs are ER, $\Delta$\%HR$_{max}$ and $\Delta$RespR, and output is $\Delta$\%VO$_2$ of six subjects achieved from the PRBS input signal of a cycling and a rowing exercise were used for the identification of the linear model. The estimated linear model was used as an initial guess for the Hammerstein system. Since the selection of a good model structure was very important to ensure maximum model performance, therefore two famous model selection criteria,  Minimum Description Length (MDL) and AIC, were used to select the model structure. It was interesting to note that both selection criteria suggested the similar ARX structure for both types of exercises. The Discrete-time IDPOLY model: $A(z)y(n) = B(z)u(n) + e(n)$ and these polynomials of the ARX models for cycling and rowing exercises are given in Table \ref{table4}. The quality of fit between measured $\Delta$\%VO$_2$ and estimated $\Delta$\%VO$_2$ is $82\%$ for cycling exercise and $80\%$ for rowing exercise, as shown in Fig. \ref{estcyc} and  Fig. \ref{estrow}, respectively.
\begin{ourfigure}
\begin{centering}
\includegraphics[scale=0.75]{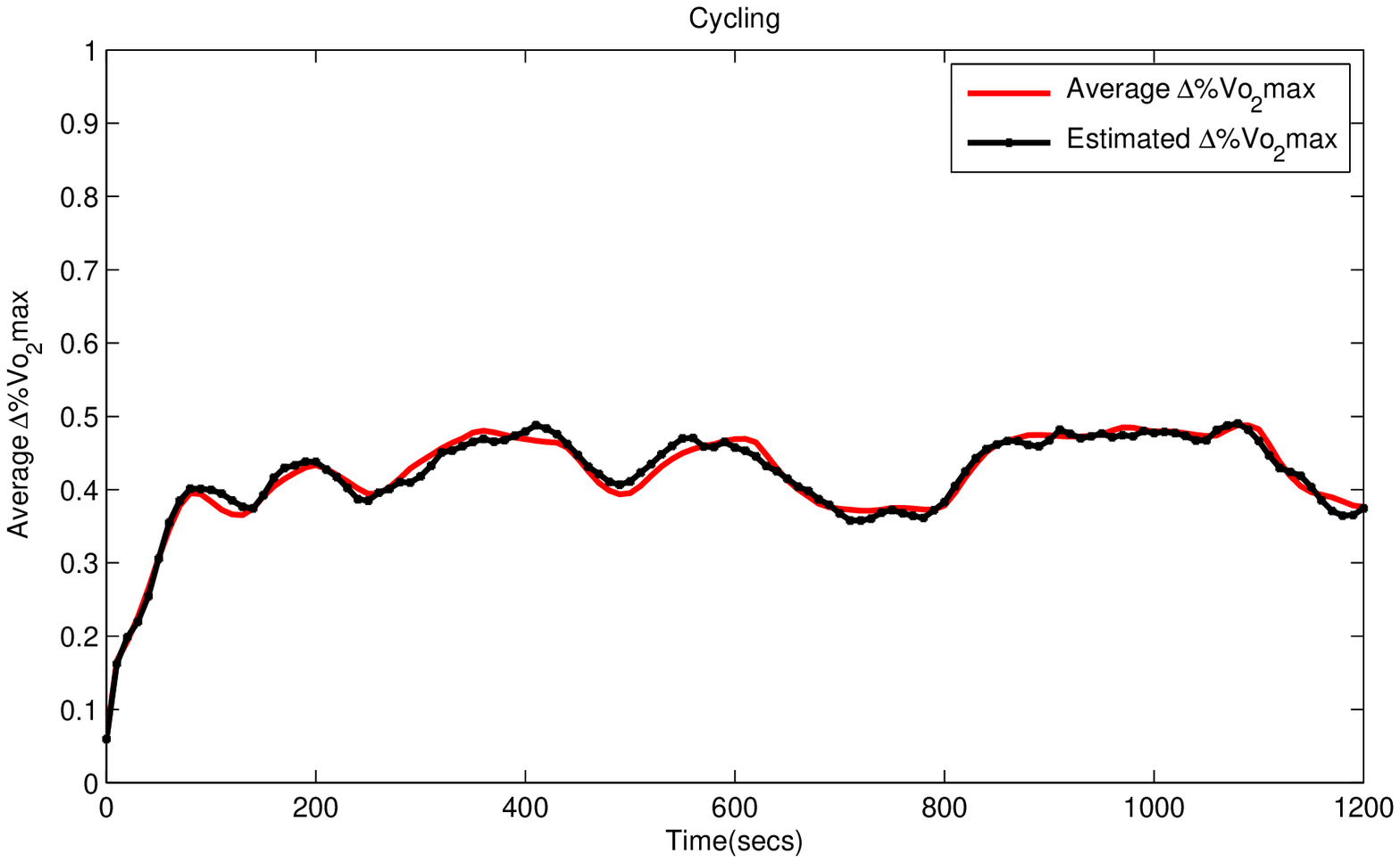}
\caption{Measured and estimated ($arx$:model) VO$_2$ during cycling.}\label{estcyc}
\end{centering}
\end{ourfigure}
\begin{ourfigure}
\begin{center}
\includegraphics[scale=0.75]{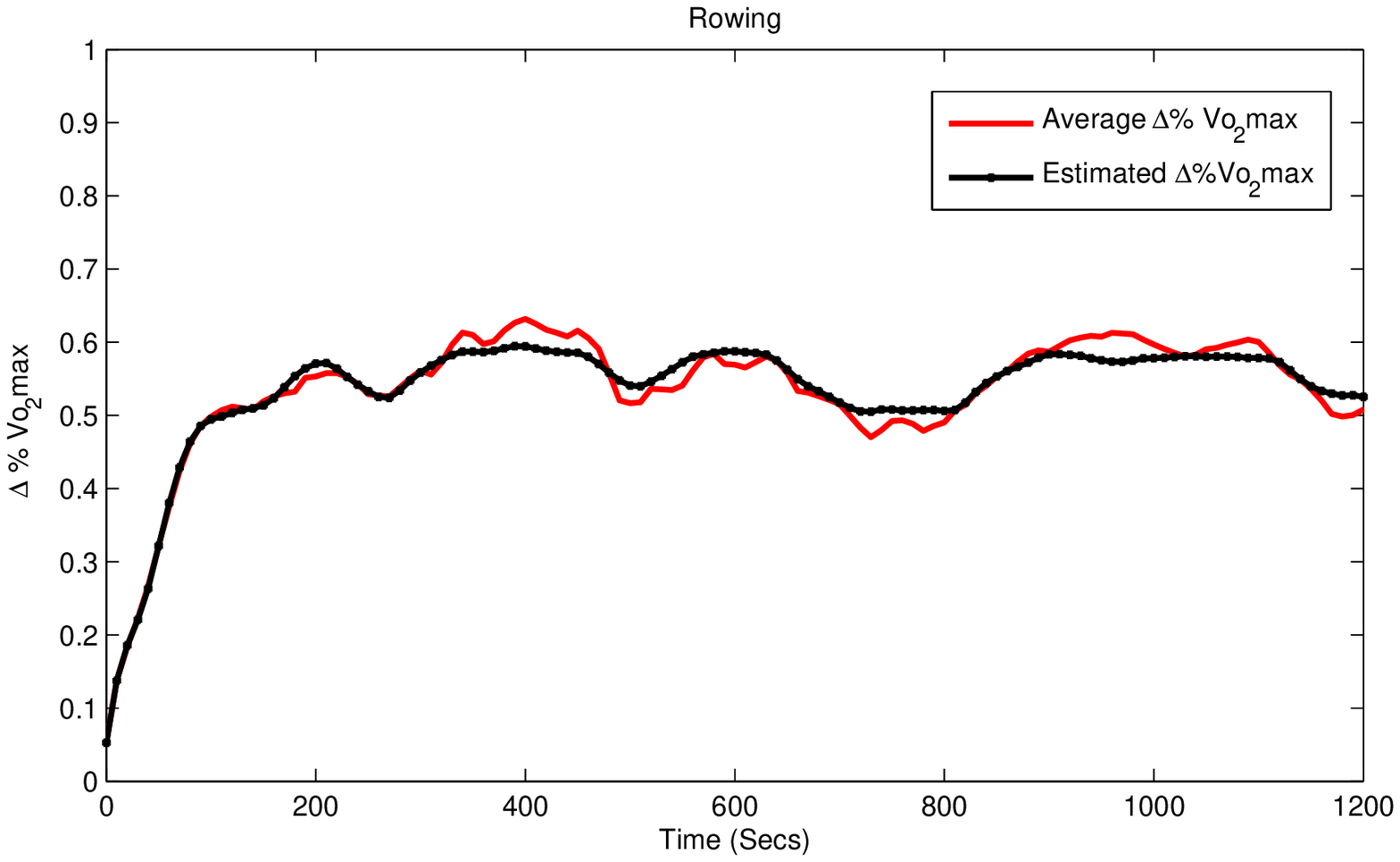}
\caption{Measured and estimated ($arx$:model) VO$_2$ during rowing.}\label{estrow}
\end{center}
\end{ourfigure}
%
\subsection{Initialisation and Estimation of the Hammerstein Model}
The developed Hammerstein models for the estimation of VO$_2$ required a good starting point to achieve a global minimum. The configuration of the Hammerstein system was achieved by using the MATLAB command IDNLHW. The obtained ARX model in Table \ref{table4} was assigned to INDLHW as an input argument (as an initial guess) in terms of the numerator and denominator. This enabled the Hammerstein system to initialise the $G(z)$. A variety of input nonlinearities were configured in the Hammerstein system in order to maximise the quality of fit between the measured and estimated values of VO$_2$. The nonlinear and linear dynamic blocks of the Hammerstein structure were estimated recursively by using the MATLAB PEM command.  This PEM command used the same cost function as described in  \ref{Eq6}. The obtained Hammerstein model for each type of exercise was used for system validation based on the individual datasets. Three best fitting Hammerstein models were used for model selection. These models were obtained by varying the units of an input function $f(u)$ as Sigmoid-net for ER, at the units of r= 3, 4 and 4, respectively (where $r$ represents the number of units in each nonlinearity function); piecewise linear for $\Delta$\%HR$_{max}$ at the units of r= 2, 3 and 3, respectively; and $\Delta$RespR at the units of r= 3, 2 and 3, respectively. The quality of fit obtained from the three best fit Hammerstein models are tabulated in Table \ref{table3} and Table \ref{table33} for cycling and rowing exercises, respectively. The maximum quality of fit was achieved for a cycling and a rowing estimator by assigning the input nonlinearity as Sigmoidnet (r=4) for ER, piecewise linear (r=3 and 3) for $\Delta$\%HR$_{max}$ and $\Delta$RespR.
 It is an important finding that the maximum quality of fit for cycling and rowing exercises is achieved with same Hammerstein structure. The estimated coefficients for cycling and rowing exercises for an  identified  Hammerstein model of the selected structure having the nonlinear function f (u) are given in Table \ref{tablea} and the linear dynamics of similar Hammerstein models is tabulated in Table \ref{table44}.
\begin{ourtable}
\caption{Estimated $G(z)$ during cycling and rowing exercise.}
\label{table44}
\begin{tabular}{c c}
\hline
\hline
Cycling&\\
\hline
\hline
$B_1(z) = z^{-2} - 0.4828 z^{-3} + 0.322 z^{-4}$\\
$B_2(z) = z^{-1} - 0.7866 z^{-2} - 0.2077 z^{-3}$\\
$B_3(z) = z^{-3}$\\
$F_1(z) = 1 - 1.091 z^{-1}+ 0.545 z^{-2} - 0.2335 z^{-3} + 0.01524 z^{-4}$\\
$F_2(z) = 1 - 1.033 z^-1 + 0.3887 z^{-2} - 0.3396 z^{-3}$\\
$F_3(z) = 1 - 0.3598 z^-1 + 0.1091 z^-2 - 0.09779 z^-3$\\
\hline
\hline
Rowing&\\
\hline
\hline

$B_1(z) = 0.4753 z^{-2} + q^{-3} - 1.448 q^{-4}$\\
$B_2(z) = z^-1 - 0.0004589 z^{-2} - 0.893 q^{-3}$\\
$B_3(q) = z^{-3}$\\
$F_1(z) = 1 - 1.823 z^{-1} + 1.295 z^{-2} - 0.6901 z^{-3} + 0.2226 z^{-4}$\\
$F_2(z) = 1 - 0.09835 z^{-1} - 0.7398 z^{-2} + 0.01827 z^{-3}$\\
$F_3(z) = 1 - 0.1031 z^{-1} - 0.08355 z^{-2} - 0.05281 z^{-3}$ \\
\hline
\hline
\end{tabular}
\end{ourtable}
\newpage

\begin{ourtable}
\caption{Estimated coefficient f(u) for ER, $\Delta$\%$HR$, $\Delta$RespR during a cycling and a rowing exercises.}
\label{tablea}
\begin{center}
\tabcolsep 1.0pt
\begin{tabular}{c c c}
\hline
\hline
&Cycling&Rowing\\
\hline
Sigmoid Coeff\\
for $ER$&&\\
\hline
No's Of Units &4 &4\\
Regressor Mean& 0.3583& 0.2808\\
Non Linear Subspace& 1&1\\
Linear Subspace& 1&1\\
Linear Coeff &-0.0231&0.0256\\
Dilation& $[368.64~368.64~3.19~15.21]$&$[372.48~372.48~0.22~168.81]$\\
Translation&$[132.08~ 132.08~-0.17~-7.66]$&$[104.59~104.59~-0.44~-94.94]$\\
Output Coeff& $[0.02~0.02~0.12~0.04]^T$& $[0.01~0.01~-0.013~-0.02]^T$\\
Output Offset & $-0.0573$&$-0.0357$ \\
\hline
Piecewise Linear\\
for $\Delta$RespR&\\
\hline
No of Units&  3&3\\
Breakpoints        &  $5.4216~10.7831~16.2103$& $12.5411~12.9880~38.6236$\\
                   &   $0.1205~0.1292 ~0.1544$& $0.1925~0.1955~0.4031$\\
\hline
Piecewise Linear &\\
for $\Delta$\%HR$_{max}$&\\
\hline
No's of Units&  3&3\\
Breakpoints           &$0.1133~0.1702~0.5731$&$0.1082~0.3057~0.3254$\\
                      & $-0.0032~0.0147~0.0288$&$0.1213~0.2989~0.3133$\\
\hline
\hline
\end{tabular}
\end{center}
\end{ourtable}
\newpage
\begin{ourtable}
\centering
\caption{Quality of fit (\%) between estimated and actual measurement VO$_2$ for cycling exercise.}
\label{table3}
\begin{center}
\tabcolsep 2pt
\begin{tabular}{c c c c c}
\hline
\hline
Subject&  \\
(Pedals/min)&36&48&60&72\\
Quality of fit (\%)\\
\hline
\hline
1&63.2097&  88.0938&  90.6072&   91.0554\\
2&63.3955& 71.6196&   84.6975&   86.3124\\
3&13.6159& 28.5308&   67.1115&   79.7304\\
4&77.7804& 64.6527&   66.3543&   79.0698\\
5&65.3201& 72.8358&   74.0862&   69.9959\\
6&61.2987& 78.1795&   59.5531&   79.1935\\
\hline
Mean& 57.4367 &  67.3187 &73.7350& 80.8929\\
\hline
\hline
1&75.4176 &  88.7289 &  92.0881&  92.5283\\
2&73.8378 & 74.3709  & 84.3008 &  86.6366\\
3&70.6766 &  78.4224 &  88.6455&  92.0582\\
4&80.2803 &  75.4968 & 86.3922 &  90.0938\\
5&67.5128 &  76.8926 & 76.1194 &  84.4881\\
6&61.0681 &  83.8245 &  60.7363&  79.1355\\
\hline
Mean&   71.4655&   79.6227&   81.3804&   87.4901\\
\hline
\hline

1&78.9775   & 89.1209  & 91.9406 &  90.8252 \\
2&76.3993   & 76.8050  & 85.7965 &  87.7043 \\
3&70.6034   & 76.2375  & 87.0541 &  91.2540 \\
4&79.1819   & 74.1882  & 84.6957 &  88.0535 \\
5&68.0779   & 73.4597  & 73.9350 &  81.8243 \\
6&63.4398   & 85.3739  & 86.8097 &  88.9736 \\
\hline
MEAN &  72.7800&   79.1975&   85.0386&   88.5665\\
\hline
\hline

\end{tabular}
\end{center}
\end{ourtable}
\begin{ourtable}
\centering
\caption{Quality of fit (\% ) between estimated and actual measurement VO$_2$ for rowing exercise.}
\label{table33}
\begin{center}
\tabcolsep 2pt
\begin{tabular}{ c c c c c}
\hline
\hline
(strokes/min)&20&24&28&32\\
\hline
\hline
Subject& \\
1& 51.6644  & 59.3733 &  71.4495&   69.6635\\
2&74.8981  & 56.1420 &  83.7805&   83.4840\\
3&78.0220  & 85.3848 &  85.9985&   83.6371\\
4&70.3125  & 77.0378 &  76.0676&   81.9562\\

5& 71.3722  & 76.5799 &  82.7029&   81.8571\\
6& 81.2850 &  84.0796 &  87.1163&   85.6079\\
\hline
Mean&   71.2590&   73.0996&   81.1859&   81.0343\\
\hline

\hline
1&  63.7548&   81.3212&   79.9988&   83.9645\\
2&81.8169&   77.9620&   82.3453&   80.7358\\
3& 82.4695&   88.6229&   88.3980&   84.8886\\
4& 67.0413&   76.9994&   73.9400&   84.7978\\
5& 72.5542&   78.0469&   85.2342&   86.1452\\
6&  78.8676&   84.8713&   85.7528&   88.1537\\
\hline
Mean&   74.4174&   81.3040&   82.6115&   84.7809\\
\hline
\hline
1&63.9516  & 79.0932 &  77.5251 &  83.1175\\
2& 79.6736  & 75.5400 &  82.9806 & 78.4864\\
3& 84.5469  & 85.4603  &  88.5285& 88.4202\\
4& 69.8190  & 79.0265 &   75.5634&  86.2967\\
5& 73.7377 &  76.2217 &   85.2115&  87.3167\\
6&80.7515 &  87.8870 &   85.2763   &  88.9118\\
\hline
MEAN &  75.4134&   80.5381&   82.5142&   85.4249\\
\hline
\hline
\end{tabular}
\end{center}
\end{ourtable}
\begin{ourfigure}
\begin{centering}
\includegraphics[scale=0.5]{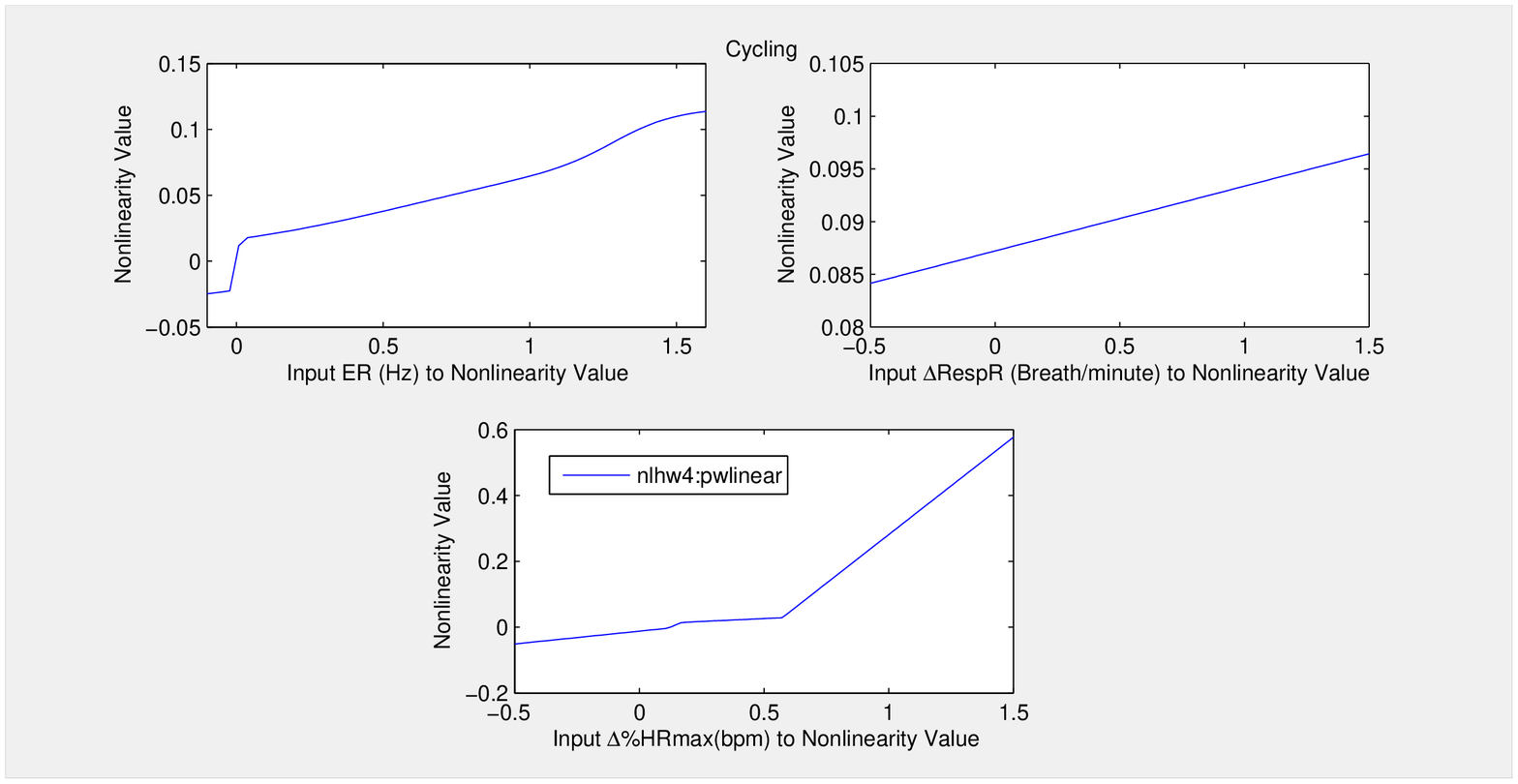}
\caption{Plot of the input function f(u) for VO$_2$ estimator during cycling.}\label{cycnon}
\end{centering}
\end{ourfigure}
The obtained function $f(u)$ given in Table \ref{tablea} was used to calculate the behaviour of nonlinearities by varying the values of ER, $\Delta$\%HR$_{max}$ and  $\Delta$\%HR$_{max}$ in the linear range. The estimated behaviour of each nonlinearity function is shown in Fig. \ref{cycnon} and Fig. \ref{rownon} for cycling and rowing exercises, respectively. The nonlinearity function is a sigmoid function for ER which behaves entirely differently for both types of exercises. However, in case of cycling, the estimated Hammerstein model shows the bandwidth between 0.25-1.5 Hz, and the input nonlinearity function of Hammerstein system behaves curvilinearly. In case of rowing exercise, the bandwidth lies between 0.25-0.55 Hz, and the input nonlinearity function  of Hammerstein system behaves curvilinear, which also pointed a fact that varying ER linearly does not give linear steady state values of VO$_2$. The behaviour of the nonlinearity function for RespR and HR is almost same for both types of exercises. It is hence concluded that the ER varies the demand of VO$_2$ only depending upon a type of exercise. Different exercises evolve a different muscle group of activities, which limits the extremity of the body movements. This means that each exercise has its own maximum ER which is related with the demands of VO$_2$.
\begin{ourfigure}
\begin{centering}
\includegraphics[scale=0.5]{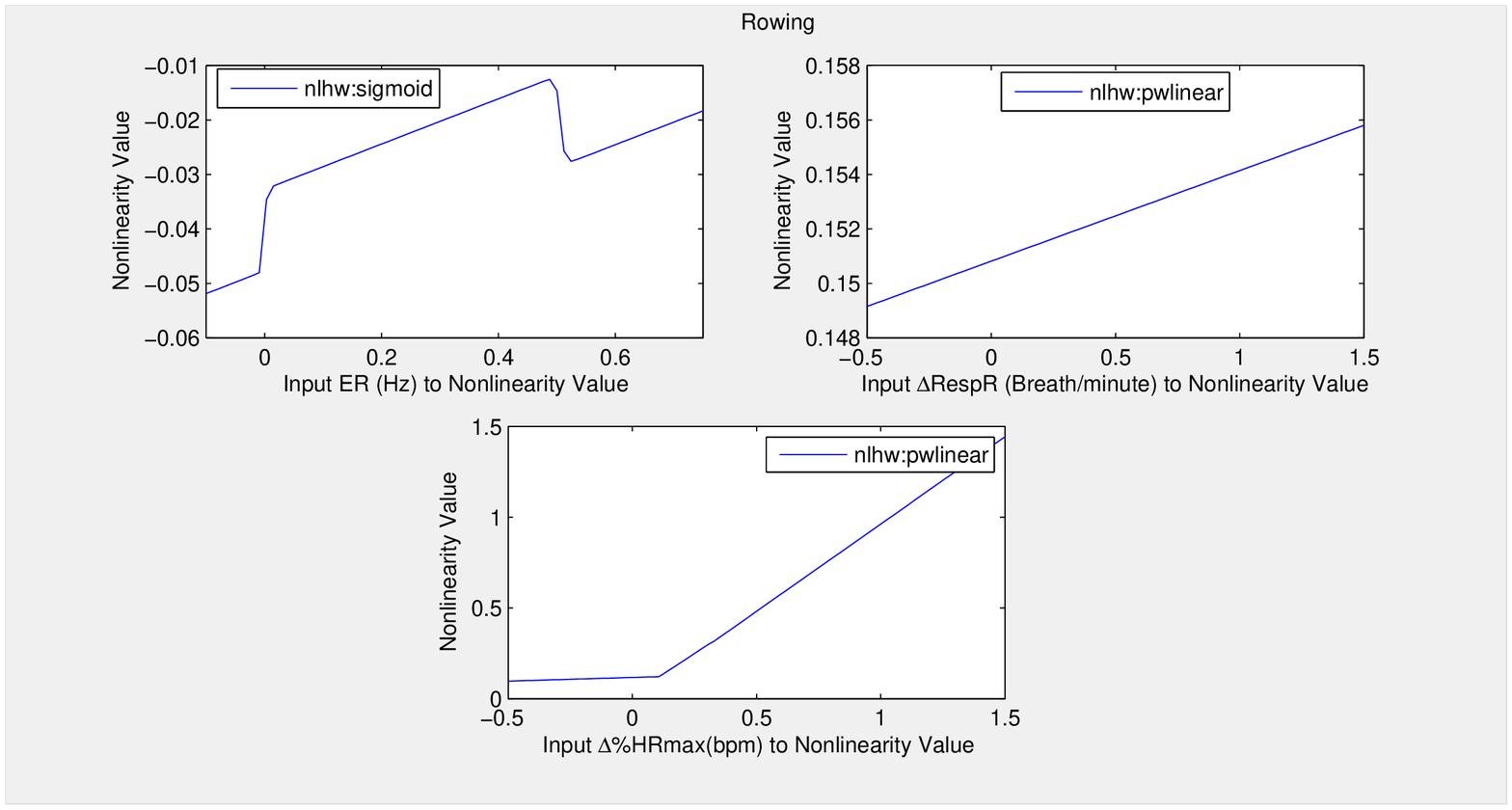}
\caption{Plot for the inputs function f(u) for VO$_2$ estimator during rowing.}\label{rownon}
\end{centering}
\end{ourfigure}

\subsection{Validation of Hammerstein Estimators for Cycling and Rowing Exercises} \label{validchap5}
This chapter presents an innovative methodology for estimating VO$_2$ during cycling and rowing exercises. The developed approach does not require calibration of VO$_2$ due to the differences in individual fitness for both types of exercises. The performance of the developed estimator was measured in terms of the quality of fit against VO$_2$ measurement at various levels of intensity during cycling and rowing exercises. This goodness of fit was calculated from the normalised root mean square error (NRMSE), i.e., quality of fit = $(1- NRMSE)\times100~\%$. The performance of the proposed estimator during cycling and rowing for all subjects is given in Table \ref{table3} and Table \ref{table33}, and shown in Fig. \ref{Fig4} and Fig. \ref{Fig5}. The obtained results illustrate that the proposed estimator in Fig. \ref{Fig2} is capable of estimating VO$_2$ by predicting $\Delta$\%VO$_2$ and by using prior estimates of HR$_{max}$, VO$_{2rest}$ and VO$_{2max}$ for each type of exercise. The obtained results for estimation show that the average quality of fit at different values of ER (36, 48, 60 and 72 pedals/min) is 73\%, 79\%, 81\% and 86\% during cycling. In case of rowing, this quality of fit at different values of ER (20, 24, 28 and 32 strokes/min) is 75\%, 80\%, 82\% and 85\%. It can be observed that the average quality of fit is improved with increased exercise intensity in both types of exercises. %
\begin{ourfigure}
\begin{centering}
\includegraphics[scale=0.7]{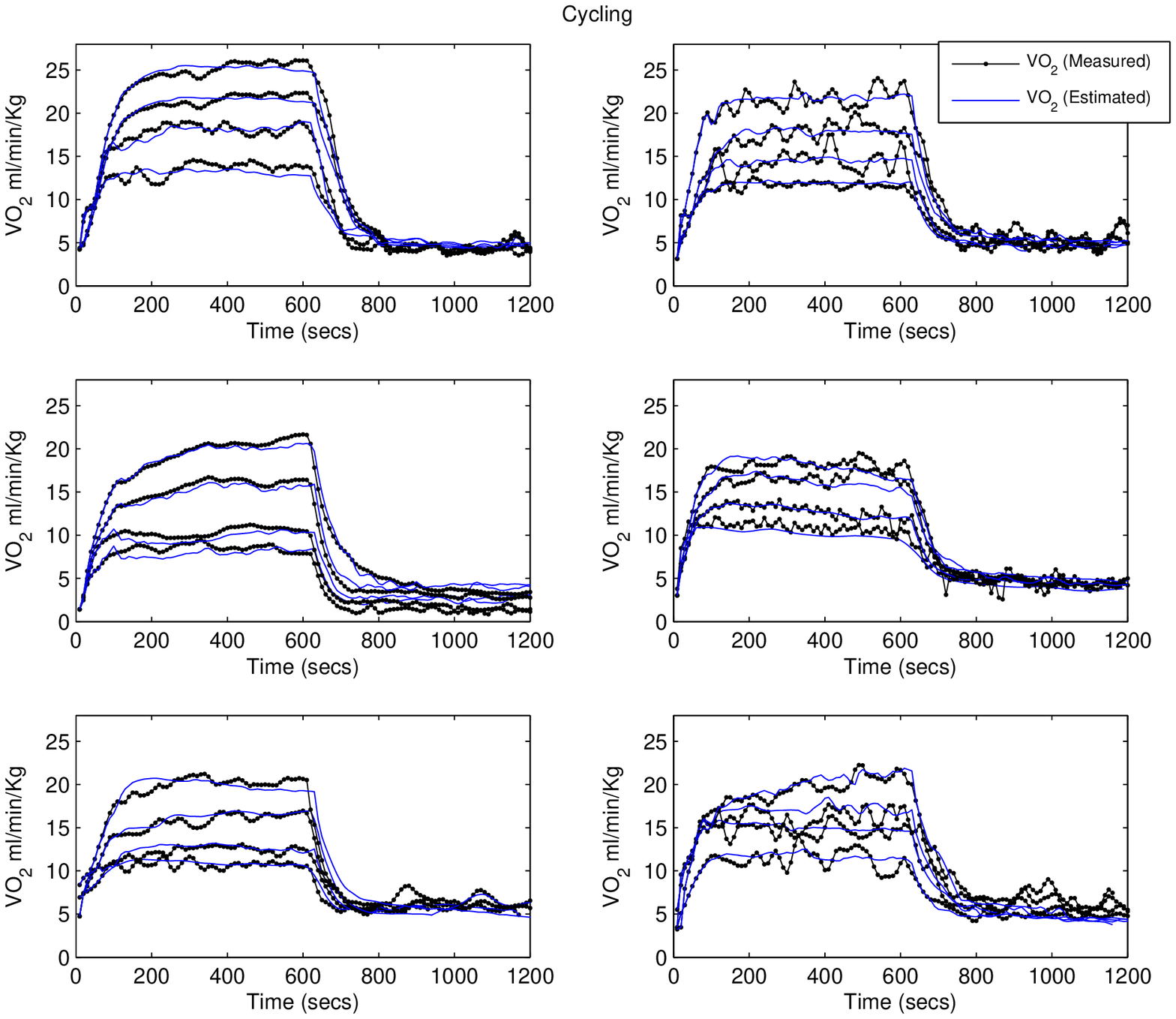}
\caption{Measured and estimated output of VO$_2$ during cycling.}\label{Fig4}
\end{centering}
\end{ourfigure}
\newpage
\begin{ourfigure}
\begin{centering}
\includegraphics[scale=0.7]{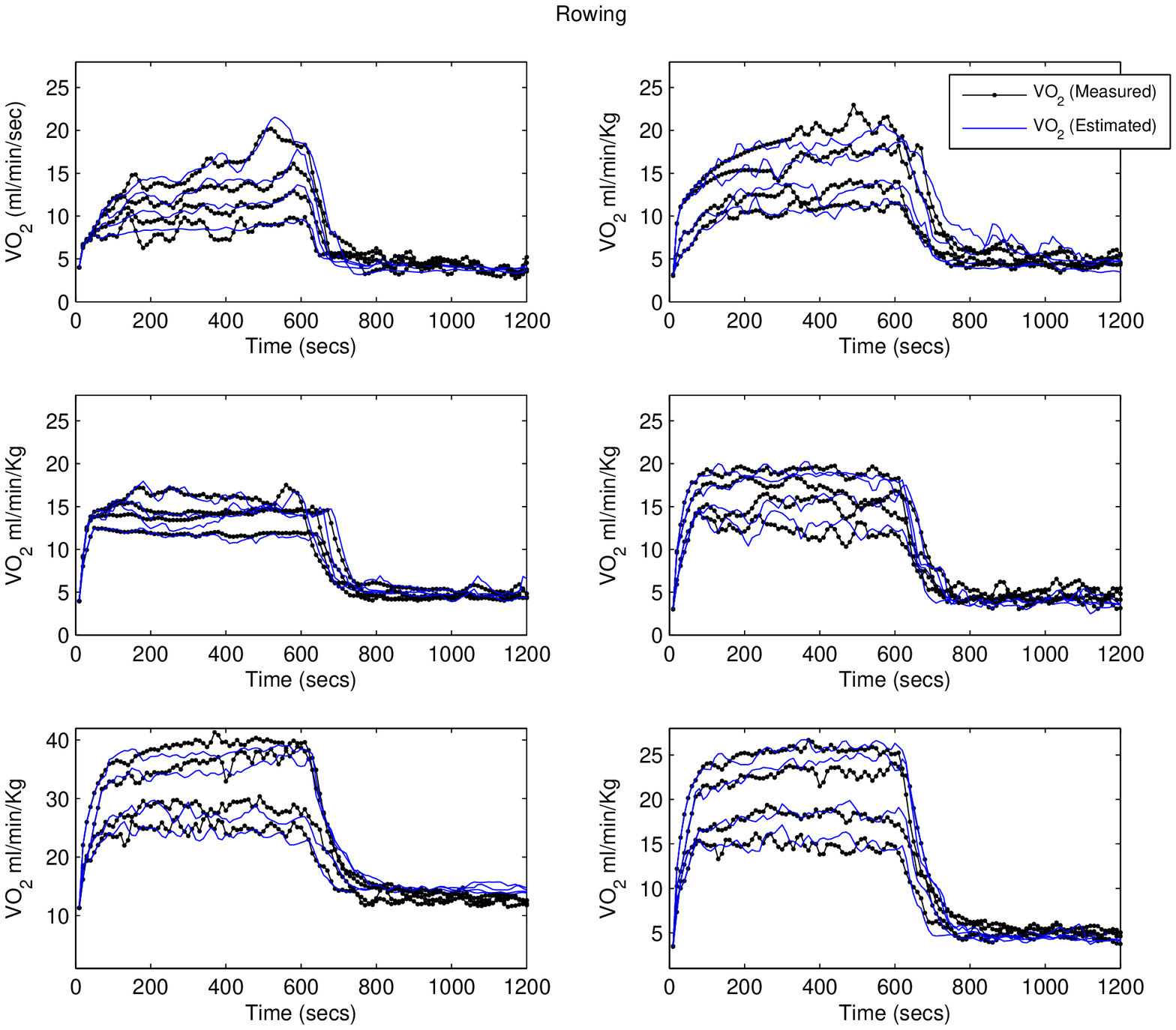}
\caption{Measured and estimated output of VO$_2$ during rowing.}\label{Fig5}
\end{centering}
\end{ourfigure}

\section{Real-time Implementation of Estimation based Self Biofeedback Control of VO$_2$ During Rhythmic Exercises}\label{sbf}
\begin{figure}
\begin{centering}
\includegraphics[scale=1]{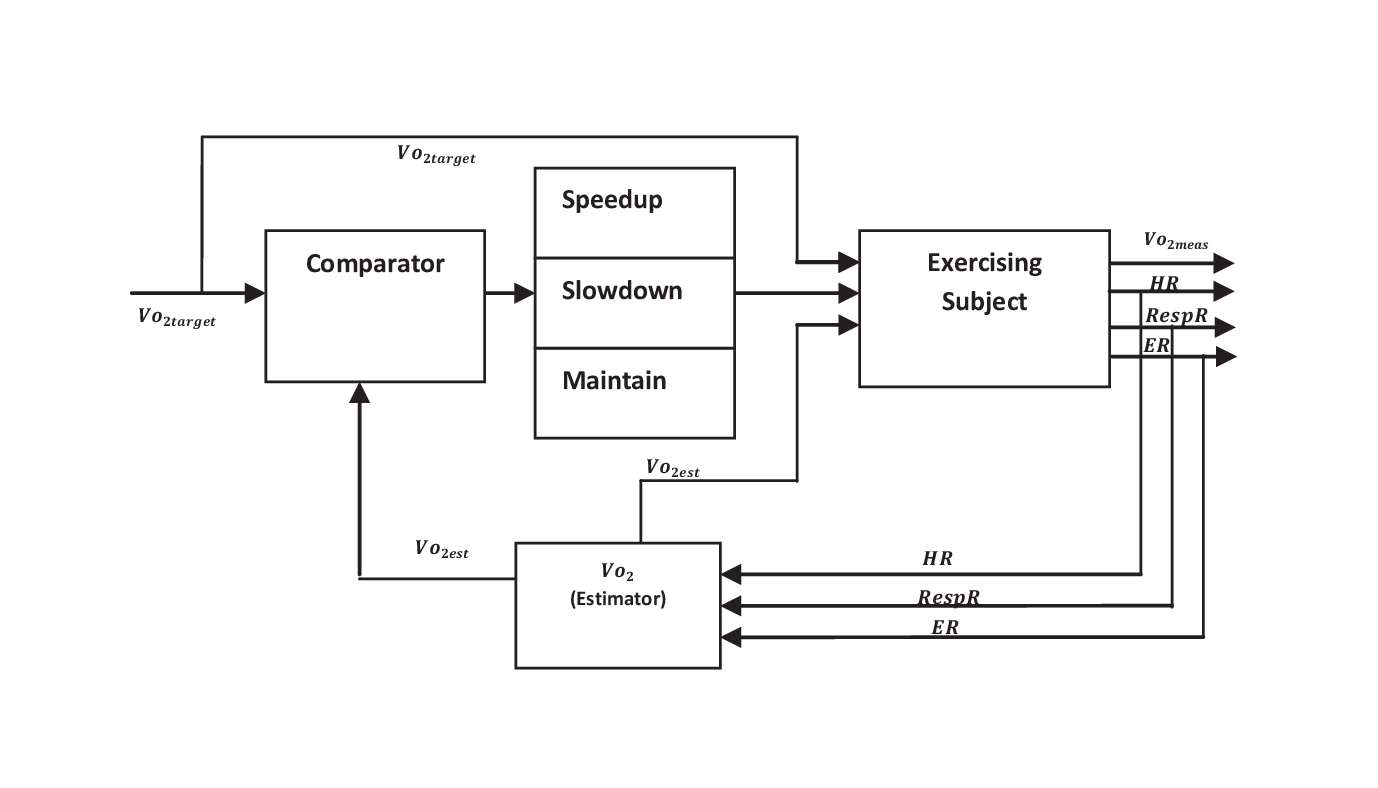}
\caption{Block diagram of the Self Biofeedback Control of VO$_2$.}\label{Fig3}
\end{centering}
\end{figure}
The SBF control of any physiological variable is normally dependent on the individual perception to achieve the targeted profile. This perception is received by an individual using the biofeedback alarming signal. This biofeedback is given to the exercising subject based on the measured and the targeted values of the control physiological variable. The SBF control of VO$_2$ is not being investigated due to an inexpensive and unwieldy measurement of VO$_2$. In this section, we present a SBF control design of VO$_2$ based on its estimate. The developed estimator predicts VO$_2$ using the measurements of HR, RespR and ER and can be manipulated as the inputs to VO$_2$ estimator. Moreover, these measurements are easily achievable in real-time. The developed estimator for each type of exercise predicts the $\Delta$\%VO$_2$ using identified Hammerstein system which was obtained independently for cycling and rowing exercises.\\
To date, the performance of SBF control is basically analysed using HR as a controlling variable; see section \cite{Goldstein1977,Dardik1991a} . However, we investigate the performance of SBF system using VO$_2$ as control variable during rhythmic activities, i.e., cycling and rowing. The block diagram of the developed SBF control of VO$_2$ is shown in Fig. \ref{Fig3}. The biofeedback control of VO$_2$ was generated by an alarm by flashing an indicator on the computer screen. The purpose of this indicator is to display the desired action to be achieved by the subject. e.g., VO$_{2target}$.  It displayed the VO$_{2est}$ and VO$_{2target}$ also  so that the subject could decide the precision of action by himself. The command of action was decided by the SBF system based on VO$_{2est}$ and VO$_{2target}$. The control command $(U(n))$ switching is based on the values of VO$_{2target}$ and VO$_{2est}$ defined as follows:\\

 U(n)=$\begin{cases}\\
Speedup~~~\mbox{if}~VO_{2est}<0.95VO_{2target}\\
~~~~~~~~~Maintain~~~\mbox{if}~0.95VO_{2target}\leq VO_{2est}\leq 1.05VO_{2target}\\
~~~~~~~~~Slowdown~~~\mbox{if}~VO_{2est}> 1.05VO_{2target}\\
\end{cases}$
Three biofeedback commands are Speedup, Slow Down and Maintain. The SBF system displays the desired action by considering the associated constraints for switching amongst the biofeedback commands. A selected command of a desired action gets turned ON, and is communicated to the exercising individual via a respective flashing indicator on the computer screen. The real-time implementation of the estimation based SBF control of VO$_2$ requires the accurate measurements of the sensory data for estimating VO$_{2est}$. These sensors were deployed on the exercising subjects and were connected to the CS using National Instruments (NI) LabVIEW software. Data received from the sensors were used by the VO$_2$ estimator to estimate VO$_2$ (VO$_{2est}$ ) in real-time. The LabVIEW software was installed on the CS, and received sensory data was filtered properly and consequently used for the mathematical manipulations for the prediction of VO$_2$. Based on the estimated output of VO$_2$, the SBF system decides which command need to be turned ON. During the resting phase, recordings of HR and RespR were used to determine the baseline measures of these quantities. The obtained resting values are used for calculating the VO$_{2max}$, VO$_{2rest}$, $\Delta\%HR$, and $\Delta$RespR which were used as the inputs of the Hammerstein system for the estimation of $\Delta$\%VO$_{2}$.  The linear block of the Hammerstein system for each type of exercise was implemented as the linear state space model and polynomials form as tabulated in Table\ref{tablea}.  The polynomial model was transformed into the state space model and assigned in LabVIEW. The nonlinearity is evaluated by using the EVALUATE MATLAB command. This command was interfaced with the help of MATLAB script, presented in LabVIEW environment.  All measurements of the inputs (HR, RespR and ER) were interfaced and synchronised at a sampling rate of 10 seconds.\\
The working principle of the proposed system can be explained by the following steps:
\begin{enumerate}
\item{ The current measurements of HR, RespR and ER are used to manipulate the inputs of an estimator by using the physical characteristics, i.e., age, weight and height, and additionally  the estimated values of VO$_{2rest}$  and VO$_{2max}$}.
\item{The manipulated inputs $\Delta$\%HR$_{max}$, $\Delta$RespR, and ER are used for getting the current value of VO$_{2est}$}.
\item{Based on the current values of VO$_{2est}$ and VO$_{2target}$, the comparator interprets the desired command of action. The required biofeedback command is communicated to the exerciser via flashing an indicator of a desired action such as slow-down, speed up or maintain the frequency of the body movement.}
\item{The exercising subject adjusts the body movements based on the visual command and the values of VO$_{2est}$ and VO$_{2target}$.These adjustments of the body movements determine the targeted ER for the targeted VO$_2$ profile of an exercising subject.}
\end{enumerate}
\subsection{Experimental Results}\label{bioresults}
This section discusses the results and performance of estimation based SBF control of VO$_2$ for the subject engaged in cycling and rowing exercises. The VO$_2$ estimator was identified from the datasets of the six healthy male subjects performing cycling and rowing exercises. The performance of the SBF control of VO$_2$ was dependent on the accurate estimation of VO$_2$. This performance was assessed in terms of a Root Mean Square Error (RMSE) between estimated and measured values of VO$_2$ using the SBF system. System validation is based on VO$_2$ measured through the gas analyser (COSMED Kb4) breath by breath, and is averaged every 10 seconds during a exercise. Table \ref{tableerr}, shows RMSE of the VO$_2$ of six subjects in real-time.
\begin{ourtable}
\centering
\caption{RMSE between VO$_{2meas}$ and VO$_{2est}$.}
\label{tableerr}
\begin{center}
\tabcolsep 1.5pt
\begin{tabular}{c c c c c c c c}
&Subject~1& Subject~2& Subject~3& Subject 4& Subject 5& Subject 6 & Mean\\
\hline
Rowing&4.8759&    1.4830&    0.3423&    0.1080&    0.9074&    0.2252& 1.3263\\
Cycling&1.0101&   1.1700&    1.7776&    1.9486&    1.0166&    1.1560& 3.4650\\
\\
\hline
\hline
\end{tabular}
\end{center}
\end{ourtable}

\begin{figure}
\begin{centering}
\includegraphics[scale=0.45]{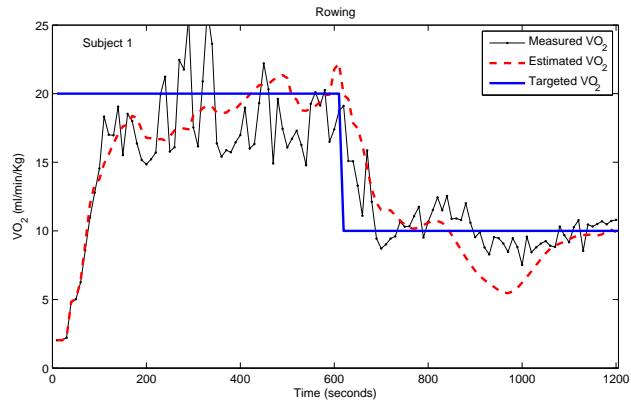}
\caption{ VO$_{2meas}$ (dashed line), VO$_{2est}$ (dotted line) and VO$_{2target}$ (solid line) during cycling.}\label{Subjects11}
\end{centering}
\end{figure}

\begin{figure}
\begin{centering}
\includegraphics[scale=0.45]{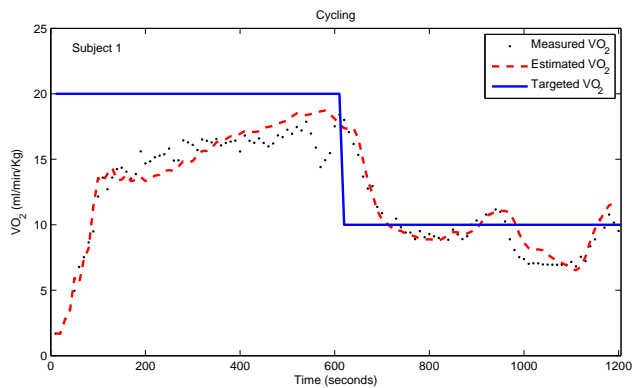}
\caption{VO$_{2meas}$ (dashed line), VO$_{2est}$ (dotted line) and VO$_{2target}$ (solid line) during rowing.}\label{Subject1}
\end{centering}
\end{figure}

\begin{figure}
\begin{centering}
\includegraphics[scale=0.5]{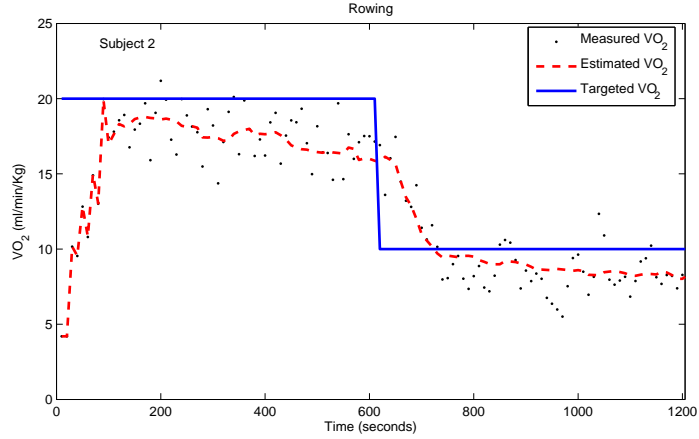}
\caption{VO$_{2meas}$ (dashed line), VO$_{2est}$ (dotted line) and VO$_{2target}$ (solid line) during cycling.}\label{Subjects12}
\end{centering}
\end{figure}

\begin{figure}
\begin{centering}
\includegraphics[scale=0.5]{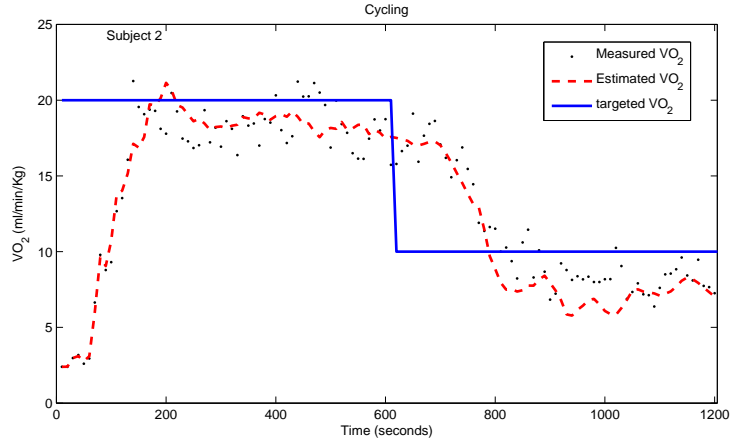}
\caption{VO$_{2meas}$ (dashed line), VO$_{2est}$
 (dotted line) and VO$_{2target}$ (solid line) during rowing.}\label{Subject2}
\end{centering}
\end{figure}

\begin{figure}
\begin{centering}
\includegraphics[scale=0.5]{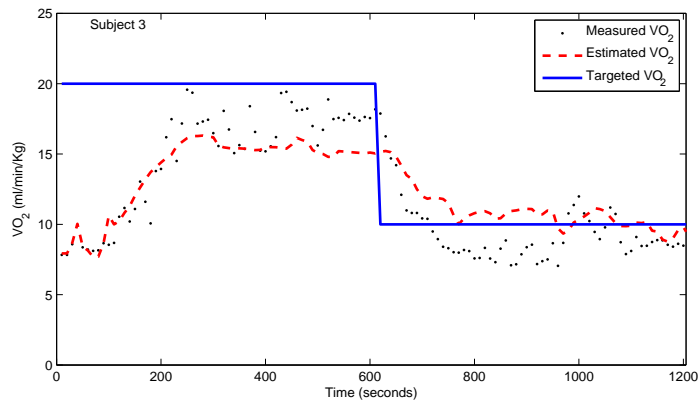}
\caption{ VO$_{2meas}$ (dashed line), VO$_{2est}$ (dotted line) and VO$_{2target}$ (solid line) during cycling.}\label{Subjects13}
\end{centering}
\end{figure}

\begin{figure}
\begin{centering}
\includegraphics[scale=0.5]{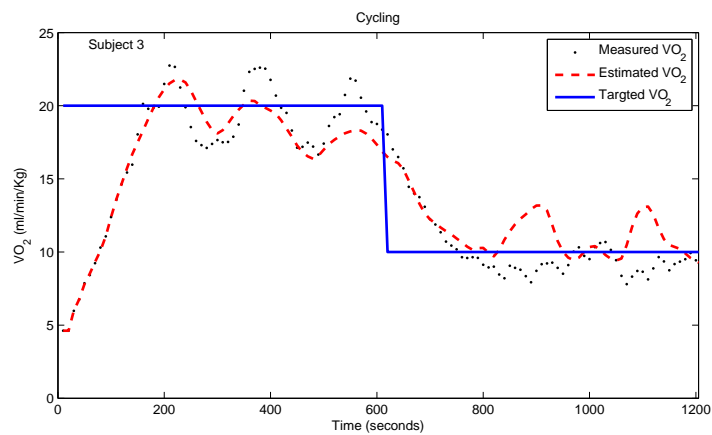}
\caption{VO$_{2meas}$ (dashed line), VO$_{2est}$ (dotted line) and VO$_{2target}$ (solid line) during rowing.}\label{Subject3}
\end{centering}
\end{figure}

\begin{figure}
\begin{centering}
\includegraphics[scale=0.5]{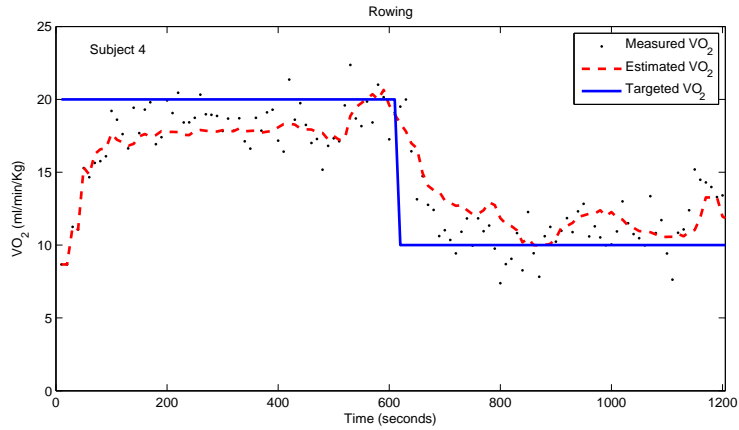}
\caption{VO$_{2meas}$ (dashed line), VO$_{2est}$ (dotted line) and VO$_{2target}$ (solid line) during cycling.}\label{Subjects14}
\end{centering}
\end{figure}

\begin{figure}
\begin{centering}
\includegraphics[scale=0.5]{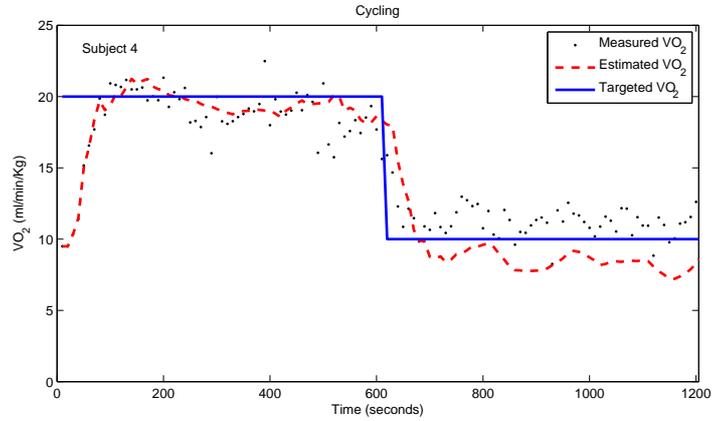}
\caption{VO$_{2meas}$ (dashed line), VO$_{2est}$ (dotted line) and VO$_{2target}$ (solid line) during rowing.}\label{Subject4}
\end{centering}
\end{figure}

\begin{figure}
\begin{centering}
\includegraphics[scale=0.5]{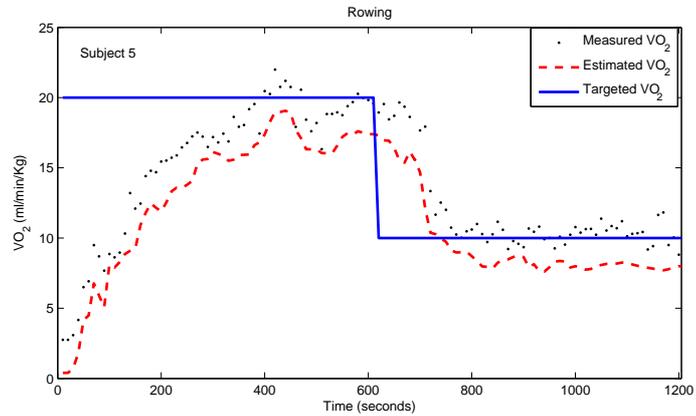}
\caption{VO$_{2meas}$ (dashed line), VO$_{2est}$ (dotted line) and VO$_{2target}$ (solid line) during cycling.}\label{Subjects15}
\end{centering}
\end{figure}

\begin{figure}
\begin{centering}
\includegraphics[scale=0.5]{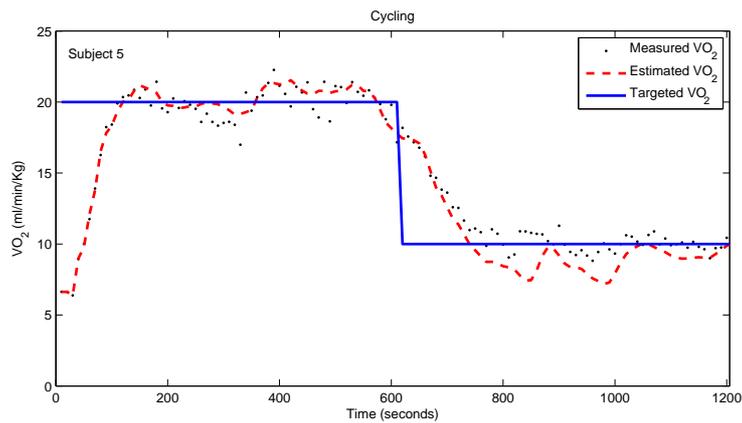}
\caption{VO$_{2meas}$ (dashed line), VO$_{2est}$ (dotted line) and VO$_{2target}$ (solid line) during rowing.}\label{Subject5}
\end{centering}
\end{figure}

\begin{figure}
\begin{centering}
\includegraphics[scale=0.5]{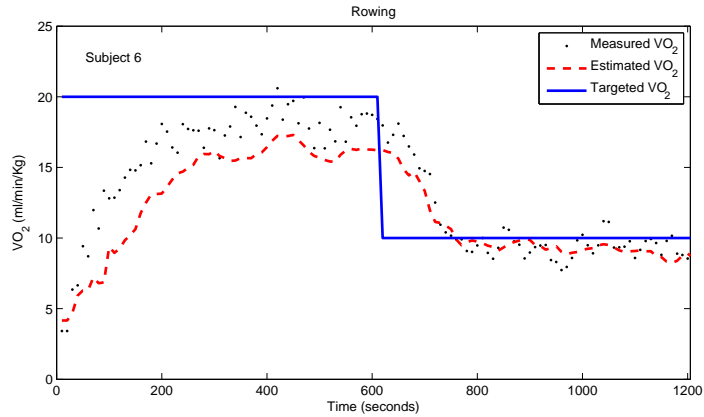}
\caption{ VO$_{2meas}$ (dashed line), VO$_{2est}$ (dotted line) and VO$_{2target}$ (solid line) during cycling.}\label{Subjects16}
\end{centering}
\end{figure}

\begin{figure}
\begin{centering}
\includegraphics[scale=0.5]{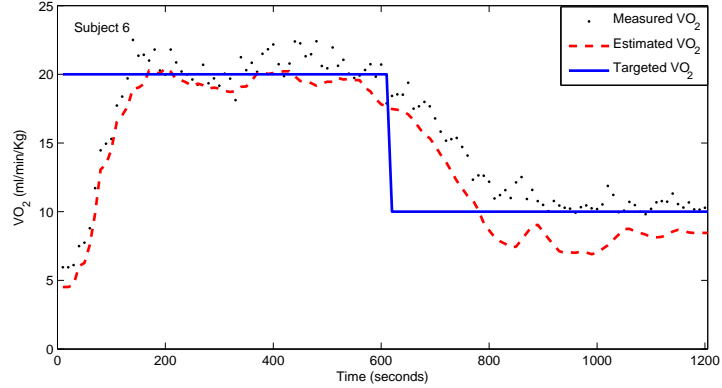}
\caption{ VO$_{2meas}$ (dashed line), VO$_{2est}$ (dotted line) and VO$_{2target}$ (solid line) during rowing.}\label{Subject6}
\end{centering}
\end{figure}

Figures (\ref{Subjects11}, \ref{Subjects12}, \ref{Subjects13}, \ref{Subjects14}, \ref{Subjects15} and \ref{Subjects16}) show that all the subjects achieved VO$_{2target}$ to some extent during the high intensity and low intensity phases of cycling. Similar finding were observed during rowing exercises as shown in figures (\ref{Subject1}, \ref{Subject2}, \ref{Subject3}, \ref{Subject4}, \ref{Subject5} and \ref{Subject6}). It can be clearly observed from these figures that the developed VO$_2$ estimator is capable of predicting the dynamics of VO$_2$ accurately using our proposed inputs, and the estimated and measured values of VO$_2$ are approximately  matching.\\
 The performance of the SBF system was assessed in terms of a steady state error (SSE) between VO$_{2target}$ and VO$_{2est}$ and transition times. The SSE is calculated in terms of RMSE which is defined as RMSE1 for high intensity exercise and RMSE2 for low intensity exercise. During high and low intensity phases of  cycling and rowing exercises, VO$_{2target}$ is 20 ml/min/Kg and 10 ml/min/Kg respectively. In real-time, developed estimators give average RMSE$~=~1.3465~$ ml/min/Kg and RMSE$~=~ 1.3236$ ml/min/Kg for a cycling and a rowing exercise, while controlling VO$_2$ of exercising subject. The detailed analysis of each subject is given in the following section.\\
\begin{ourtable}
\centering
\caption{Steady state error and transient time, from low to high and high to low intensity during cycling and rowing exercises.}
\label{table11a}
\begin{center}
\tabcolsep 0.75pt
\begin{tabular}{c c c c c c c c c c c}
\hline
\hline
&&Cycling&&&&&Rowing\\
\hline
Subject & RMSE1 & RMSE2 & TS & T$_{HL}$ && RMSE1 & RMSE2 & TS & T$_{HL}$\\
&(ml/min/Kg)&(ml/min/Kg)&(secs)&(secs)&&(ml/min/Kg)&(ml/min/Kg)&(secs)&(secs)\\
\hline
S1  &	3.8400 &1.6445& 120 & 100&&2.445 & 0.9180& 100&  120\\
S2	&   1.7573 &1.7033& 100 & 130&&2.3547& 1.6965& 120&  210\\
S3	&   1.2495 &0.9000& 250 & 120&&3.0230 &  1.7568 &160&  150\\
S4	&   1.366  &1.2680& 100 & 120&&2.0376 & 2.0297 & 90 &  90\\
S5	&   0.6187 &0.3086& 260 & 180&&1.9135&  0.7727 & 110 & 170\\
S6  &   1.0235 &0.8966& 280 & 120&&1.0235 & 0.9734 & 140 & 140\\
Mean &  1.6425 &1.1202& 185 & 128&&2.1329 & 1.3579 &  120& 147\\
\hline
\hline
\end{tabular}
\end{center}
\end{ourtable}
\textbf{Subject S1 }

\textbf{Cycling}\\
During the high intensity phase of a cycling exercise, the Subject S1 achieved a steady state (SS) in 120 seconds, as shown in Fig. \ref{Subjects11} with the RMSE1 $= 3.840$ ml/min/Kg. This is highest SSE amongst all the subjects; see Table \ref{table11a}. However, during low intensity phase of an exercise, he achieved a SS in 700 seconds. The transition time between high to low intensity $(T_{HL})$ was about 100 seconds, which is less than a value of TS to achieve a SS with RMSE2 $= 1.6445$ ml/min/Kg,  and  less than a value of RMSE1 as well. These results indicated that a steady state control of VO$_2$ using a SBF system for the Subject S1 performed more efficiently during the low intensity of a cycling exercise.

\textbf{Rowing}\\
During the high intensity phase of a rowing exercise, the Subject S1 achieved a SS in 200 seconds, as shown in Fig. \ref{Subject1} with the RMSE1$=2.445$ ml/min/Kg. The Subject S1 achieved a SS in TS seconds is the longest duration amongst all subjects during the high and low intensity phases of cycling and rowing exercises; see  Table \ref{table11a}.  However, during low intensity exercise, the Subject S1 achieved SS in 720 seconds, T$_{HL}$ is 120 seconds, which is less than a value of TS to achieve SS with the RMSE2 $=0.9180$ ml/min/Kg, and  also less  than the RMSE1. These results indicate that a steady state control of VO$_2$ using a SBF system for the Subject S1 was performed more efficiently during the low intensity of a rowing exercise.
\newpage
\textbf{Subject S2}

\textbf{Cycling}\\
During the high intensity phase of cycling exercise, the Subject S2 achieved a steady state (SS) in 100 seconds, as shown in Fig. \ref{Subjects12} with the RMSE1 $= 1.7573$ ml/min/Kg. This is highest SSE amongst all the subjects; see Table \ref{table11a}. However, during low intensity phase of an exercise, Subject S2 achieved a SS in 730 seconds. The T$_{HL}$ was about 130 seconds, which is greater than TS to achieve SS with the RMSE2 $=1.7033$ ml/min/Kg which is slightly less than a value of RMSE1. These results indicated that the steady state control of VO$_2$ using a SBF system with the Subject S2 performed adequately during both the low and high intensity phases of a cycling exercise.\\
\textbf{Rowing}\\
During the high intensity phase of rowing exercise,  the Subject S2 achieved a SS in 120 seconds, as shown in Fig. \ref{Subject2} with the RMSE1 $= 2.3547$ ml/min/Kg. However, during low intensity exercise, the Subject S2 achieved a SS in 820 seconds. The $(T_{HL})$ is 210 seconds, which is greater than a value of TS to achieve a SS. The RMSE2 $=1.6965$ ml/min/Kg is less than RMSE1. Similar results based on the SBF system were observed with the Subject S2 during a cycling exercise.

\textbf{Subject S3}

\textbf{Cycling}\\
During the high intensity phase of a cycling exercise, the Subject S3 achieved a SS in 250 seconds, as shown in Fig. \ref{Subjects13}. The TS of the Subject S3 is the third longest duration among all the subjects with the RMSE1 $= 1.249$ ml/min/Kg. However, during low intensity phase of an exercise, the Subject S3 achieved SS in 720 seconds. The T$_{HL}$ was about to 120 seconds, which is less than TS with the RMSE2  = 0.9005 ml/min/Kg. The RMSE2 is less than a value of RMSE1.  These results indicated that a steady state control of VO$_2$ using a SBF system for the Subject S3 performed more efficiently during the low intensity phase of a cycling exercise.

\textbf{Rowing}\\
During the high intensity phase of rowing exercise, the Subject S3 achieved the SS in 160 seconds, as shown in Fig. \ref{Subject3} with the RMSE1 $= 3.023$ ml/min/Kg.  However, during low intensity exercise, the Subject S3 achieved SS in 750 seconds, T$_{HL}$ $= 150$ seconds, which is greater than the TS to achieve SS with the RMSE2 $=1.7568$ ml/min/Kg, which is less than a value of RMSE1. These results indicated that a steady state control of VO$_2$ using a SBF system for the Subject S3 performed more efficiently during the low intensity phase of a rowing exercise.
\newpage
\textbf{Subject S4}

\textbf{Cycling}\\
During the high intensity phase of cycling exercise, the Subject S4 achieved a SS in 100 seconds with the RMSE1 $=1.3666$ ml/min/Kg, as shown in Fig. \ref{Subjects14}. However, during low intensity phase of an exercise, the Subject S4 achieved SS in 720 seconds. The T$_{HL}$ was about to 120 seconds, which is greater than TS to achieve SS with the RMSE2 $= 1.2687$ ml/min/Kg, which is  less than a value of RMSE1. Results showed that the steady state control of VO$_2$ using a SBF system for the Subject S4 performed more efficiently for the low intensity phase of a cycling exercise.

\textbf{Rowing}\\
During the high intensity phase of rowing exercise, the Subject S4 achieved a SS in 90 seconds, as shown in Fig. \ref{Subject4} with the RMSE1 $= 2.0376$ ml/min/Kg.  However, during low intensity exercise, the Subject S4 achieved a SS in 690 seconds, T$_{HL}$$~=~ 90 seconds$, which is equivalent to TS to achieve SS with RMSE2 $= 2.0297$ ml/min/Kg, which is less than a value of RMSE1. Thus, the steady state control of VO$_2$ using a SBF system for the Subject S4 is adequate for both the low and high intensity phases of a rowing exercise.

\newpage
\textbf{Subject S5 }

\textbf{Cycling}\\
During the high intensity phase of cycling exercise, the Subject S5 achieved a SS in 260 seconds with the RMSE1 $ =0.6187$ ml/min/Kg, as shown in Fig. \ref{Subjects15}. However, during low intensity phase of an exercise, the Subject S5 achieved a SS in 780 seconds. The T$_{HL}$ was about to 180 seconds, which is greater than TS to achieve a SS with the RMSE2 $= 0.3086$ ml/min/Kg, and less than a value of RMSE1. These results indicated that the SBF system performed well for the Subject S5 during the low intensity phase of a cycling exercise.

\textbf{Rowing}\\
During the high intensity phase of rowing exercise, the Subject S5 achieved a SS in 110 seconds, as shown in Fig. \ref{Subject5} with the RMSE1 $= 1.9135$ ml/min/Kg.  However, during low intensity exercise, the Subject S5 achieved SS in 770 seconds, T$_{HL}$  $= 170 seconds$, which is less than a value of TS to achieve SS with the RMSE2 $= 0.7727$ ml/min/Kg. This RMSE2 is less than RMSE1. These results indicated that the steady state control of VO$_2$ using SBF system performed well for the Subject S5 during the low intensity phase of a cycling exercise.\\
\newpage
\textbf{Subject S6 }

\textbf{Cycling}\\
During the high intensity phase of cycling exercise, the Subject S6 achieved a SS in 185 seconds with RMSE1 $=1.6425$ ml/min/Kg, as shown in Fig. \ref{Subjects16}. The TS to achieve SS is the longest duration amongst all subjects during cycling and rowing exercises. However, during the low intensity phase of an exercise, the Subject S6 achieved a SS in 728 seconds. The T$_{HL}$ was about 128 seconds, which is greater than TS to achieve SS with the RMSE2 $= 1.1202$ ml/min/Kg, which is less than a value of RMSE1.  These results indicated that a steady state control of VO$_2$ using SBF system performed well for the subject the Subject S6 during the low intensity phase of a cycling exercise.\\
During the high intensity phase of rowing exercise, the Subject S6 achieved a SS in 140 seconds, as shown in Fig. \ref{Subject6} with the  RMSE1 $= 1.0235$ ml/min/Kg.  However, during low intensity exercise, the Subject S6 achieved SS at t=740 seconds, T$_{HL}$  = 140 seconds, which is less than TS to achieve a SS with the RMSE2 = 0.9734  ml/min/Kg, which is less than a value RMSE1. These results indicate that a steady state control of VO$_2$ using SBF system were performed well for   the Subject S6 during the low intensity phase of a cycling exercise.

Finally, the analyses of the results indicate that all the subjects performed well during low intensity phase in both types of exercises.  However, the transient response was not properly controlled using the SBF control of VO$_2$ in both types of exercise during onset and offset activity. In fact,  the control  of own physiological parameter, based on the biofeedback, is highly dependent on subject psychological factors. Therefore, it can be concluded that the use of SBF system to control VO$_2$ is more efficient in steady state only for low intensity phase of the exercises.
\section{Discussions}\label{chap5discussion}
The first part of this chapter presented a method for VO$_2$ estimation for predicting its transient and steady state behaviour accurately during rhythmic exercises. The developed method of VO$_2$ estimation could be easily applied against the measurement of VO$_2$. The proposed method was successfully developed and implemented in real-time. This development was based on the fact that (i) the ER represents the intensity of the rhythmic exercise, (ii) VO$_2$ varies with the ER until the subject achieves his/her VO$_{2max}$\cite{10,110}, (iii) HR also varies until it achieves HR$_{max}$. Based on these facts, VO$_2$ estimator uses the ER, $\Delta$RespR and $\Delta$\%HR$_{max}$ as the input, and $\Delta$\%VO$_2$ as an output. It appears that the average quality of fit gets better with the increase in exercise intensity as reported in the literature \cite{Smolander2008, Smolander2011}.  However, our proposed method of VO$_2$ estimation, using Hammerstein model, is unique in the sense that it can predict the dynamic and steady state responses of VO$_2$ during exercises of moderate and high intensity very accurately, in contrary to the HR-HRV method; see \cite{Smolander2011}. Furthermore, our method yields an adequate estimate of VO$_2$ during low intensity exercises and it does not require any calibration of the estimator in contrary to the TRAD(HR-VO$_2$) curve calibration method; see \cite{Spurr1988,Ceesay1989}.

In this study, the $\Delta$\%VO$_2$ is proposed as a unique variable for predicting the VO$_2$ during exercise. The proposed estimator can cater for subject to subject variations for predicting the VO$_2$ but cannot estimate VO$_2$ with changes in the physical activity like switching from cycling to rowing exercise.  Real-time experimental results proved that the developed VO$_2$ estimation method can be used as a good plate platform for designing a control system without taking the measurements of VO$_2$ in its feedback loop. Since our developed estimation approach uses the estimate of
VO$_{2max}$ and VO$_{2rest}$, it may be possible that the actual measurements of these quantities could improve the performance of the developed methodology further. The validation study of the designed estimator shows that the response of VO$_2$ is varied amongst the subjects while regulating the VO$_2$ using the SBF. Therefore, the VO$_2$ transient response is not effectively controlled by the SBF system.  However, the steady state control of VO$_2$ is effectively achieved by using the SBF system. In addition, the safety and efficiency of an exercise prescription can be improved further, possibly in future research, by using a model based VO$_2$ controller capable of predicting the desired body movements,  i.e., ER$_T$ to achieve the targeted VO$_2$ profile.
\section{Concluding Remarks}\label{chap5con}
The proposed method estimates VO$_2$ by using non-invasive and easily measurable quantities such as HR, RespR and HR using the prior information about the individual parameters, i.e., VO$_{2rest}$ and  VO$_{2max}$.The designed estimator predicts  the steady state and dynamic behaviour  of the VO$_2$ very accurately  without the need of any expensive individual calibration of the estimator in contrary to traditional methods. The developed method of estimation of VO$_2$ is useful for control applications and it can also be used for the measurement of VO$_2$. The $\Delta$\%VO$_2$ is found as a unique variable for predicting VO$_2$ during exercises specifically to cope with the variations among the subjects.  Since the performance of the developed VO$_2$ estimator is limited only for one type exercise, an estimation based self-biofeedback system is developed and validated in real-time. The analysis of the present study shows that the transient response of VO$_2$ varies amongst the subjects irrespective of low or high intensity exercises. These results show that the VO$_2$ transient response is not controlled by the SBF system effectively and its efficiency is only limited to steady state control. The comparison between low and high intensity exercises show that the performance of the steady state control using the SBF system is more efficient than for low intensity phase of the exercise. Based on this analysis, the future research should focus on the design of a robust controller for regulating VO$_2$ during cycling and rowing exercises.
\chapter{Conclusion and Future Directions}
The research work described in this report is divided into following two main categories:
\subsection{Concluding Remarks for HR Regulation System}
 The first part demonstrated the modelling and control of human HR regulation system for an unknown rhythmic exercise. The experimental findings showed that the relation between ER and HR was different for walking, cycling and rowing exercises. Also, the behaviour of HR varied due to interpersonal variability. Based on these observations, the universal model structure was obtained using the AIC model selection criteria.
The AIC suggested that the LTI 2$^{nd}$-order model structure was suitable for low- and moderate-intensity exercise. However, only four cases of high-intensity exercise suggested the $p=3$ LTI model.  Because we required a universal structure that could model the HR response during low- and high- intensity exercises. Therefore, the model structure $p=2$ was selected for all LTI models at different intensity levels during exercises. This 2$^{nd}$-order LTI model has four parameters: $a_1$, $a_2$, $b_1$ and $b_2$. These were identified for 2 subjects using a simple LS algorithm. The results show that the parameters $a_1$ and $a_2$ do not change significantly (SD = 0.32, SD = 0.263, respectively) among subjects and type of exercise as compared with parameters $b_1$ and $b_2$ (SD = 9.559, SD = 8.336, respectively). This suggested that parameters $b_1$ and $b_2$ are more sensitive to changes in subject and exercise type.  Moreover, the LTI model gives better HR response during low- and moderate-intensity exercise, irrespective of the type of exercise. During high-intensity exercise the HR response varies with time because of the cardiovascular drift. Hence, a 2$^{nd}$-order LTV  model structure was used to model the HR response during walking, cycling and rowing exercises.
The LTV model parameters were estimated using RLS based on a KF as an adaptation gain. The estimated LTV models for each exercise type at each intensity level were compared with LTI models for subjects S1 and S2. In both cases the coefficient of correlation between the estimated and measured values of $\Delta$ HR were improved. These results indicate that the LTV model for each intensity level gives a better description of the HR response than the LTI model.
Therefore, the control of HR response during unknown rhythmic activity is possible by using the adaptive control approach. The literature on the physiological control of HR also indicated that an indirect adaptive H$_\infty$ control approach is suitable as a control approach for the development of an HR control system during an unknown type of rhythmic activity. This proposed approach only requires the measurement of the HR and ER for estimating the ER$_T$ as a control input. An indirect adaptive H$_\infty$ control approach was successfully applied and simulated by using the 2$^{nd}$-order LTI  model structure using an individual switch model that switches the model parameters based on ER$_T$. The simulation study of the subjects S1 and S2 showed that the proposed control approach can estimate ER$_T$ in a desired bandwidth which accommodates the body biomechanics/ bandwidth of ER in the required limits and also copes with subject-to-subject variations.

Simulation results show that the average RMSE between the HR$_T$ and HR simulated of two subjects is calculated 5.58 bpm, 6.52 bpm  and 3.5 bpm for a walking, a cycling and a rowing, respectively. Some of the cases of walking, cycling and rowing experienced the average approx. overshoot of 2.52\% and 3.12\% while tracking the HR$_T$ profile. Hence, the overshoot within the range of 0 to 4.5\%  and RMSE between 0 to 6.52 bpm as a steady state error is considered as the acceptable range for the purpose of the validation in the real-time HR regulation system.

The designed and simulated indirect adaptive H$_\infty$ control for HR regulation was implemented and validated in real time. The HAS was integrated with the developed control approach and used to communicate the ER{$_T$} to the subjects. The developed HR control system was tested with six healthy subjects performing cycling and rowing exercises. The participating subjects have different physical characteristics which vary the controller demand i.e., ER$_T$. The developed indirect H$_\infty$ control approach tracked the physical variations amongst the subjects and the types of exercise to achieve the HR$_T$. The real-time experimental study shows that the ER$_{T}$ was closely followed by the exercising subject using the HAS and achieved the HR$_{T}$ with average root mean square (RMS) tracking error of 3.186 bpm for cycling and 2.934 bpm for a rowing. These results illustrate that the developed system can successfully track the changes in HR to a reference profile during any type of rhythmic exercise activity. The controller can adapt smoothly to changes in the set point of HR. To assess how well the human actuation system (HAS) worked by using the beep signal, the correlation coefficient $R$ was calculated between the ER$_{T}(t)$ and ER$_{est}(t)$ . A high value of $R$ indicated that all subjects had responded well to follow the ER$_{T}(t)$, with an average correlation coefficient of 0.914 for a cycling and  0.842 for a rowing exercise. However, the transient response of HR varied from subject to subject; some achieved HR$_T$ in less than 200 seconds while some required more than 200 seconds. Moreover, some cases of the human subjects experienced the overshoot within acceptable range during cycling and rowing exercises. However, steady state was maintained through the course of exercise irrespective of the type of exercise.

As the real time experimental results during HR control and simulated results show the comparable performance in all cases, it can be  concluded that the developed system for HR control during unknown type of rhythmic activities achieves the desired performance in  real-time.
In summary, the major findings of the first part are:
 \begin{enumerate}
 \item{Analyses of the experimental data concluded that the walking, cycling and rowing exercises exhibit different HR response in terms of steady state and transient response. Moreover, the HR response varies due to subject's physical characteristic, i.e., age, weight and height. Therefore, it is impossible to obtain a single stationary ER-HR model that can accommodate multiple rhythmic activities. Based on these observations, this type of the modelling problem is considered as known model structure and unknown model parameters.}
\item{The 2$^{nd}$-order LTI structure was found to be suitable  for modelling HR response during various type of activities for low- and moderate-intensity exercises. However, the order of the LTI model gets higher for high intensity exercises  because HR varies due to cardiovascular drift which produces further delay to reach the steady state and hence the performance of 2$^{nd}$-order LTI model gets severely affected.}
\item{The obtained 2$^{nd}$-order LTI universal structure with unknown time varying parameters was considered LTV  model. It was found that the estimated response of HR using the LTV model gives a better description than that of the LTI model at all intensity levels for walking, cycling and rowing.
This approach to parameter estimation does not require any information about the type of rhythmic activity. It only requires the measurement of ER and HR. Therefore, we found that the HR response can be modelled in real-time by using a 2$^{nd}$-order LTV model structure and an indirect adaptive control approach  can be used to control HR during unknown type of rhythmic activity.}
\item{An indirect adaptive H$_\infty$ control approach was successfully simulated and applied for rhythmic exercise using a switch model for HR regulation during walking, cycling and rowing. The simulation study indicated that indirect adaptive control can cope with the variations of the model parameters.}
\item{To help the subject follow the controller demand, i.e., ER$_T$ in real-time, a novel human actuating system (HAS) has been investigated and developed. It hhas been observed that this type of actuating system does not communicate the actual measurement of HR to the subject, thereby prevents the HR response from any variations due to psychological factors, i.e., motivation and perception.}
\item{The experimental validation of the HR regulation system shows that all subjects achieve the ER$_T$ successfully, and consequently the HR$_T$ as well by following the HAS.  However, the transient behaviour of the HR varies from subject to subject. However, most importantly all subjects faithfully achieved the steady state value of the HR$_T$.}
  \end{enumerate}

\subsection{Concluding Remarks for VO$_2$ Estimation and Control}
This section of our research work is further divided into two parts: In the first part of the research work, we developed a new method for estimating VO$_2$, and experimentally verified that it could serve the purpose of an estimator design. The VO$_2$ estimation methodology relies on the ER representing the intensity of any rhythmic exercise. It utilizes the concept that the VO$_2$ varies with the exercise intensity until the subject achieves his/her VO$_{2max}$\cite{10,110}. Similarly for HR, the HR also varies until it achieves HR$_{max}$. Based on this observation, we developed a Hammerstein model to estimate VO$_2$. Its inputs are ER, $\Delta RespR$ and $\Delta$\%HR$_{max}$, and its output is $\Delta$\%VO$_2$. These input and output datasets were obtained for six subjects during cycling and rowing and the Hammerstein model was identified for these exercises by merging the datasets of all exercising subjects. The obtained exercise-dependent models were then evaluated on the individual subjects. It appears that the average quality of modelling for both types of exercise improves as the exercise intensity increases. Similar findings were reported by other researchers \cite{Smolander2008, Smolander2011}.  However, our new method for estimating VO$_2$ using a Hammerstein model can accurately predict the dynamic and steady state behaviour of VO$_2$ during exercise of moderate and high intensity.  The developed VO$_2$ estimation method does not require individual calibration of the estimator like the TRAD(HR-VO$_2$). In \cite{Smolander2011}, the HR-HRV method was tested in free living conditions. The pooled regression analysis showed that the estimated VO$_2$ values accounted for the variability in the actual VO$_2$. At group level, the HR-HRV method underestimated the measured VO$_2$ slightly. While some of the estimated VO$_2$ values at lower exercise intensities were markedly underestimated, the performance of this method improved during moderate and high intensities of an activity. Thus, this method cannot be considered sufficiently accurate to determine the actual steady state value of VO$_2$ as compared to the TRAD(HR-VO$_2$) method. From a control application point of view these methods are therefore not suitable to apply against the measurement of VO$_2$ during exercise. On the other hand, the HR-HRV method can cope with an imprecise estimation of the steady state values of VO$_2$. \cite{Su2007C} developed a Hammerstein estimator for oxygen uptake (oxygen consumption) during moderate treadmill exercise. To obtain the Hammerstein VO$_2$ estimator, six healthy male subject participated in the experimental studies. The results obtained showed that the relationship between steady-state VO$_2$ and treadmill speed is nonlinear, which was successfully modelled using support vector regression methods. This type of model is applicable to the design of a controller to regulate VO$_2$ during treadmill walking exercise but the accuracy of the estimation is not always acceptable. The estimator developed in this work has been shown to accurately predict the steady state and dynamic behaviour of the VO$_2$ without the need for expensive individual calibration. $\Delta \%VO_2$ is proposed as a unique variable for predicting VO$_2$ during exercise. The proposed estimator can cope with subject-to-subject variation in predicting the VO$_2$ but cannot estimate absolute VO$_2$ with changes in physical activity.
 The developed estimators for cycling and rowing were implemented in real-time for the development of the SBF control of VO$_2$ for cycling and rowing exercise. Real-time experimental analyses proved that the developed estimation approach for VO$_2$ served as a good platform for designing the control system without considering the measurements of VO$_2$ in feedback. The results obtained also show that the estimated transient response of VO$_2$ was close to the actual measurements, and the steady state values were also estimated adequately during low-and high-intensity exercise.
The second part of this project treated the performance of the SBF control of VO$_2$ based on its estimate. Our analysis shows that the dynamic response of VO$_2$ varies amongst subjects for both low- and high-intensity exercise. Thus, the VO$_2$ transient response is not effectively controlled by the SBF system.  From the VO$_2$ regulation point of view, the SBF control of VO$_2$ is more efficient for low-intensity exercise in which humans can easily achieve the target profile with their body movements.

In conclusion, the major findings of the second part of this report are as follows:
\begin{enumerate}
\item{The developed approach predicts percentage $\Delta VO_2$ using the Hammerstein model and the output \% $\Delta VO_2$ is converted into a VO$_2$ estimate using the individual estimate of the VO$_{2rest}$ and VO$_{2max}$. This allows the estimation of VO$_2$ to be performed without expensive calibration of the estimator.}
\item{The efficiency of the estimation approach was validated in real-time using the estimation-based self-biofeedback control system for VO$_2$ during cycling and rowing. The experimental results show that the real-time estimate of VO$_2$ accurately estimated the dynamic and steady state behaviour of VO$_2$.}
\item{The SBF control of VO$_2$ during low- and high-intensity cycling and rowing exercise was also analysed in this part of the report. We showed that the self-biofeedback control of VO$_2$ is more efficient during low-intensity exercise irrespective of the type of exercise.}
  \end{enumerate}

\section{Future Research Directions}
The future research directions are also divided into following two distinct dimensions:
\subsection{HR Regulation System for Rhythmic Activities}
The possible future research in the area of HR control during rhythmic exercises is listed as follows:
\begin{enumerate}
\item{The efficiency of LTV model can be studied further with the increase in the number of subjects to observe the global tendency of the model. The modelling of HR response can be to improved using the robust KF approach \cite{Savkin1995, Savkin1996b, Savkin1997, Savkin1998, Petersen1999, Pathirana2005}.}
\item{The developed HR regulation system can be applied to regulate HR at the constant level of HR$_T$ while engaging the subject in multiple exercising activities as suggested by the American College of Sports and Medicine \cite{Balday2000}. Therefore, the developed control methodology opens the doors for gaited and arthritic patients to perform their exercises effectively by allowing the body joints with enhanced capability without experiencing any fatigue.}
\item{The experimental analyses show that all subjects achieved the desired profile. However, their transient behaviour varied with the type of exercise, and showed inter-subject variability. In future, the efficiency of the developed control approach can be improved further by using advanced techniques for robust control of uncertain systems with structured uncertainties \cite{Savkin1995a,Savkin1999}.}
\end{enumerate}
\subsection{VO$_2$ Estimation and Control for Rhythmic Activities}
The possible future directions for the estimation and control of VO$_2$ during rhythmic activities are summarised as follows:
\begin{enumerate}
\item{In our research, we developed a methodology using the Hammerstein system to estimate VO$_2$ during rhythmic exercise, the efficiency of the designed estimator can be further improved using robust KF estimation \cite{Savkin1995, Savkin1996b, Savkin1997, Savkin1998, Petersen1999, Pathirana2005}. This approach can easily cope with parameter variations during rhythmic exercises.}
\item{The estimators developed for cycling and rowing are exercise dependent. Therefore, a possible future research direction will be to investigate a VO$_2$ predictor independent of the type of rhythmic exercise, as ER is a versatile measure of exercise intensity \cite{Cheng2009}.}
\item{The estimation was based self-bio-feedback control of VO$_2$ for cycling and rhythmic exercise. The performance of the SBF system is intensity dependent; during low-intensity exercise the SBF system was more efficient, irrespective of the type of exercise. To improve the safety and efficiency of an exercise prescription, the future research should focus on the control of VO$_2$ based on these estimates using the robust control approaches \cite{Su2007a, Cheng2008} and a model-predictive control approach \cite{Su2010}.}
\item{The developed estimation approach uses the estimate of VO$_{2max}$ and VO$_{2rest}$. Actual measurements of these quantities can further improve the estimation.}
\end{enumerate}

\bibliographystyle{IEEEtran}
\bibliography{References6}
\end{document}